\documentclass[11pt]{article}
\pdfoutput=1

\usepackage{jheppub}
\usepackage{amsmath,amssymb,amsfonts}
\usepackage{amsthm}
\usepackage{breqn}

\usepackage{esint}
\usepackage{graphicx,tikz}
\usetikzlibrary{decorations.pathreplacing}
\usetikzlibrary{decorations.pathmorphing}
\usetikzlibrary{arrows}
 \usepackage{epstopdf}
\usepackage{framed}
\usepackage{xcolor}
\colorlet{shadecolor}{gray!20}
\usepackage{mathtools}
\usepackage{physics}
\usepackage{wrapfig}
\usepackage{enumitem}

\usepackage{mathrsfs}
\usepackage{setspace}
\definecolor{hyperref}{RGB}{026,028,087}

\usepackage{placeins}
\numberwithin{equation}{section}

\usepackage{bbm}
\renewcommand{\vec}[1]{\mathbf{#1}}

\newcommand{\sgn}{\operatorname{sgn}}

\newcommand{\cpi}{c_{{\pi}}}

\definecolor{light-gray}{gray}{0.95}

\renewcommand{\cpi}{c_s}
\newcommand{\bfk}{\mathbf{k}}

\usetikzlibrary{patterns}
\usetikzlibrary{shapes}
\usetikzlibrary{arrows,shapes,positioning}
\usetikzlibrary{decorations.pathmorphing}
\usetikzlibrary{decorations.markings}
\tikzset{cross/.style={cross out, draw=black, minimum size=2*(#1-\pgflinewidth), inner sep=0pt, outer sep=0pt},
cross/.default={1pt}}

\tikzset{zigzag/.style={decorate, decoration=zigzag}}

\tikzstyle arrowstyle=[scale=1]
\tikzstyle directed=[postaction={decorate,decoration={markings,
    mark=at position .65 with {\arrow[arrowstyle]{stealth}}}}]
\tikzstyle reverse directed=[postaction={decorate,decoration={markings,
mark=at position .65 with {\arrowreversed[arrowstyle]{stealth};}}}]

\usepackage{tikz-feynman}
\tikzfeynmanset{compat=1.0.0}

\begin{document}

\title{Inflation in motion: \\
unitarity constraints in effective field theories with (spontaneously) broken Lorentz symmetry
}

\author[a]{Tanguy Grall}
\affiliation[a]{Department of Applied Mathematics and Theoretical Physics, University of Cambridge, Wilberforce Road, Cambridge CB3 0WA, U.K.}

\author[a,b]{and Scott Melville}
% \affiliation[b]{DAMTP, University of Cambridge, Wilberforce Road, Cambridge CB3 0WA, U.K.}
\affiliation[b]{Emmanuel College, University of Cambridge, St Andrew’s Street, Cambridge CB2 3AP, U.K.}

\emailAdd{tg418@cam.ac.uk, scott.melville@damtp.cam.ac.uk}

\abstract{
During inflation, there is a preferred reference frame in which the expansion of the background spacetime is spatially isotropic. 
In contrast to Minkowski spacetime, observables can depend on the velocity of the system with respect to this cosmic rest frame. 
We derive new constraints from radiative stability and unitarity on effective field theories with such spontaneously broken Lorentz symmetry. In addition to a maximum energy scale, there is now also a critical velocity at which the theory breaks down. The theory therefore has different resolving power in time and in space, and we show that these can only coincide if cubic Lorentz-violating interactions are absent. 
Applying these bounds to the Effective Field Theory of Inflation, we identify the region of parameter space in which inflation can be both single-field and weakly coupled on subhorizon scales.  
This can be implemented as a theoretical prior, and we illustrate this explicitly using Planck observational constraints on the primordial bispectrum.
}

%\keywords{EFT, Inflation, Unitarity, Power Counting}

\maketitle

%\newpage
%\setcounter{tocdepth}{2}
%\tableofcontents

%%%%%%%%%%%%%%%%
\section{Introduction}
%%%%%%%%%%%%%%%%

Unitarity is a fundamental pillar of quantum field theory. 
The requirement that probability amplitudes correctly normalise is essential if we are to make sense of our theoretical computations. 
Studying the restrictions that this places on a physical theory has proven invaluable in guiding past efforts to construct viable models of the unknown---for example constraining the pion cross section before the development of QCD \cite{Chew, Eden}, and bounding the Higgs mass before the LHC \cite{Lee:1977eg}. 
 In cosmology, we are faced with the challenge of modelling the earliest moments of our Universe. Can we similarly leverage unitarity to guide our efforts, and improve our understanding of the physics responsible for inflation?

The Effective Field Theory (EFT) approach to inflation \cite{Cheung:2007st} provides a model-independent framework with which to analyse the non-Gaussianities produced during inflation.
Rather than specify the matter content (e.g. a particular scalar potential) responsible for an FLRW expansion in the early Universe, instead it describes fluctuations about this background from the point of view of the spontaneously broken Lorentz symmetry---this allows the EFT of Inflation to capture a wide variety of different theories which all share the same background and linear cosmology, without committing to a particular underlying (UV complete) model. 
However, this also means that existing (Lorentz-invariant) techniques from particle physics, including the usual implementation of unitarity, cannot be applied directly to the EFT of Inflation. 
As a step towards implementing unitarity in the EFT of Inflation, in this work we study the scattering of sufficiently subhorizon modes (whose propagation is not affected by the expanding spacetime, but whose interactions need not be Lorentz-invariant), and are able to identify the region of parameter space in which this scattering is unitary.

Our main result is the construction of unitarity bounds for EFTs with broken boosts, using a new partial wave expansion for $2 \to 2$ scattering which accounts for a non-zero centre-of-mass velocity. We also show how simple scaling arguments, from the power counting of loops and the optical theorem for $n \to n$ scattering, can be extended to include the effects of broken boosts. This equips us with the necessary tools to analyse the scattering of subhorizon modes during inflation, and we identify the region of parameter space in which subhorizon physics can be approximately single-field (and weakly coupled), which can be compared with Planck's observational constraints on primordial non-Gaussianity.

%%%%
\subsubsection*{Inflationary Correlators}
%%%%
Metric fluctuations about an FLRW background, $ds^2 = -dt^2 + a^2 (t) d \mathbf{x}^2/ c_T^2$ (where $c_T$ is the tensor sound speed\footnote{
Rather than work in units in which $c_T = 1$, we will keep factors of $c_T$ explicit so that dimensional analysis can be carried out in time and space separately. 
}), can be described at leading order in derivatives by the EFT \cite{Cheung:2007st}, 
\begin{align}
S_{\rm LO} [\delta g^{\mu\nu} ] = 
\int d^4 x \sqrt{-g}  \Bigg\{ &
\frac{M_P^2}{2} \left(  R - 6 H^2 \right) + M_P^2 \dot H \left(  - 2  +   \delta g^{00}  \right)  +  \sum_{n=2}^{\infty} \frac{M_n^4}{n!} ( \delta g^{00} )^n   
\Bigg\} \, ,
\label{eqn:SLO1}
\end{align}
where the scales $M_0$ and $M_1$ have been fixed so that the background is stable. 
%The coefficients on the first line have been fixed in terms of the background scales $H$ and $\dot H$, and in terms of $c_s = c_\pi / c_T$ (the ratio of the scalar sound speed to the tensor sound speed). 
Since temporal diffeomorphisms have been spontaneously broken, $\delta g^{\mu\nu}$ now propagates a scalar mode, $\pi$ (in addition to the usual tensor modes of General Relativity). 
%The dynamics of this scalar at quadratic order is completely fixed by the symmetry breaking up to an overall scale, much like the effective chiral Lagrangian for pions. 
The decay constant $f_\pi$ associated with this symmetry breaking is set by $M_1$, namely $f_\pi^4 = 2 c_s M_P^2 | \dot H|$, while the $\pi$ sound speed is set by $M_2$, namely $M_2^4 = \frac{1}{4} f_\pi^4 (1-c_s^2 ) / c_s^3$,  where $c_s = c_\pi / c_T$ is the ratio of the scalar sound speed to the tensor sound speed. 
%it propagates with a sound speed $c_\pi$ determined by $M_2$, and has a decay constant $f_\pi^4 = 2 c_s M_P^2 | \dot H|$ associated Noether current.
$\pi$ is related to the usual scalar curvature perturbations, $ \zeta = -H \pi/f_\pi^2$, whose power spectrum and higher order correlations seed the structure observed in the CMB.  

%This is made manifest by introducing a St\"{u}ckelberg field, $\pi$, which non-linearly realises temporal diffeomorphisms, producing a new action $S_{\rm LO} [ \delta g^{\mu\nu} , \pi ]$ with the symmetry, $t \to t + \xi$ and $\pi \to \pi - \xi f_\pi^2$. 
%The decay constant associated with this symmetry is $f_\pi^4 = 2 \cpi M_P^2 | \dot H|$. 
%
Although $\pi$ mixes non-trivially with $\delta g^{00}$, if we focus on scalar modes with sufficiently large energies they decouple from the tensors,   
\begin{align}
\text{Decoupling Limit:} \quad
 \frac{\omega}{f_\pi}  \gg  \frac{f_\pi}{M_P} \qquad \qquad ( \text{equivalently,} \;\; M_P \to \infty \;\; \text{with} \;\;  f_\pi \;\; \text{fixed} ).  
 \label{eqn:DL_intro}
\end{align}
In this limit, the leading order interactions for the scalar perturbations produced during inflation can be written succinctly in terms of $Z_{\pi \pi} = Z^{\mu\nu} \partial_\mu \pi \partial_\nu \pi$, 
\begin{align}
S_{\rm LO} [ \pi ] = \int d^4 x \, \sqrt{-Z}  \; \left\{ 
-\frac{1}{2} Z_{\pi \pi} 
+ \frac{ \alpha_1 }{ f_\pi^2 }   \;  \dot \pi^3 
- \frac{ \alpha_2 }{ f_\pi^2 }  \; \dot \pi  Z_{\pi\pi}   
+ \frac{ \beta_1 }{ f_\pi^4 }  \; \dot \pi^4
- \frac{ \beta_2 }{  f_\pi^4 }   \; \dot \pi^2  Z_{\pi\pi}  
+  \frac{ \beta_3 }{ f_\pi^4 }   \; Z_{\pi\pi}^2  
\right\} 
\label{eq:EFT_action_decoupling_intro}
\end{align}
where $Z^{\mu\nu} = \text{diag} ( -1 , c_\pi^2 / a^2 \delta^{ij}  )$ is the kinetic matrix for fluctuations that propagate with sound speed $c_\pi$ on the FLRW background, and the coefficients $\{ \alpha_i , \beta_i \}$ are fixed in terms of $c_s$ and the scales $\{ M_3, M_4 \}$ from \eqref{eqn:SLO1}. 
%There is also a small mass term, $m_\pi^2 \pi^2$, which is the same order as the time-dependence of the coefficients.
The decay constant $f_\pi$ has been used to normalise the interactions so that inflationary correlators are given by\footnote{
The factors of $H$ in the correlators can be most easily seen by switching to conformal time, in which the mode functions of $\pi$ scale as $\sim H/k^{3/2}$, $\sqrt{-g} \sim ( H \eta)^{-4}$ and $g^{\mu\nu} \sim (H \eta )^2$, so the interaction coefficients scale as 
$ \sqrt{-g} \, \alpha_i  \left( g^{\mu\nu} \right)^3 / f_\pi^2 \sim  \alpha_i H^2 / f_\pi^2$ and $\sqrt{-g} \, \beta_i  \left( g^{\mu\nu} \right)^4 / f_\pi^4 \sim  \beta_i H^4 / f_\pi^4$. 
},  
\begin{align}
 \label{eq:power_spectrum}
 &\langle \zeta_{\bfk_1} \zeta_{\bfk_2} \rangle_{\text{in-in}}' 
% &=  \frac{H^2}{f_\pi^4}  \langle \pi_{\bfk} \pi_{-\bfk} \rangle    
 \sim  \left( \frac{H^4}{f_\pi^4} \right) \, p_{\bfk_1}    \;\; , \;\;\\ 
  \label{eq:bispectrum}
 &\langle \zeta_{\bfk_1} \zeta_{\bfk_2} \zeta_{\bfk_3} \rangle_{\text{in-in}}' 
 %&=  \frac{H^3}{f_\pi^6}  \langle \pi_{\bfk_1} \pi_{\bfk_2} \pi_{-\bfk_1-\bfk_2} \rangle   
 \sim  \left( \frac{ H^4 }{f_\pi^4} \right)^2 \,  \alpha \, f_{\bfk_1 \bfk_2}   \;\;,\;\;   \\
   \label{eq:trispectrum}
& \langle \zeta_{\bfk_1} \zeta_{\bfk_2} \zeta_{\bfk_3} \zeta_{\bfk_4} \rangle_{\text{in-in}}'
%%&=   \frac{H^4}{f_\pi^8}  \langle \pi_{\bfk_1} \pi_{\bfk_2} \pi_{\bfk_3} \pi_{-\bfk_1-\bfk_2-\bfk_3} \rangle   
\sim  \left( \frac{ H^{4} }{f_\pi^4} \right)^3 \, \left(\beta g_{\bfk_1 \bfk_2 \bfk_3} + \alpha^2  \, h_{\bfk_1 \bfk_2 \bfk_3}\right)   \, ,  \, 
\end{align} 
where $p_{\bfk_1}, \;f_{\bfk_i \bfk_2}  , \; g_{\bfk_1\bfk_2\bfk_3}$ and $h_{\bfk_1 \bfk_2 \bfk_3}$ are known functions of the momenta \cite{Chen:2006nt,Cheung:2007sv,Huang:2006eha} (with an overall momentum-conserving delta-function removed).
%In terms of the original EFT coefficients, 
%\begin{align}
%\alpha_1 &= - \frac{1}{2} ( 1 - c_s^2 )^2 - \frac{4}{3} c_3     \;\; , \;\;  &\alpha_2 &= \frac{1}{2} ( 1-  \cpi^2) \;\;,  \;\; \nonumber \\
%\beta_1 &=  \frac{1}{8} (1 - c_s^2 )^3 + 2 c_3 (1 - c_s^2) + 2 c_4   \;\; , \;\; &\beta_2 &= \frac{3}{2} \alpha_1  + \frac{1}{2} (1-\cpi^2)^2  \;\; , \;\; \beta_3 = \frac{1}{8} ( 1- \cpi^2 ) 
%\end{align}
Measurements of the power spectrum (\ref{eq:power_spectrum}) fix $f_\pi = (58.64 \pm 0.33 ) H$ \cite{Aghanim:2018eyx}, while bounds on primordial non-Gaussianities, the bispectrum \eqref{eq:bispectrum} and trispectrum \eqref{eq:trispectrum}, place constraints on the coefficients $\alpha_i$ and $\beta_i$ respectively.
Our goal in this work is to assess under what conditions $S_{\rm LO} [ \pi]$ mediates unitary scattering amplitudes between $\pi$ quanta, and hence apply unitarity as a theoretical constraint on the coefficients $\alpha_j$ and $\beta_j$ to complement observational searches for primordial non-Gaussianity.

%%%%
\subsubsection*{Inflationary Amplitudes}
%%%%
There are a number of obstacles which prevent directly applying Lorentz-invariant amplitude techniques to inflation. Even working within the decoupling limit \eqref{eqn:DL_intro}, the expanding spacetime background spontaneously breaks the Poincar\'{e} symmetry which underpins our usual definition of a scattering $\hat{S}$-matrix (energy is no longer conserved, particle production can occur, plane waves are no longer well-defined asymptotic states, there is no known LSZ procedure to relate field correlators to in-out observables, etc.). 
Overcoming these issues would be a monumental task. 
Here, we propose to take but the first step beyond the usual Lorentz-invariant Minkowski spacetime setting. 
By focussing on \emph{subhorizon modes}, i.e. fluctuations whose time derivatives are much larger than any time derivative of the background, it is possible to define in-out scattering amplitudes. Formally, the FLRW mode function for an on-shell $\pi$ particle\footnote{
$\pi$ fluctuations propagating on an FLRW background are canonically quantised as \cite{Birrell:1982ix},
\begin{align}
 \pi (t , \mathbf{x} )  =  \int \frac{d^3 \mathbf{k}}{ (2 \pi a / c_\pi )^{3/2} }  e^{i \mathbf{k} \cdot \mathbf{x} } \left[   \hat{a}_{\mathbf{k}} f_{\bfk} (t) + \hat{a}_{- \mathbf{k}}^{\dagger} f_{-\bfk}^* (t)    \right]  \nonumber
\end{align}
where $[ \hat{a}_{\bfk} , \hat{a}_{\bfk}^{\dagger} ] = -i$ fixes the normalisation of the above mode function as $\sqrt{ \frac{\pi}{4 H} } e^{- i \pi \nu /2}$.
} is
$
 f_{\mathbf{p}} (t) \propto H_\nu^{(1)} \left(    c_\pi | \mathbf{p} | / a H  \right)
 $, a Hankel function of order $\nu$ ($= 3/2$ for a light scalar on quasi-de Sitter), but on sufficiently subhorizon scales the effects of the expanding spacetime become unimportant and $\pi$ behaves like a scalar on flat space, with plane-wave mode functions (that conserve energy and momentum). The $|\mathbf{p}|$ at which this happens is set by $\nu$, since $H_\nu^{(1)} ( z ) \sim   e^{ i z } / \sqrt{z}$ when $z \gg | \nu^2 - \frac{1}{4} |$ (up to an overall phase factor). When considering scattering over timescales $H t \ll 1$ (so that $c_\pi | \mathbf{p} |/a H = \text{const} - c_\pi | \mathbf{p} | t/a + ...$), a $\pi$ particle with this momentum is described by a plane-wave mode function, 
\begin{align}
\text{Subhorizon Modes:} \;\;\;\;  \frac{ c_\pi | \mathbf{p} |}{a H} \gg \left| \nu^2 - \frac{1}{4}  \right|  \;\;\;\;\;\;\;\; \Rightarrow \;\;\;\; \;\;\;\;   f_{\mathbf{p}} (t) \propto e^{ i c_\pi | \mathbf{p} | t /a}  \, . 
\label{eqn:subhorizon}
\end{align}
%where we have neglected $m_\pi^2 / H^2$ since it is slow-roll suppressed. 
These approximate plane waves will characterise our in- and out- states, and allow us to apply the LSZ procedure and compute scattering amplitudes as if on flat space. Throughout this work, we will refer to \eqref{eqn:subhorizon} as ``subhorizon scales'', i.e. the regime in which $\pi$ modes behave as approximately plane waves.

%%%
\begin{figure}[htbp!]
\centering
\includegraphics[width=0.75\textwidth]{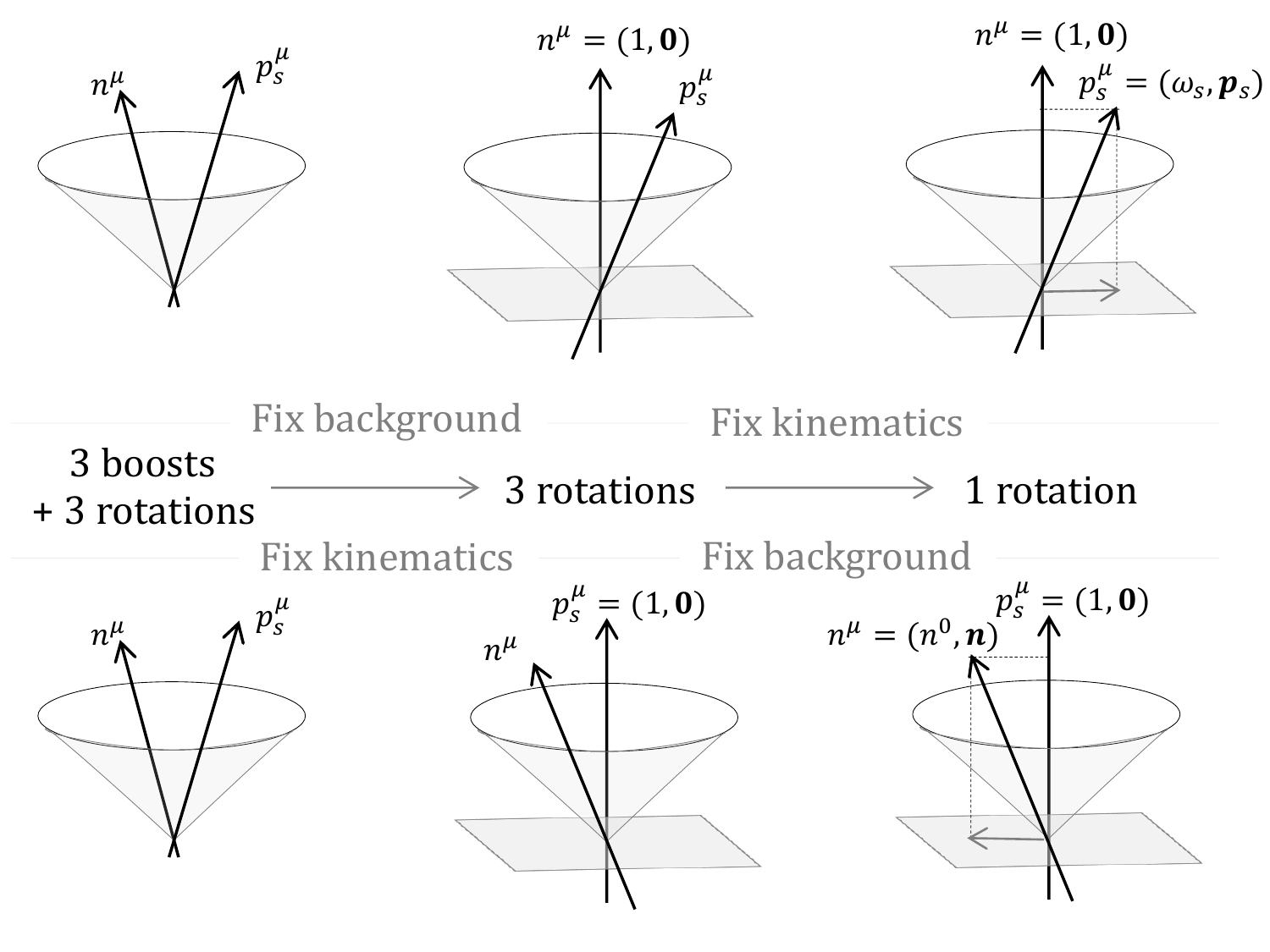}
\caption{Cartoon of the symmetry breaking. The scattering process can be described covariantly using two time-like vectors $n^\mu$ (the rest frame of the background) and $p_s^\mu$ (the center of mass motion). We then proceed along the top row, by first fixing our coordinates such that the background is at rest $n^\mu = (1, \mathbf{0})$ (removing $n^\mu$ in this way leaves a theory which is only manifestly invariant under spatial rotations), and then specifying the kinematics of the particles (the spacelike part of the CoM motion then breaks the 3 spatial rotations down to just 1). Since the underling physics is Lorentz invariant, a completely equivalent description is shown in bottom row, in which first one fixes coordinates such that the CoM is at rest, but at the price of now having a background spacetime which appears to expand anisotropically. In either case, once both the background and the CoM motion are fixed, there is only 1 rotational symmetry remaining.}
\label{fig:cosmic_motion}
\end{figure}
%%%

However with $f_\pi$ fixed this EFT retains the effects of symmetry breaking: in the decoupling limit \eqref{eqn:DL_intro} and for subhorizon modes \eqref{eqn:subhorizon}, the action $S_{\rm LO} [\pi]$ in \eqref{eq:EFT_action_decoupling_intro} describes a single scalar field on a flat background with (spontaneously) broken boosts. 
Unlike in Lorentz-invariant EFTs, the scattering amplitude may now depend on the Lorentz-frame in which it is evaluated. That is to say, once coordinates have been fixed so that the background expansion is isotropic (the cosmic rest frame), there is no longer any freedom to boost away the centre-of-mass motion---this is depicted in Figure~\ref{fig:cosmic_motion}.
As a result, the scattering amplitudes computed from $S_{\rm LO} [ \pi ]$ will depend on both the total incoming energy, $\omega_s$, \emph{and} the total incoming momentum, $\mathbf{p}_s$. For instance, the tree-level $2 \to 2$ amplitude has the form,
\begin{align}
 \mathcal{A}_{2 \to 2}^{\rm tree}  =  \frac{\omega_s^4}{f_\pi^4} \, F \left(  \frac{ c_\pi | \mathbf{p}_s | }{\omega_s}   \right) 
 \label{eqn:A22_intro}
\end{align}  
where $F$ is a dimensionless function of the ratio $\rho_s = c_\pi | \mathbf{p}_s |  / \omega_s$ (related to the velocity of the centre-of-mass, $\rho_s = v_{\rm CoM} / c_\pi$) and depends linearly on the coefficients $\beta_i$ and $\alpha_i \alpha_j$ in \eqref{eq:EFT_action_decoupling_intro}. 
We stress that this breaking of Lorentz symmetry occurs spontaneously at low energies due to the spacetime background, and does not correspond to any fundamental Lorentz-violation on small scales.
The precise question we wish to address is: when is $\mathcal{A}_{2\to 2}^{\rm tree}$ a viable (unitary) description of the subhorizon physics during inflation? Since this amplitude stems from the EFT of Inflation, which has assumed only that there is a single light degree of freedom (arising from the symmetry breaking), this is equivalent to asking: 
\emph{when is inflation approximately single-field (and weakly coupled) on subhorizon scales?}

To answer this question, we derive new unitarity constraints at a finite centre-of-mass velocity. The existing Lorentz-invariant implementation of unitarity (the partial-wave expansion) relies on using boosts to set $|\mathbf{p}_s|=0$, which can no longer be done for \eqref{eqn:A22_intro}---instead, we have developed a more general ``spherical-wave expansion'' which can accommodate $|\mathbf{p}_s| \neq 0$. Using these spherical-wave amplitudes, we identify the region of $\{ c_s, \alpha_1 \}$ parameter space in which the $S_{\rm LO} [ \pi ]$ interactions are unitary (for subhorizon modes in the decoupling limit). 
Since Lorentz-boosts are spontaneously broken, the EFT cutoffs in energy and momentum need no longer be the same (for instance the UV completion has states characterised by both a mass and a sound speed). 
In particular, there is a maximum energy and a maximum momentum, 
\begin{align}
 \omega_{\rm max} = 2 f_\pi \left(  \frac{30 \pi \, c_s^4}{ 1- c_s^2 }  \right)^{1/4}    \;\;\;\; , \;\;\;\;  c_\pi | \mathbf{p}_s |_{\rm max} =    2 f_\pi \left(  \frac{2 \pi }{ 15 \alpha_1^2}  \right)^{1/4} 
\end{align}
at which the EFT breaks down. 
Demanding that both of these cutoff are large enough for the EFT to resolve subhorizon scales \eqref{eqn:subhorizon}  can either be used as a theoretical prior---for instance assuming that subhorizon physics is single-field and weakly coupled improves the Planck $68 \%$ confidence interval on equilateral and orthogonal $f_{NL}$ by a factor of $\approx 3$---or can be used as a way to test with future data whether other light fields play an important role on subhorizon scales. 
In contrast to previous estimates of the strong coupling scale in the EFT of inflation \cite{Shandera:2008ai, Leblond:2008gg, ArmendarizPicon:2008yv, Baumann:2011su, Assassi:2013gxa, Cannone:2014qna, Adshead:2014sga} (see also \cite{Koehn:2015vvy, deRham:2017aoj}), we have studied the perturbative unitarity of 2-to-2 scattering with a non-zero $|\mathbf{p}_s|$, and found numerically precise expressions for the cutoff in both energy and momentum (which in general are very different). 

We also point out the curious feature that Lorentz-violating cubic interactions (such as $\dot \pi^3$) lead to a \emph{minimum} interaction energy, since otherwise their exchange contributions to $\mathcal{A}_{2\to 2}^{\rm tree} \sim \omega_s^6/s$ grow faster at small $s$ than is allowed by unitarity. For $\alpha_1 \dot \pi^3 / f_\pi^2$, scattering at fixed $|\mathbf{p}_s|$ leads to the unitarity requirement,
\begin{align}
s > s_{\rm min} = \frac{ | \mathbf{p}_s |^6 }{ f_\pi^4 }  \; \frac{25}{64} \, \frac{\alpha_1^2}{4 \pi}  \, . 
\end{align}
Since $s = \omega_s^2 - c_\pi^2 | \mathbf{p}_s |^2$, this means that it is only possible for the EFT to have comparable cutoffs in energy and momentum if $\alpha_1$ vanishes (or is made smaller than the $\pi$ mass and other slow-roll suppressed corrections which we have neglected).

%%%%
\subsubsection*{Loops and Power Counting}
%%%%
Since our analysis relies on studying the high-energy behaviour of amplitudes from $S_{\rm LO}$, it is essential that higher-order EFT corrections can be neglected. 
%Finally, in order to trust bounds obtained from the high-energy behaviour of $S_{\rm LO}$, one must make some assumption about the size of higher-order EFT corrections, which we will denote $S_{\rm HO}$. 
%Although the unitarity of $2 \to 2$ scattering (iii) is our main result, we will devote discussion to a systematic estimate of all loop diagrams (i) because it allows us to address the question of \emph{radiative stability}. 
%The leading order action $S_{\rm LO}$ is inevitably corrected by higher order terms, $S_{\rm HO}$.
The precise form and size of these corrections depends on the underlying UV physics, and cannot be determined within the EFT. 
However, given knowledge of $S_{\rm LO}$ (e.g. measurements of $c_s, \, \alpha_1, ...$), it is possible to place lower bounds on these corrections from \emph{radiative stability}---i.e. that quantum corrections to $S_{\rm HO}$ (in particular the running induced by loops of $S_{\rm LO}$) are at most an order one effect. Placing such bounds on $S_{\rm HO}$ is important because $S_{\rm LO}$ can only ever describe energy scales at which $S_{\rm HO}$ can be neglected, and in general it is not possible to simply tune $S_{\rm HO}$ to zero (over a wide range of scales) because they are regenerated by loops of $S_{\rm LO}$. 

We therefore complement our analysis of unitarity in theories with broken boosts with an exploration of radiative stability, and show how to power count the higher-order corrections to the EFT of Inflation. In particular, the background scale $H$ is separated from the decay constant $f_\pi$ by less than two orders of magnitude, so it may seem at first sight that describing subhorizon physics (processes with $\omega \gg H$) is not possible. However, the EFT cutoff (and the scale at which symmetry is restored) is related to $f_\pi$ by a factor of the field coupling, and in particular can be as large as $\Lambda^4 \sim 16 \pi^2 f_\pi^4 \sim ( 200 H )^4$. This is precisely the analogue of chiral perturbation theory ($\chi$PT) for pions, in which the decay constant ($f_\pi^{\rm pion} = 93$ MeV) is very close to the pion mass ($m_\pi^{\rm pion} \approx 140$ MeV), and so it is only possible to describe pion scattering consistently thanks to the fact that the chiral symmetry  breaking scale is $\Lambda_{\chi SB} = 4 \pi f_\pi  \sim 1.2$GeV \cite{Georgi:1989xy}
%, and so corrections like $\partial^2 / \Lambda^2_{\chi SB}$ are around $1\%$ at energies $\partial \sim m_\pi^{\rm pion}$ 
(see e.g. \cite{Manohar:2018aog} for further discussion of power counting in $\chi$PT).
Our discussion is complementary to the recent power counting developed in \cite{Adshead:2017srh, Babic:2019ify} (which treats fluctuations about a covariant theory)\footnote{
See also \cite{Colladay:1998fq, Kostelecky:2003fs, Kostelecky:2000mm} for an EFT construction of higher-dimension operators in the Standard Model when boosts are spontaneously broken.
}.

%%%
\subsubsection*{Synopsis}
%%%
In section~\ref{sec:2}, we derive constraints from radiative stability on the higher-order EFT corrections, beginning with the familiar Lorentz-invariant examples of a single scalar field and a massive vector field before moving on to our new analysis of a scalar with broken boosts and then the EFT of Inflation. 
In section~\ref{sec:3}, we relate the EFT interaction coefficients to the scale at which (perturbative) unitarity is lost, again providing a review of the well-known Lorentz-invariant case before turning to how this can be extended to theories with broken boosts, and finally comparing our results with the Planck constraints on $f_{\rm NL}$. 
We derive constraints from radiative stability and unitarity independently, so that sections \ref{sec:2} and \ref{sec:3} may be read in either order. 
In section~\ref{sec:disc} we summarise and discuss possible future work.

%%%%%%%%%%%%%%%%%%%%%%%%%%%%%%%%%%%%%%%%%%%%%%%%%%%%%%%%%%%%%%%%%%%%%%%%%%%%%%%%%%%%%%%%%%%%%%%%
%%%%%%%%%%%%%%%%%%%%%%%%%%%%%%%%%%%%%%%%%%%%%%%%%%%%%%%%%%%%%%%%%%%%%%%%%%%%%%%%%%%%%%%%%%%%%%%%

% 								POWER COUNTING

%%%%%%%%%%%%%%%%%%%%%%%%%%%%%%%%%%%%%%%%%%%%%%%%%%%%%%%%%%%%%%%%%%%%%%%%%%%%%%%%%%%%%%%%%%%%%%%%
%%%%%%%%%%%%%%%%%%%%%%%%%%%%%%%%%%%%%%%%%%%%%%%%%%%%%%%%%%%%%%%%%%%%%%%%%%%%%%%%%%%%%%%%%%%%%%%%
 
\FloatBarrier
%%%%%%%%%%%%%%%%
\section{Radiative Stability Bounds}
\label{sec:2}
%%%%%%%%%%%%%%%%

In order to construct a useful effective field theory, one must employ a consistent ``power counting scheme'' (a set of rules for deciding which operators are the most important) which is ``radiatively stable'' (preserved under quantum corrections). 
In particular, when we truncate our EFT and include only a particular set of leading order interactions, $S_{\rm LO}$, they will inevitably generate higher order corrections, $S_{\rm HO}$. 
In this section, we first review radiatively stable power counting schemes in Lorentz-invariant theories, and then discuss how these can be extended to theories in which boosts are broken, finally arriving at a consistent power counting scheme for the EFT of Inflation in which the interaction coefficients are naturally bounded in terms of the sound speed of scalar fluctuations, $\cpi$.

%%%%%%%%%
\subsection{EFTs with Lorentz Invariance}
%%%%%%%%%

We begin by briefly reviewing the constraints imposed by radiative stability in effective field theories with Lorentz invariance, building up from a single scalar field to a massive vector field. This provides a simple example of estimating loop corrections in a theory with a nonlinearly realised gauge symmetry (which parallels our approach to the EFT of Inflation in section~\ref{sec:23}). 

%%%%%%%%
\subsubsection*{Single Scalar Field}
%%%%%%%%

The first step in constructing an effective field theory action is to identify a basis of local operators built from the desired degrees of freedom and consistent with the desired symmetries. 
For example, to construct a Lorentz-invariant theory using a single scalar field with a shift symmetry, such a basis would be all possible Lorentz scalars built from the field and its derivatives,
%\footnote{
%Since the action (and on-shell correlation functions) is not affected by total derivatives or operators containing the classical equations of motion, it is sufficient to use a reduced basis where these have been discarded. 
%}:
\begin{equation}
 \mathcal{L} [ \phi \; , \;  \partial_\mu ]  = Z ( \partial \phi )^2  + \frac{ ( \partial \phi )^4 }{ (M_4)^4 } + \frac{ (\partial \phi )^6 }{ (M_6)^6 } + ... + \frac{ (\partial \partial \phi )^2 }{ ( M_2' )^2 } + \frac{ (\partial \partial \phi )^4 }{ ( M_4' )^8 } + ... 
 \label{eqn:egL}
\end{equation}
%where the high energy details have been absorbed into the various scales $M_i , \, ... $ etc. 
The second step is to consider which choices of the various scales, $\{ M_n, M_n' , ...\}$, are radiatively stable. This is important because while large hierarchies between the scales can seem acceptable classically, in the quantum theory these interaction coefficients run and hierarchies are typically washed out. For example, a loop of two $(\partial \phi )^4/M_4^4$ vertices requires a counterterm that schematically looks like $\sim (\partial \partial \phi )^4 /M_4^8 $, causing $M_4'$ to flow down to $M_4$ upon renormalisation---this means that once a $(\partial \phi)^4/M_4^4$ interaction is included in our theory, it is not possible to switch off higher order terms like $(\partial \partial \phi)^4$ (over a range of scales) since they are generated by loops.
 %Quantum mechanically the Wilson coefficients run. 
 
To systematically assess the impact of quantum corrections, it is convenient to adopt a particular ``power counting scheme'', in which every scale $\{ M_n, M_n', ... \}$ is replaced with a dimensionless (order unity) constant according to a set of rules: for example, in the well-known ``single-scale-single-coupling'' scheme\footnote{
This scheme is also known as SILH, after its original use for a ``Strongly Interacting Light Higgs'' \cite{Giudice:2007fh}. Setting $g_\phi  = 4\pi$ recovers the ``Naive Dimensional Analysis'' introduced in \cite{Manohar:1983md} (see also \cite{Georgi:1989xy, Cohen:1997rt, Jenkins:2013sda,Buchalla:2013eza}), and $g_\phi = 1$ is the traditional Weinberg power counting for EFTs \cite{Weinberg:1978kz}.  
} every derivative is suppressed by a single scale $\Lambda$ and every field is suppressed by a second scale $\Lambda_{\phi} = \Lambda/g_\phi$ (which can be written in terms of $\Lambda$ and a single dimensionless coupling $g_\phi$), such that  
%\begin{equation}
$S [ \phi ] = \int d^4 x \,    \Lambda^4 / g_\phi^2 \;  \mathcal{L}  \left[  g_\phi \phi / \Lambda \; , \;    \partial /  \Lambda  \right] \, , $
%\label{eqn:SILH}
%\end{equation}
where the overall scale in front of $\mathcal{L}$ is chosen to give a canonical kinetic term.  
Adopting any particular power counting scheme always represents a slight loss of generality from \eqref{eqn:egL}, since not all UV-complete theories produce an action of this form in the IR (e.g. single-scale-single-coupling only captures those characterised by a single heavy mass, $\Lambda$, and a single coupling strength $g$ between heavy and light physics),
but with the significant gain that any loop correction may now be systematically estimated. 
In the case of the single-scale-single-coupling scheme, all interactions are of the form $\mathcal{O}_{ab} = g_\phi^{a-2 }\partial^{2b} \phi^a / \Lambda^{a+2b-4}$ (multiplied by an order unity Wilson coefficient), and the $L$-loop diagram made from $V_{ab}$ such vertices produces a counterterm\footnote{
The total number of $\phi$ lines in such a graph is $\sum_{ab} a V_{ab} = 2 I + E$, where $I$ is the number of internal $\phi$ propagators and $E$ is the number of outgoing $\phi$ legs. Since the Euler formula for planar graphs gives the number of internal lines $I = L - 1 + \sum_{ab} V_{ab}$, this fixes $E$ in terms of $V_{ab}$ and $L$. In a regularisation scheme which respects the EFT counting (e.g. dimensional regularisation), the $\Lambda_n$ scales can arise only from each vertex factor. The typical size of each loop integral is $d^4 k /(2 \pi)^4 \sim 1/ (4\pi)^2$, and then finally the total number of derivatives, $D$, is determined by the total dimension of the operator being $4$. See e.g.  \cite{Burgess:2007pt} for a review.
% See e.g. \cite{Burgess:2007pt} for more details and \cite{Goon:2016ihr} for applications of this counting to non-renormalization theorems in various scalar field theories. We'll make repeated use of this power couting in what follows.  
},
\begin{align}
\Delta \mathcal{O}_{ E D}  &\sim   \left(  \frac{\Lambda^4}{g_\phi^2}  \right)^{\sum V_{ab} } \left(  \frac{ g_\phi  }{ \Lambda  \partial  }  \right)^{\sum a V_{ab} } \left(  \frac{\partial^2}{\Lambda^2}  \right)^{\sum b V_{ab}}  \left(  \frac{\partial^4}{16 \pi^2} \right)^L   ( \partial \phi )^{2 -2L + \sum V_{ab} (a-2) }  \nonumber  \\
&= \left( \frac{ g_\phi }{4 \pi}   \right)^{2L}  \frac{ g_\phi^{E-2} \partial^D \phi^E }{ \Lambda^{E + 2D - 4} }   \, .
\end{align}
In this power counting scheme, radiative stability (i.e. that running from loop corrections is at most an order one correction to the tree-level Wilson coefficient) requires that $g_\phi \lesssim 4 \pi$. 
This demonstrates something important: the scale suppressing the fields, $\Lambda/g_\phi$, cannot be made arbitrarily lower than the scale suppressing the derivatives, $\Lambda$, because a loop can trade $\phi$'s for derivatives. Note that since derivatives can never be turned back into $\phi$'s, a hierarchy like $\Lambda/g_\phi \gg \Lambda$ is perfectly consistent with radiative stability.   

The strategy which we shall adopt in this paper is to separate out the leading-order interactions of interest and treat them like \eqref{eqn:egL}, assuming nothing about their overall size or hierarchies (which is determined by the underlying UV physics), but then to include higher-order corrections for which we will assume a simple power counting scheme, determined by a single scale and a small number of dimensionless parameters (this is arguably the most agnostic one can be, since one must always assume something about the higher-order corrections if one is to trust leading-order computations within the EFT). 
For example, if we were to focus on a leading order interaction like $(\partial \phi)^4$ for our single scalar field, we would write,
\begin{align}
S [ \phi ] = \int d^4 x \, \left(   - \frac{1}{2} ( \partial \phi )^2  + \frac{C}{f_\phi^4} (\partial \phi )^4    + \frac{ \epsilon_{\phi} \Lambda^4 }{g_\phi^2}  \mathcal{L}_{\rm HO} \left[   \frac{ g_\phi \phi }{ \Lambda } \; , \; \frac{\partial}{\Lambda}  \right]      \right)
\label{eqn:dphi4eg}
\end{align}
where $C/f_\phi^4$ is a free Wilson coefficient, higher order interactions are suppressed in derivatives by $\Lambda$ and in fields by $\Lambda / g_\phi$, and $\epsilon_{\phi}$ controls any overall hierarchy between $\mathcal{L}_{\rm LO}$ and $\mathcal{L}_{\rm HO}$. Loops from $\mathcal{L}_{\rm HO}$ generate corrections to $\mathcal{L}_{\rm HO}$ which $\sim \epsilon_\phi^{V-1} \left( g_\phi / 4 \pi \right)^{2L}$, and so radiative stablility requires $g_\phi \lesssim 4 \pi$ and $\epsilon_{\phi} \lesssim 1$. 
But since loops of $(\partial \phi)^4$ can also generate higher order interactions, radiative stability also requires that the scale $\Lambda$ cannot be made arbitrarily small (it must be at least as large as the scale $C/f_\phi^4$) and that the hierarchy $\epsilon$ cannot be made arbitrarily small (it must be within at least one loop factor of $C/f_\phi^4$),
\begin{align}
 \frac{  g_\phi^2 }{\Lambda^4}  \gtrsim   \frac{ C }{f_\phi^4}  \;\;\;\; \text{and} \;\;\;\; \epsilon^2 \frac{ g_\phi^2 }{ \Lambda^4 }  \gtrsim  \frac{ g_\phi^2 }{16 \pi^2}  \frac{ C }{f_\phi^4} \, .
 \label{eqn:radstabLI}
\end{align}
%if the Lagrangian \eqref{eqn:dphi4eg} is to be radiatively stable. 
Conversely, loops from $\mathcal{L}_{\rm HO}$ can never renormalise either $(\partial \phi)^2$ or $(\partial \phi)^4$ in any scheme which respects the EFT power-counting\footnote{
On dimensional grounds, loops which are regularised in a scheme which do not introduce a new heavy scale  (e.g. dimensional regularisation) can only produce interactions with at least an additional four derivatives (or a factor of $m_\phi^2 / \Lambda^2$, which we assume to be negligible), and so all leading order interactions of form $(\partial \phi )^n$ and $\partial^2 (\partial \phi)^n$ are not renormalised within the EFT. We refer to \cite{Goon:2016ihr} for further discussion of this type of non-renormalization in various scalar and gravitational field theories.
}, so a hierarchy like $C/f_\phi^4 \ll g_\phi^2/\Lambda^4$ is radiatively stable.
These bounds will be important in Section~\ref{sec:3} when we compute high-energy scattering amplitudes using only leading-order interactions like $(\partial \phi )^4$, as they tell us under what conditions the higher order corrections may be safely neglected (in a radiatively stable way).

%%%%%%%%
\subsubsection*{Massive Vector Field}
%%%%%%%%

Before moving on to theories with broken boosts, let us exemplify how radiative stability is implemented when gauged symmetries are spontaneously broken. Consider a massive vector field $A^\mu$, with a power counting of the form,
%
%Gauge symmetries are responsible for decoupling unwanted degrees of freedom from the system---for instance to describe the two transverse polarisations of a massless spin-1 particle, we use the vector field $A_\mu$ plus a gauge symmetry which removes the longitudinal mode. When restoring these degrees of freedom (e.g. by giving $A_\mu$ a mass so that it now propagates three degrees of freedom) care must be taken with the power counting since there is a limit in which the gauge invariance is restored (e.g. the massless limit).  
%
%We will exemplify this with the example of a single vector field, $A_\mu$. 
%Naively, one might imagine that the SILH scheme would correspond to an action,
\begin{align}
S_U [ A^\mu ] = \int d^4 x \; \left\{ 
-\frac{1}{4} F^2 - \frac{1}{2} m^2 A^2 
+ \frac{ M^4 }{g_A^2} \mathcal{L}_{\rm gauge} \left[  \frac{ g_A F^{\mu\nu} }{M^2} , \frac{\partial_\mu}{M} \right]
+ \frac{\epsilon_*  M^4 }{g_A^2} \mathcal{L} \left[  \frac{ g_A A^\mu}{M} , \frac{\partial_\mu}{M} \right]
   \right\}
   \label{eqn:S1U}
\end{align}
where $F^{\mu\nu} = \partial^{\mu} A^{\nu} -  \partial^{\nu} A^{\mu} $ and we have separated the interactions into those that respect the would-be symmetry $A_\mu  \to A_\mu + \partial_\mu \xi$ in the absence of the mass term, and those that explicitly break this symmetry. 
%, which are controlled by a small parameter $\epsilon_*$.
It is tempting to conclude, by analogy with the scalar example above, that this action is radiatively stable for any $\epsilon_* \lesssim 1$ and $g_A \lesssim 4\pi$. However this is not the case. 
Since the propagator for $A^\mu$ does not have a canonical $1/p^2$ fall-off, naive estimates of loop divergences must be considered more carefully. 

To bring the propagator of $A^\mu$ into a canonical form, we can perform the so-called ``Stuckelberg procedure'' to replace $S_U [A^\mu ]$ with an equivalent action $S[ \tilde{A}^\mu, \phi ]$ in which $\tilde{A}^\mu$ enjoys a gauge invariance.
% (and hence has canonical $\sim 1/p^2$ propagator). 
This is achieved via the replacement, 
\begin{align}
 A^\mu = \tilde{A}^\mu + \frac{ \partial^\mu \phi }{m} \; , 
\end{align} 
where the normalisation of $\phi$ has been fixed so that the mass term $-\tfrac{1}{2} m^2 A^2$ leads to a canonical kinetic term $-\frac{1}{2} (\partial \phi )^2$. There is also a mixing term, $m \tilde{A}^\mu \partial_\mu \phi$, but at sufficiently high energies, $\partial \gg m$, the two fields decouple (as required by the Goldstone Equivalence Theorem). 
The action \eqref{eqn:S1U} in the decoupling limit ($m \to 0$ with $m M$ fixed) then corresponds to an effective description of $\phi$ only,
\begin{align}
S [  \phi ] = \int d^4 x \; \left\{ 
 - \frac{1}{2} ( \partial \phi )^2 
+ \frac{\epsilon_*  M^4 }{g_A^2} \mathcal{L} \left[  \frac{ g_A \partial \phi }{m M} \;\; , \;\; \frac{\partial}{M} \right]
   \right\} \, . 
   \label{eqn:S1DL}
\end{align}
It is now clear that $g_A \sim 4 \pi$ is \emph{not} radiatively stable, since an $L$-loop diagram with $V_a$ vertices of $\mathcal{O}_a = \epsilon_* M^4 / g_A^2 \, ( g_A \partial \phi / m M )^a$ requires a counterterm
%,
%\begin{align}
%\Delta \mathcal{O} \sim  \left(  \frac{ \epsilon^2_* M^4 }{g_A^2}   \right)^{\sum V_a}   \left(  \frac{ g_A }{ m M}   \right)^{\sum a V_a}   \left( \frac{ \partial^{4} }{16 \pi^2} \right)^L   (\partial \phi )^{ 2 - 2 L +  \sum (a-2) V_a}
%\end{align}
which is only smaller than the tree-level vertices in $S[\phi]$ if,
\begin{align}
  \left(  \epsilon_* \frac{ M^2 }{ m^2}    \right)^{-1 + \sum V_a} \left( \frac{ g_A^2 }{16 \pi^2}  \frac{M^2}{m^2}  \right)^{ L }    \lesssim  1  \; . 
\end{align}
This example highlights the difficulties with power counting directly in the unitary gauge \eqref{eqn:S1DL}---radiative stability actually requires that $g_A \lesssim  m / M $ and $\epsilon_*  \lesssim \left( m / M \right)^2  \,  . $
This is made transparent by defining new couplings,  $g_\phi = ( M / m ) g_A$ and $\epsilon_\phi = ( M / m )^2  \epsilon_*$, which brings the decoupling limit action into the single-field-single-coupling form, and is radiatively stable for all $\epsilon_\phi \lesssim 1$ and $g_\phi \lesssim 4\pi$. 
%We will also see in Section~\ref{sec:3} that this is the tuning required if the tree-level scattering amplitudes of \eqref{eqn:S1U} are to be unitary at all energies below the cutoff $M$. 

%%%%%%%%%%%%%%%%%%
\subsection{EFTs with Broken Boosts}
\label{sec:22}
%%%%%%%%%%%%%%%%%%

We will now move on to the first main aim of this paper: to explore how radiative stability constrains an effective field theory with a preferred time-like direction, $n^\mu$, which explicitly breaks Lorentz boosts. The inclusion of $n^\mu$ introduces a new set of operators into the EFT basis, for example explicit time derivatives $n^\mu \partial_\mu$ may now appear. 
%We first focus on a single scalar field, $\pi(\vec x,t)$, before turning to the EFT of Inflation in Section~\ref{sec:23}.  
For a single scalar field, we will consider the following power-counting scheme for the higher-order interactions,
%the appropriate generalisation of the power counting \eqref{eqn:SILH} for a single scalar field is,
\begin{align}
S [ \pi  ]  = \int d^4 x \, \sqrt{-Z} \left\{
 - \frac{1}{2} Z^{\mu\nu} \partial_\mu \pi \partial_\nu \pi   + \mathcal{L}_{\rm LO}
+  \frac{ \epsilon_\pi \Lambda^4}{g_\pi^2}  \mathcal{L}_{\rm HO} \left[  \frac{g_\pi  \pi }{\Lambda}  \; , \; \frac{\partial_\mu}{\Lambda} \; , \; g_n n^\mu   \; , \;  Z^{\mu\nu}  \right] 
 \right\} \, . 
 \label{eqn:Spi}
\end{align}
where $\mathcal{L}_{\rm LO}$ is a set of (leading order) interactions which are not renormalised by any interaction in $\mathcal{L}_{\rm HO}$.
The two key differences are the addition of a new coupling, $g_n$, which controls the symmetry-breaking (taking $g_n \to 0$ recovers invariance under boosts), and the introduction of a kinetic matrix $Z^{\mu\nu}$ (which can differ from the background metric $g^{\mu \nu}$ in the direction $n^\mu n^\nu$) which is implicitly used to perform all index contractions. Note that although boosts are broken, we continue to treat spatial derivatives as suppressed\footnote{
Although it is possible to include large spatial derivative corrections in the EFT of inflation \cite{Gwyn:2012mw}, which would give a non-linear dispersion relation.
} (in particular this retains a linear dispersion relation $\omega \propto |\mathbf{p}|$ at leading order). 
We will assume that $\pi$ has an approximate shift symmetry\footnote{
The shift symmetry is softly broken by a small mass, $m_\pi$, which we take to be smaller than every other relevant scale in the problem---its only role is to regulate any potential IR divergences and allow for a convergent partial wave expansion in section~\ref{sec:3}.
} and the kinetic matrix is diagonal, $ Z^{00} = - 1$ and $Z^{ij} = c_\pi^2 \delta^{ij}$ in the frame $n_\mu = \delta_{\mu}^0 $, where $c_\pi$ is the sound speed of $\pi$ and determines its dispersion relation,
%\footnote{
%We will work throughout with all dimensionful quantities expressed in units of \emph{energy}, using the appropriate factors of $\cpi$ to convert spatial into temporal units---for example all terms in the dispersion relation \eqref{eqn:pi_disp} have units of (energy)$^2$. Similarly, factors of $\cpi^{-3} f_\pi^4$ which routinely appears in the action are an energy density \cite{Baumann:2011su}.
%}, 
\begin{align}
 Z^{\mu\nu} p_\mu p_\nu  := - \omega^2  + c_\pi^2 \mathbf{p}^2   = - m_\pi^2 \, . 
 \label{eqn:pi_disp}
\end{align} 
Since this forms the free propagator of $\pi$, treating $Z^{\mu\nu}$ as an effective metric simplifies the power counting of the interactions\footnote{
Note that the effective volume element is $\sqrt{-Z} := \sqrt{ \text{det} \left( Z^{\mu\nu}  \right)^{-1}  } = 1/\cpi^3$. To avoid confusion, we will never use $Z_{\mu\nu}$ with lowered indices (since if indices are raised/lowered with the background $g_{\mu\nu}$, then $Z_{\mu\nu} \neq ( Z^{\mu\nu} )^{-1}$). On the other hand, since $g^{\mu\nu} n_\nu = Z^{\mu\nu} n_{\nu}$, we will use both $n^\mu$ and $n_\mu$.
}. 

There are now effectively three different scales in the power counting \eqref{eqn:Spi}: in addition to the the scale $\Lambda$ which suppresses derivative contractions $Z^{\mu\nu} \partial_\mu \partial_\nu$ and the scale $\Lambda_\pi = \Lambda / g_\pi$ which suppresses $\phi$ insertions, there is now a separate scale $\Lambda_n = \Lambda / g_n$ which suppresses time derivatives $n^\mu \partial_\mu$. 
%Unlike in the Lorentz-invariant case, there is now a different scale for time derivatives. 
However, since $n_\mu n_\nu Z^{\mu\nu} = -1$, loops can remove factors of $n_\mu$, e.g.
%\begin{align}
% n^\mu n^\nu \int \frac{d^d k }{ (2 \pi )^d \sqrt{-Z}}  \;  \frac{ k_\mu k_\nu}{Z^{\mu\nu} ( k+p)_\mu ( k + p)_\nu  + m_\pi^2 }  =  n^\mu n^\nu \left(  \frac{\left( Z^{\mu\nu} \right)^{-1}}{d}     -   \frac{ p_\mu p_\nu }{m_\pi^2}    \right)  \frac{ \Gamma ( 1 - d/2 ) }{ (4\pi)^{d/2} } m_\pi^d
%\end{align}  
\begin{align}
 n^\mu n^\nu \int \frac{d^d k }{ (2 \pi )^d \sqrt{-Z}}  \;  \frac{ k_\mu k_\nu }{ Z^{\mu\nu} k_\mu k_\nu  + m_\pi^2 }  =  n^\mu n^\nu   \left( Z^{\mu\nu} \right)^{-1}  \frac{ \Gamma ( 1 - d/2 ) }{ d (4\pi)^{d/2} } m_\pi^d
 \label{eqn:nn}
\end{align}  
and so radiative stability does not allow the scale $\Lambda / g_n$ to be be made arbitrarily smaller than $\Lambda$. 
We can see this concretely by focusing on the interactions in $\mathcal{L}_{\rm HO}$ with fewest derivatives per field (since including more derivatives will not change the conclusion), 
\begin{align}
 \mathcal{O}_{ab} = \epsilon_\pi \frac{\Lambda^4}{ g_\pi^2 } \left(   \frac{ g_n g_\pi \dot \pi }{ \Lambda^2 }  \right)^a \left(   \frac{ g_\pi^2 Z_{\pi\pi}  }{ \Lambda^4 }    \right)^b \, ,
 \label{eqn:piOab} 
\end{align}
where we have introduced the notation $\dot \pi = n^\mu \partial_\mu \pi$ and $Z_{\pi\pi} = Z^{\mu\nu} \partial_\mu \pi \partial_\nu \pi$.
An $L$-loop diagram with $V_{ab}$ insertions of each $\mathcal{O}_{ab}$ leads to a counterterm,
\begin{align}
\Delta \mathcal{O}   \sim  \left(  \epsilon_\pi \frac{\Lambda^4}{g_\pi^2}   \right)^{\sum V_{ab} }   
\left(   \frac{ g_\pi  g_n    }{ \Lambda^2 }  n  \right)^{\sum a V_{ab} } 
\left(   \frac{ g_\pi^2  }{ \Lambda^4 }    \right)^{\sum b V_{ab} } 
    \left( \frac{\partial^{ 4 } }{ 16 \pi^2  }   \right)^L (\partial \pi )^{2 -2L + \sum ( a+2b-2 )V_{ab} }
\end{align}
where the factors of $n^\mu$ may either be contracted into the derivatives (producing $\dot \pi$) or into each other ($n_\mu n^\mu = -1$) as in \eqref{eqn:nn}, and all derivatives are implicitly contracted using $Z^{\mu\nu}$. For radiative stability we require that this is smaller than the tree-level terms already in $\mathcal{L}_{\rm HO}$,
%\begin{align}
%\Delta \mathcal{O}  \lesssim  \epsilon^2 \frac{\Lambda^4}{g_\pi^2}   \left( \frac{\partial}{\Lambda} \right)^{4L} 
%\left(   \frac{ g_n g_\pi  \dot \pi }{ \Lambda^2 }  \right)^{\sum a V_{ab} - 2N } 
%\left(   \frac{ g_\pi^2 Z^{\mu\nu} \partial_\mu \pi \partial_\nu \pi  }{ \Lambda^4 }    \right)^{ 1 - L + N + \sum ( b-1 ) V_{ab} } 
%\end{align}
%which results in a bound on the couplings, 
\begin{align}
\left( \epsilon_\pi  \right)^{-1 + \sum V_{ab} }   \left(  \frac{ g_\pi^2 }{16 \pi^2}  \right)^{L} \left(  g_n^2  \right)^N    \lesssim 1   \, , 
\label{eqn:piradstab}
\end{align}
where 
%$L$ is the number of loops, $\sum V_{ab}$ is the total number of vertices and 
$2N < \sum a V_{ab}$ is the number of $n_\mu n^\mu$ contractions. Since this must hold for all possible choices of $L$, $V_{ab}$ and $N$, we must have $g_\pi \lesssim 4\pi $, $\epsilon_\pi \lesssim 1$ and $g_n \lesssim 1$ separately. 

Since loop corrections (in a scheme which preserves the EFT power counting, like dimensional regularisation) always introduce at least 4 additional derivatives (or a factor of $m_\pi^2 / \Lambda^2$, which we assume to be negligible), one may have a radiatively stable hierarchy between interactions $\partial^p (\partial \pi )^n$ with $p<4$ and $p \geq 4$ \cite{Goon:2016ihr}.  
For instance, consider an action of the form \eqref{eqn:Spi} with,
\begin{align}
 \mathcal{L}_{\rm LO} =  \sum_{a,b} C_{ab} \frac{ \dot \pi^a Z_{\pi\pi}^b}{ f_{\pi}^{2a+4b-4} }
\end{align}
where $C_{ab}$ are a set of constant coefficients (and $f_\pi$ is an arbitrary scale). 
Although the $C_{ab}$ do not receive large renormalisation within the EFT (so can take any value from the point of view of radiative stability), they do generate terms in $\mathcal{L}_{\rm HO}$. Radiative stability then places lower bounds on the parameters 
%$\epsilon^2, g_n$ and $\Lambda$ 
appearing in $\mathcal{L}_{\rm HO}$, analogous to \eqref{eqn:radstabLI},
\begin{align}
 \frac{g_\pi^{a+2b-2} g_n^{a} }{\Lambda^{2a+4b-4}}  \gtrsim  \frac{ C_{ab} }{ f_\pi^{2a+4b-4}}  \;\;\;\; \text{and}\;\;\;\;  \frac{\epsilon_\phi \Lambda^4}{g_\pi^2}    \frac{g_\pi^{a+2b} g_n^a }{\Lambda^{2a+4b}}   \gtrsim \frac{g_\pi^2}{16 \pi^2} \frac{ C_{ab} }{f_\pi^{2a+4b-4}} \, . 
 \label{eqn:Cablower}
\end{align}
Compared with the ``natural'' value $C_{ab} / f_\pi^{2a+4b-4} \sim \epsilon_\phi g_\pi^{a+2b-2} g_n^a/ \Lambda^{2a+4b-4}$ inferred from the counting in \eqref{eqn:Spi}, the leading-order coefficients may be either a factor of $1/\epsilon_\phi$ larger or a factor of $16 \pi^2 / g_\pi^2$ larger without spoiling radiative stability. 
%Once a particular $C_{ab}$ is measured, \eqref{eqn:Cablower} places lower bounds on both the $g_\pi/ \Lambda^2$ and $\epsilon^2$ which control the higher order corrections in $\mathcal{L}_{\rm int}$. For example, from the quartic action \eqref{eqn:S3S4}, a measurement of $\beta_1$ would require,
%\begin{align}
% \frac{g_\pi^2 g_n^4}{ \Lambda^4} \gtrsim \frac{\beta_1}{f_\pi^4} \;\;\;\; \text{and} \;\;\;\; \epsilon^2 \frac{ g_\pi^2 g_n^4}{ \Lambda^4}  \gtrsim \frac{g_\pi^2}{16 \pi^2} \frac{\beta_1}{f_\pi^4}  \, , 
%\end{align}
%i.e. the EFT cutoff cannot be significantly larger than the measured scale $( f_\pi^4/  \beta_1 )^{1/4}$ (up to factors of $g_\pi$ and $g_n$ associated with the operator $\dot \pi^4$), and the higher order terms cannot be suppressed by more than a one-loop suppression of $g_\pi^2/16\pi^2$ relative to the leading $\beta_1 / f_\pi^4$.   

We reiterate that the underlying UV physics may not produce a low-energy action of the form \eqref{eqn:Spi}, 
%(i.e. higher order corrections cannot be guaranteed to have this form), 
but for the purposes of computing low-energy observables using $\mathcal{L}_{\rm LO}$ we must make some assumption about when $\mathcal{L}_{\rm HO}$ can be neglected, and here we have shown how to do this in a radiatively stable way. 
It may seem that, without independently measuring the corrections in $\mathcal{L}_{\rm HO}$, one cannot use the bound \eqref{eqn:Cablower} to say anything about the size of the $C_{ab}$.  However, we will now show that when boosts are only broken spontaneously there is an additional relation between $g_\pi/\Lambda^2$ and $g_n/f_\pi^2$ which translates \eqref{eqn:Cablower} into a bound on $C_{ab}$.

%For instance, suppose that we single out the $\alpha_1 \dot \pi^3 / f_\pi^2$ interaction\footnote{
%It is similarly possible to single out $\beta_1 \dot \pi^4 / f_\pi^4$ interaction, since an accidental $Z_2$ symmetry allows for a naturally small $\alpha_1$ \cite{Senatore:2010jy}.    
%}, and wish to make $\alpha_1$ as large as possible (e.g. to produce an observable bispectrum). For an $L$-loop diagram of $V$ such vertices to be a radiatively stable correction to the remaining terms in $\mathcal{L}$, we require,
%\begin{align}
%\left( \frac{g_\pi}{ 4 \pi }   \right)^{2L} \left(  \frac{\alpha_1 \Lambda^2}{ g_\pi g_n^3 f_\pi^2 }   \right)^V  \lesssim \epsilon^2 \, . 
%\end{align} 
%Ensuring this for all $V > 1$ and $L > 0$ requires,
%\begin{align}
%|  \alpha_1 | \Lambda^2  \lesssim f_\pi^2 \times  g_n^3 \, \text{min} \left( \epsilon^2 (4\pi)^2 , g_\pi^2   \right)
%\label{eqn:a1_neccessary}
%\end{align}
%(which can be seen from taking $V=2, \, L=1$ or from taking $V \gg L$). 
%Equation \eqref{eqn:a1_neccessary} is therefore a necessary restriction on $\alpha_1 \Lambda^2$ imposed by radiative stability alone, while the stronger bounds \eqref{eqn:ab_radstab} contain an additional assumption that there are no hierarchies between the different interactions (which may be ``natural'' from the point of view of known UV completions, but is not required by EFT consistency). In either case, if $\alpha_i$ and $\beta_i$ are to be $\mathcal{O} (1)$ then the largest the EFT cutoff can be is $\Lambda^2 \sim 4\pi f_\pi^2$. 

%%%%
\subsubsection*{Spontaneously Broken Boosts}
%%%%
%%%%%
%\paragraph{Gauged Shift Symmetry:}
%%%%%
While \eqref{eqn:Spi} is a consistent power counting for any scalar field theory with broken Lorentz boosts,
there is a special tuning of the Wilson coefficients 
%in \eqref{eqn:S3S4} 
which promotes the global shift symmetry, $\pi (x) \to \pi (x) + c$, to a local one,
% (since Z^{\mu\nu} = cpi^2\, g^{\mu\nu} -  (1-\cpi^2 ) n^\mu n^\nu )
\begin{align}
 \pi (x) \to   \pi (x)  +  f_\pi^2 \, \xi ( x )   \;\; \text{and} \;\; 
 n_\mu \to n_\mu +  \partial_\mu \xi ( x )   \; \;\text{with} \;\;  Z^{\mu\nu}  + (1 - c_s^2) n^\mu n^\nu \;\; \text{fixed},
 \label{eqn:local_shift}
\end{align}
%. In terms of the dimensionless field $\hat{\pi} = g_\pi \pi / \Lambda$, the desired local symmetry (with charge $q$) is,
%\begin{align}
%\hat{\pi} (x) &\to  \hat{\pi} (x)  + \frac{ q \, \xi ( x ) }{ \cpi^2 }    \;\;\;\; , \;\;\;\;
%&g_n n_\mu &\to g_n n_\mu + \frac{ \partial_\mu \xi ( x ) }{\Lambda}  
% \label{eqn:local_shift}
%\end{align}
where $f_\pi^2$ sets the scale of the associated Noether current (and is no longer arbitrary), $c_s$ is a fixed constant, and the preferred time-like direction $n_\mu$ is now allowed to be different at each spacetime point\footnote{
This allows $n_\mu$ to ``eat'' the scalar $\pi$, and in the unitary gauge $\pi = 0$ the dynamics is encoded entirely in $n_\mu (x)$. When we connect with gravitational theories below, we will see that this is precisely the symmetry required if $\pi$ is to represent the scalar fluctuations of a spacetime metric which has been foliated using $n_\mu$---it allows $\pi$ to be traded for a geometric description in which the surface normal to $n_\mu$ fluctuates. 
}. This symmetry corresponds to non-linearly realised Lorentz boosts which leave the speed $c = c_\pi / c_s$ invariant---it mixes the time-like direction $n_\mu$ with spatial coordinates, but locally preserves the metric $c_s^2 g^{\mu\nu} =  Z^{\mu\nu} + (1-c_s^2) n^\mu n^\nu $ with associated $ds^2 = -dt^2 + d \mathbf{x}^2 / c^2$. The constant $c_s$ in \eqref{eqn:local_shift} describes the mismatch between the $Z^{\mu\nu}$ cone and the cone which is preserved by the (non-linearly realised) boosts, i.e. the ratio of the scalar speed $c_\pi$ at low energies to the invariant speed $c$ at high energies. 
%
%To explicitly find the tuning of $\{ \hat{\alpha}_1 ... \hat{\beta}_3 \}$ which realises the local symmetry \eqref{eqn:local_shift}, we vary the action \eqref{eqn:S34} to first order in $\xi$. 

This non-linearly realised symmetry has two important consequences. Firstly, it fixes higher-order Wilson coefficients in terms of lower-order ones. For instance, consider expanding $\mathcal{L}_{\rm LO}$ up to quartic order in fields, producing an action of the form \eqref{eq:EFT_action_decoupling_intro}.
For \eqref{eqn:local_shift} to be a symmetry of this action, it is enough for the variation proportional to $ Z_{\xi \pi} = Z^{\mu\nu} \partial_\mu \xi \partial_\nu \pi$ to vanish, since the variation proportional to $n^\mu \partial_\mu \xi$ can always be removed by adding powers of $(1 + n_\mu n^\mu)$ to the action
 (which vanish once we fix the frame $n_\mu = \delta_\mu^0$). 
Setting ${ \delta S }/{ \delta Z_{\xi \pi}} = 0 $ fixes all but one Wilson coefficient at each order in $\pi$,
\begin{align}
\alpha_2 &= \frac{1 - \cpi^2}{2 \cpi^2}    \;\; , 
&\beta_2 + \frac{3}{2 \cpi^2}  \alpha_1     &=  \frac{ (1-\cpi^2)^2 }{2 \cpi^4}   \;\; , 
 &\beta_3  &= \frac{1 - \cpi^2 }{8 \cpi^4}  \;\; .   
 \label{eqn:ahatTuning}
\end{align} 
The second important consequence of a non-linear symmetry such as \eqref{eqn:local_shift} is that the field coupling $g_\pi$ is now fixed. Since the interactions must group together into invariant combinations, these must be compatible with the power counting. For instance, terms in $\dot \pi$ and $Z_{\pi\pi}$ must now form the invariant,
\begin{align}
\dot \pi   - \frac{ \partial_\mu \pi \partial_\nu \pi}{2 c_s^2 f_\pi^4} \left( Z^{\mu\nu} + ( 1 - c_s^2 ) n^\mu n^\nu \right)     \;\;\;\; \text{c.f} \;\;\;\; \ g_n \dot \pi   +  \frac{g_\pi}{\Lambda^2} \left(  Z_{\pi\pi} + g_n^2 \dot \pi^2      \right) 
\label{eqn:gpicf}
\end{align}
and so we learn that a consistent power counting of the form \eqref{eqn:Spi} requires\footnote{
Note that $g_n^2 \gtrsim 1 - c_s^2$  is enough for the $\dot \pi^2$ term required in \eqref{eqn:gpicf} to appear as a small correction of $\mathcal{O} ( (1-c_s^2)/g_n^2 )$ to the higher order terms in $\dot \pi^n$. Meanwhile, since this is the only way to produce $Z_{\pi\pi}^n$ interactions, one must have $g_\pi \sim g_n \Lambda^2/c_s^2 f_\pi^2$ (values much greater than this would not be acceptable).
},
\begin{align}
 \frac{g_\pi}{\Lambda^2}  \sim \frac{ g_n }{c_s^2 f_\pi^2}   \;\;\;\; \text{and} \;\;\;\; g_n^2 \gtrsim 1 - c_s^2 \; .
\end{align} 
This is usually the case when $\pi$ takes on the role of a Goldstone mode---previously the scale suppressing the field was arbitrary, $g_\pi \pi / \Lambda$ could take any value (providing $g_\pi \lesssim 4 \pi$ for radiative stability), but now thanks to the shift symmetry there is a decay constant which naturally normalises $\pi/f_\pi$. For pions, $\pi$ non-linearly realises a simple shift symmetry, which results in the relation $\Lambda^{\rm pion} \sim g_\pi^{\rm pion} f_\pi^{\rm pion} \lesssim 4 \pi f_{\pi}^{\rm pion}$. For non-linearly realised boosts, we have,
\begin{align}
g_\pi \lesssim 4 \pi  \;\;\;\; \Rightarrow \;\;\;\;  \Lambda^4 \lesssim \Lambda_{\rm max}^4 =  \frac{16 \pi^2 c_s^4 f_\pi^4}{g_n^2}
\label{eqn:Lmax}
\end{align}
which involves the two additional parameters $\{ g_n, c_s \}$ which characterise the symmetry breaking. $\Lambda$ is the scale at which the EFT breaks down, and the lowest possible scale at which the Lorentz boosts \eqref{eqn:local_shift} can be restored. The bound \eqref{eqn:Lmax} follows entirely from the symmetry breaking, and does not require any particular interaction to be large (for instance \eqref{eqn:Lmax} still holds if all of the remaining coefficients vanish, $\alpha_1 =0, \beta_1 = 0,$ etc.)  
 
Since $g_\pi / \Lambda^2$ is now fixed in terms of $g_n$, the lower bound \eqref{eqn:Cablower} from radiative stability becomes a lower bound on $g_n$, 
\begin{align}
 g_n \gtrsim \left(  c_s^{2a+4b -4} C_{ab}  \right)^{ \frac{1}{2a+2b-2} } \, .
 \label{eqn:gnmin}
\end{align} 
Having measured a $C_{ab} \dot \pi^a Z_{\pi\pi}^b /f_\pi^{2a+4b-4}$ interaction in a theory with spontaneously broken boosts \eqref{eqn:local_shift}, a power counting of the form \eqref{eqn:Spi} in which the symmetry breaking is controlled by a single parameter $g_n$ is only radiatively stable providing,
\begin{align}
1 \gtrsim g_n^{2a+2b-2} \gtrsim  c_s^{2a+4b-4} C_{ab}  \;\;\;\; \Rightarrow \;\;\;\; C_{ab} \lesssim  c_s^{4-2a-4b} \, . 
\label{eqn:Cablower_boosts}
\end{align}
For instance, for the quartic action in \eqref{eq:EFT_action_decoupling_intro}, this bound gives $\alpha_i \lesssim 1/c_s^2$ and $\beta_i \lesssim 1/c_s^4$---this is consistent with \eqref{eqn:ahatTuning} for those coefficients fixed by the symmetry, and additionally constrains the free coefficients $\alpha_1$ and $\beta_1$. 
So while radiative stability in a theory with explicitly broken Lorentz boosts can only relate the leading coefficients $C_{ab}$ to lower bounds on the scale of higher order corrections, when boosts are broken spontaneously (non-linearly realised) then radiative stability requires that each $C_{ab}$ be bounded in terms of $c_s$ (the ratio of $c_\pi$ to the invariant speed $c$ which characterises the Lorentz boosts).    
 
We will now show how this theory of a single scalar $\pi$, with non-linearly realised boosts \eqref{eqn:local_shift}, emerges from the EFT of Inflation for metric perturbations about an FLRW background in the decoupling limit, and discuss how these bounds apply to the inflationary bispectrum and trispectrum.

%%%%%%%%
\subsection{EFT of Inflation}
\label{sec:23}
%%%%%%%%

We will now show how the radiative stability constraints developed above can be applied to a theory of metric fluctuations about a background which spontaneously breaks boosts. 

The action should be constructed from all local operators built from $\delta g^{\mu\nu}$ which are invariant under spatial diffeomorphisms (since the background breaks temporal diffeomorphisms), as discussed in \cite{Cheung:2007st}. 
%Once again, leading-order operators like $(\delta g^{00} )^n$ are not renormalised in a scheme which respects the EFT counting (like dimensional regularisation), and so we can separate those out explicitly as in \eqref{eqn:SLO1}. 
As above, we will separate the action into a leading-order piece, $S_{\rm LO}$, given in \eqref{eqn:SLO1}, plus higher-order corrections, $S_{\rm HO}$. In particular, while we allow for arbitrary scales $\{ M_n \}$ in \eqref{eqn:SLO1}, we will assume a particular power counting scheme for $S_{\rm HO}$,
% analogous to \eqref{eqn:S1U},
\begin{align}
S_{\rm HO} [ \delta g^{\mu\nu}  ] = \int d^4 x \; \sqrt{-g } &\Bigg\{ 
M^4  \mathcal{L}_{\rm diff} \left[  \frac{R }{M^2}  \; ,  \; \frac{\nabla_\mu}{M}  \right]  +  \epsilon_* \frac{M^4}{ g_*^2 }  \mathcal{L} \left[   \delta g^{\mu\nu}  \; , \; \frac{\nabla_\mu}{M}  \; , \; g_* n^\mu \;, \; \eta_*  t  \right] 
\Bigg\} \, , 
\label{eqn:newEFT}
\end{align}
where $R$ is the Ricci scalar of $g^{\mu\nu}$, $n^\mu$ is a constant time-like unit vector (which we take to be $\delta_\mu^0$), and we have separated the terms according to whether they are invariant under the would-be diffeomorphism symmetry in the absence of the background. 
This split into $\mathcal{L}_{\rm diff}$ and $\mathcal{L}$ parallels the massive vector power counting \eqref{eqn:S1U}, with the addition of two new couplings: $g_*$ controls the breaking of boosts, and $\eta_*$ controls the breaking of time translations\footnote{
Note that $g_*$ should be at least as large as $\eta_*/M$, since the breaking of time-translations automatically generates a preferred direction, $\eta_*/M \, \nabla_\mu t$, but may be much larger (e.g. scale invariance sets $\eta_* \to 0$ but leaves $g_*$ finite \cite{Pajer:2016ieg}).
}. Since the canonically normalised metric fluctuations are $2 \delta g^{\mu\nu}/M_P$, it is the ratio $2M/M_P$ that plays the role of the field coupling. 

If we were to compare the power counting parameters in $S_{\rm HO}$ with the scales in $S_{\rm LO}$, we might expect ``natural'' values of\footnote{
For instance, for a canonical (dimensionless) scalar field which slowly-rolls down a potential, 
\begin{equation}
 \mathcal{L} = M_P^2 (\partial \phi )^2 + M_P^2 H^2 V (\phi) 
 +  M^4 \mathcal{L} \left[ \phi  ;  \frac{\nabla_\mu}{M}  \right] 
   \;\;\; \text{with} \;\;\;  \partial_t^n \bar{\phi} \sim  ( \sqrt{\epsilon} H )^n \bar{\phi}
\end{equation}
fluctuations in unitary gauge have the form \eqref{eqn:newEFT} with,
\begin{equation}
\epsilon_* \sim  \epsilon \, H^2 / M^2   \;\; , \;\;  g_* \sim \sqrt{\epsilon} H/M \;\; , \;\;\;  \eta_* \sim \sqrt{\epsilon} H \;\; , 
\label{eqn:guess}
\end{equation}
where $M \gtrsim H$ controls the small derivative corrections to the potential. Tuning the potential to be flat corresponds to suppressing each $\lambda_n \phi^n$ interaction in $V(\phi)$ by a power of $\epsilon^{n/2}$, thus lowering $\eta_*$ to $\epsilon H$, as described in \cite{Adshead:2017srh}. 
},
\begin{align}
  M^4 \sim M_P^2 H^2 \;\;\;\; , \;\;\;\;  \epsilon_*  \sim   |\dot H | \,  /   H^2    \;\;\;\;  , \;\;\;\;  \eta_* \sim   \ddot H /  \, | \dot H | \;\;\;\; ,  \;\;\;\; g_*^2 \sim   M_2^4 / M_1^4    \,  ,  
\end{align}
but in order to determine which range of these parameters are required for a consistent EFT we must turn to radiative stability arguments.
 
As in the massive vector case, analysing the radiative stability of \eqref{eqn:newEFT} directly is difficult because the propagator of $\delta g^{\mu\nu}$ no longer has a canonical $1/p^2$ fall-off. 
It is more convenient to perform the Stuckelberg procedure, restoring the broken time diffeomorphisms at the expense of introducing an additional degrees of freedom $\pi$.
%, so that both $\delta g^{\mu\nu}$ and $\pi$ have canonical propagators. 
This is achieved by replacing the metric by $g^{\alpha \beta} = \tilde{g}^{\mu \nu} \partial_\mu \tilde{x}^{\alpha} \partial_\nu \tilde{x}^{\beta}$, as though performing the time diffeomorphism, 
\begin{align}
x^0 = \tilde{x}^0 + \frac{ \pi ( \tilde{x} ) }{ f_\pi^2 }  \;\;\;\;  , \;\;\;\;  \, x^i = \tilde{x}^i  \, . 
\label{eqn:Stu}
\end{align}
This results in a new theory, with degrees of freedom $\delta \tilde{g}^{\mu \nu}$ and $\pi$, which now nonlinearly realises time diffeomorphisms\footnote{
See \cite{Creminelli:2006xe, Cheung:2007st, Piazza:2013coa, Delacretaz:2015edn} (and also \cite{ Shapere:2012nq, Wilczek:2012jt, Castillo:2013sfa}) for further discussion of the Stuckelberg procedure in theories with broken time translations.
}.
The scale $f_\pi$ is fixed so that $\pi$ has a canonical kinetic term, 
\begin{equation}
  f_\pi^4 = - 2 M_P^2 \dot H \cpi  \;\;\;\; \text{and} \;\;\;\; \cpi^2 = \frac{1}{1 + \frac{2 M_2^4}{ M_P^2 \dot H} }  \, ,
  \label{eqn:fpics}
\end{equation}
where $c_s = c_\pi / c_T$ is the ratio of the scalar sound speed to the tensor sound speed. Equation \eqref{eqn:fpics} ensures that expanding $S_{\rm LO}$ to quadratic order in $\pi$ gives an action\footnote{
The Stuckelberg field $\pi$ also acquires a small mass from the explicit time dependence of the Wilson coefficients, $\sqrt{-g} M_0^4 ( t + \frac{\pi}{f_\pi^2} ) \sim  \sqrt{-Z}   | \dot{H} | \cpi   \pi^2$.
}, 
\begin{align}
S^{(2)}_{\rm LO} [ \pi , \delta \tilde{g}^{\mu\nu} ]  
\supset  \int d^4 x \sqrt{ - Z} \;  \left\{ 
-  \frac{1}{2} Z^{\mu\nu} \partial_\mu \pi \partial_{\nu} \pi  
-  \frac{ f_\pi^2 }{M_P}  \left( 1 + \cpi^2  \right)  \; \dot \pi  \frac{M_P \delta \tilde{g}^{00}}{2}   
\right\} \, , 
\label{eqn:L2Z}
\end{align}
in which the kinetic matrix $Z^{\mu\nu}$ has canonical form ($Z^{00} = -1$, $Z^{ij} = c_{\pi}^2/a^2 \, \delta^{ij}$).
The mixing between $\pi$ and metric fluctuations becomes unimportant at energies $\omega \gg  f_\pi^2 (1+\cpi^2) / M_P$---we will work firmly in this decoupling limit \eqref{eqn:DL_intro}, in which the Stuckelberged metric is simply,
\begin{equation}
  \delta g^{\mu\nu} = n^\mu n^\nu \left[
 - \frac{ 2  \dot \pi }{ f_\pi^2 }    +  \frac{1-\cpi^2}{\cpi^2} \frac{ \dot \pi^2 }{ f_\pi^4 }  +  \frac{ Z_{\pi \pi}  }{ \cpi^2 f_\pi^4}  
\right]
\label{eqn:StuckelbergDecoupling}
\end{equation}
and the action $S [\pi, \delta \tilde{g}^{\mu\nu} ]$ becomes a functional of $\pi$ only. Furthermore, we will neglect the time variation of the EFT coefficients, since they are slow-roll suppressed (this corresponds to taking $\eta_* \ll \epsilon_*$ so that the breaking of time translations is much weaker than the breaking of boosts, and is analogous to ``conformal limit'' studied in \cite{Pajer:2016ieg}). 
   
The higher-order corrections in $\mathcal{L}_{\rm gauge}$ do not contribute any $\pi$ interactions (since they are invariant under temporal diffeomorphisms), while the symmetry-breaking interactions can be written as, 
\begin{align}
 \int d^4 x \; \sqrt{-Z}  \; \Bigg\{  
     \;  \cpi^3 \, \frac{ \epsilon_* M^4 }{g_*^2}   \mathcal{L} \left[  g_*^2 \delta g^{00}\, , \; \frac{\nabla_\mu}{\cpi M }\, , \; \cpi g_* n_\mu \, , \; Z^{\mu\nu} \right] 
\Bigg\} \, ,
\label{eqn:newEFT2}
\end{align}
with $\delta g^{00}$ given by \eqref{eqn:StuckelbergDecoupling}.
As in section~\ref{sec:22}, it is convenient to use the kinetic matrix $Z^{\mu\nu}$ as an effective metric for the interactions, but this requires rescaling $M$ and $g_*$ so that the power counting in \eqref{eqn:newEFT2} matches that in \eqref{eqn:newEFT} (which can be seen from $g^{\mu\nu} \nabla_\mu \nabla_\nu \sim \cpi^{-2} Z^{\mu\nu} \nabla_\mu \nabla_\nu $ and $g^{\mu\nu} n_\mu \nabla_\nu = Z^{\mu\nu} n_\mu \nabla_\nu $). 
Just as with the massive vector, we can now see that it would be incorrect to conclude that radiative stability requires $g_* \lesssim 1$ and $\epsilon_* \lesssim 1$. 
Explicitly, if we consider an $L$-loop diagram containing $V_{ab}$ vertices from \eqref{eqn:newEFT2} of the form,  
\begin{align}
 \mathcal{O}_{ab}  &= \epsilon_* \cpi^3 M^4 \, \left(  \frac{g_*^2}{f_\pi^2} \dot \pi    \right)^a  \left(  \frac{g_*^2}{\cpi^2 f_\pi^4}   Z_{\pi \pi}   \right)^b   \,,
\end{align}
where we have neglected insertions of the $\dot \pi^2$ term from $\delta g^{00}$ because they will turn out to be subleading and focussed on the interactions with fewest derivatives (since adding more derivatives will not change the conclusion),
we find that radiative stability requires
\begin{align}
 \left(  \frac{ \epsilon_* M^4}{ M_P^2 \dot H}   \right)^{ -1 + \sum  V_{ab} }  \left(  \frac{ \cpi^2 g_*^2 M^4 }{ 16 \pi^2 f_\pi^4 }  \right)^L  \left(  \cpi^2 g_*^2   \right)^N      \lesssim 1  \, . 
 \label{eqn:dg00radstab}
\end{align}
This can be made transparent by defining new power counting parameters
\begin{align}
 \Lambda = \cpi M \;\;\;\; , \;\;\;\; g_n = \cpi g_*  \;\;\;\; , 
% \;\;\;\; m_\pi = \cpi m_* \;\;\;\; , 
 \;\;\;\;   \frac{g_\pi}{ \Lambda^2} =  \frac{1}{\cpi} \frac{ g_* }{ f_\pi^2}  
  \;\;\;\; , \;\;\;\; \frac{\epsilon_\pi}{g_\pi^2} = \frac{1}{\cpi}   \frac{ \epsilon_* }{ g_*^2 }     \, ,
   \label{eqn:dg00Topi}
 \end{align}
in terms of which \eqref{eqn:newEFT2} takes the form \eqref{eqn:Spi}, and so radiative stability requires $\epsilon \lesssim 1$, $g_\pi \lesssim 4\pi$ and $g_n \lesssim 1$.   

But the higher-order corrections cannot be made arbitrarily small, since they are also renormalised by loops from $S_{\rm LO}$. Repeating the exercise of estimating loops of $\pi$ on dimensional grounds, one arrives at the analogous relations for a $\Lambda_{\rm max}$ \eqref{eqn:Lmax} and for a $g_n^{\rm min}$ \eqref{eqn:gnmin}, which in terms of the original power counting parameters in \eqref{eqn:newEFT} read,
\begin{align}
 M^4 \lesssim M_{\rm max}^4 = \frac{16 \pi^2 f_\pi^4}{ \cpi^2 g_*^2}  \;\;\;\; \text{and} \;\;\;\;  1 \gtrsim  ( c_s  g_* )^{2n-2} \gtrsim   \frac{M_n^4}{f_\pi^4 c_s^{1-2n}} \; , 
 \label{eqn:Uradstab}
\end{align}
for every $n$. Just as in \eqref{eqn:Cablower_boosts}, for a power counting scheme in which the spontaneous breaking of boosts is described by a single order parameter $g_n$ to be radiatively stable, one requires that the leading-order coefficients are bounded: $M_n^4 \lesssim f_\pi^4 c_s^{1-2n}$. This suggests the definition of dimensionless coefficients,
\begin{align}
 M_n^4 = c_n \times f_\pi^4 c_s^{1-2n} \, ,  
 \label{eqn:MnTocn}
\end{align}
%where $c_2 = \tfrac{1}{4} ( 1- c_s^2 )$, 
in terms of which $g_*$ is radiatively stable providing $1 \gtrsim (c_s g_*)^{2n-2} \gtrsim c_n$. This result coincides with the naturalness arguments made in \cite{Baumann:2014cja}, but here follows from a consistency (radiative stability) of the EFT and assumes only that the UV physics produces higher-order corrections of the form \eqref{eqn:newEFT}. 
There is also an analogous bound to \eqref{eqn:Cablower} on the hierarchy $\epsilon_*$, 
\begin{align}
 1 \gtrsim \frac{\epsilon_* M^4}{ M_P^2| \dot H| }  \gtrsim  \frac{ M^4 }{ M_{\rm max}^4 } \, 
 \frac{c_n}{ (c_s g_*)^{2n-2}}   \; , 
 \label{eqn:epradstab}
\end{align}
for every $n$. Note that \eqref{eqn:Uradstab} and \eqref{eqn:epradstab} imply that $g_*^2 \gtrsim M_2^4 /M_1^4$ and $\epsilon_* \lesssim | \dot H| / H^2$ when $M \sim M_P^2 H^2$, as anticipated in \eqref{eqn:guess}.

It is particularly worth emphasising the bounds which arise from the $M_2^4$ interactions, since they place bounds on $g_*$ and $\epsilon_*$ in terms of the sound speed ratio $c_s$. In terms of the $g_n$ and $\epsilon_\pi$ defined in \eqref{eqn:dg00Topi},
\begin{align}
 1 \gtrsim g_n^2 \gtrsim 1 - c_s^2  \;\;\;\;  \text{and} \;\;\;\; 1 \gtrsim \epsilon_\pi  \gtrsim  \frac{ M^4 }{ f_\pi^4 } \, \frac{ 1 - c_s^2 }{16 \pi^2} \, . 
 \label{eqn:M2radstab}
\end{align}
In the absence of the other $M_n$ interactions, as $c_s \to 1$ radiative stability allows for the higher-order corrections to be very small. In particular, the scale suppressing time derivatives can be as high as $\Lambda / \sqrt{1 - c_s^2}$ (in the next section we will see that this is naturally interpreted as a Lorentz-dilation of $\Lambda$ when a collection of $\pi$ particles are boosted to a velocity close to $c_T$). 
Also, we remark that any scaling in which $\epsilon_\pi g_n^2 \sim 1 - c_s^2$ is particularly useful because it ensures that the true sound speed of $\pi$ really does approach $c_T$ when we tune $M_2^4 \to 0$. This is because higher derivative terms like $\partial^2 \dot \pi^2$ can contribute to the dispersion relation and redress the $c_s$ given by \eqref{eqn:fpics} to a new sound speed, 
\begin{align}
1 - \frac{\omega^2}{ | \mathbf{p} |^2} =  1 - \cpi^2  + \epsilon_\pi g_n^2 \frac{ \omega^2}{\Lambda^2} + ...  
\end{align}
when $M_2^4$ is close to zero. In order to treat the sound speed as a constant (particularly when it is close to one) when scattering at energies $\omega \sim \Lambda$, then one must ensure that the combination $\epsilon_\pi g_n^2 \sim 1 - c_s^2$ in a way consistent with \eqref{eqn:M2radstab}\footnote{
For this to be consistent with \eqref{eqn:Uradstab}, we also need all of the $M_n \sim 1 - c_s^2$ to vanish in this limit as well---see also \eqref{eqn:last} from unitarity bounds.
}.

 %%%%%%%%%%%%%%%%%%%%%%%
%%%%
\subsubsection*{Connection with Observation}
%%%%

To connect with observational constraints on the bispectrum \eqref{eq:bispectrum}, we can focus on the expansion of $S_{\rm LO}$ up to quartic order in $\pi$ only \eqref{eq:EFT_action_decoupling_intro}.  Note that the gauge symmetry we have introduced when performing the Stuckelberg procedure \eqref{eqn:Stu} is precisely the gauged shift symmetry \eqref{eqn:local_shift} for $\pi$ discussed in section~\ref{sec:22} (on identifying $n_\mu = \partial_\mu t $).  
%When expanded to quartic order in $\pi$, the leading order $\mathcal{L}_{\rm LO}$ therefore has the form \eqref{eq:EFT_action_decoupling_intro} with 
The interaction coefficients $\alpha_j$ and $\beta_j$ are therefore fixed as in \eqref{eqn:ahatTuning}, with the remaining $\{ \alpha_1 , \beta_1 \}$ given by,
\begin{align}
 	\alpha_1  =  - \frac{4}{3} \frac{c_3}{c_s^2} - \frac{1}{2} \frac{ (1- c_s^2 )^2 }{c_s^2 }   \;\;\;\; , \;\;\;\;
 	\beta_1= 
 	    \frac{2}{3} \frac{c_4}{c_s^4} + 2 \frac{c_3}{c_s^2} \frac{ 1-c_s}{c_s^2} +  \frac{1}{8} \frac{ (1 - c_s^2 )^3 }{c_s^4}  \;\; . \label{eqn:abToc} 
 \end{align} 
where the $c_n$ are given in \eqref{eqn:MnTocn}, and are required to be $\lesssim 1$ \eqref{eqn:Uradstab}, and consequently all $\alpha_j \lesssim 1/c_s^2$ and $\beta_j \lesssim 1/c_s^4$ are bounded by radiative stability.
 
\begin{figure}[htbp!]
\includegraphics[height=5.9cm]{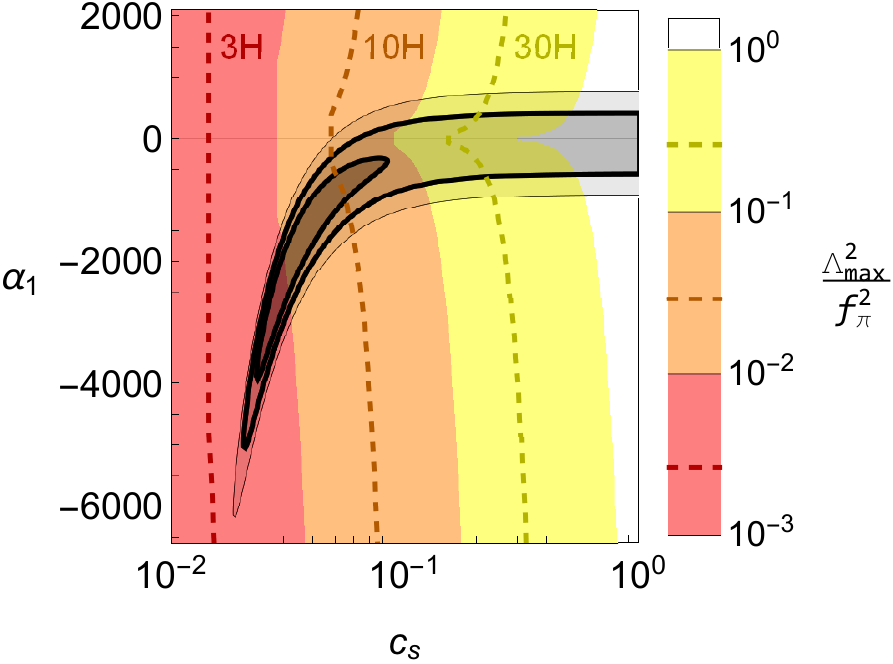}
\quad
\includegraphics[height=5.9cm]{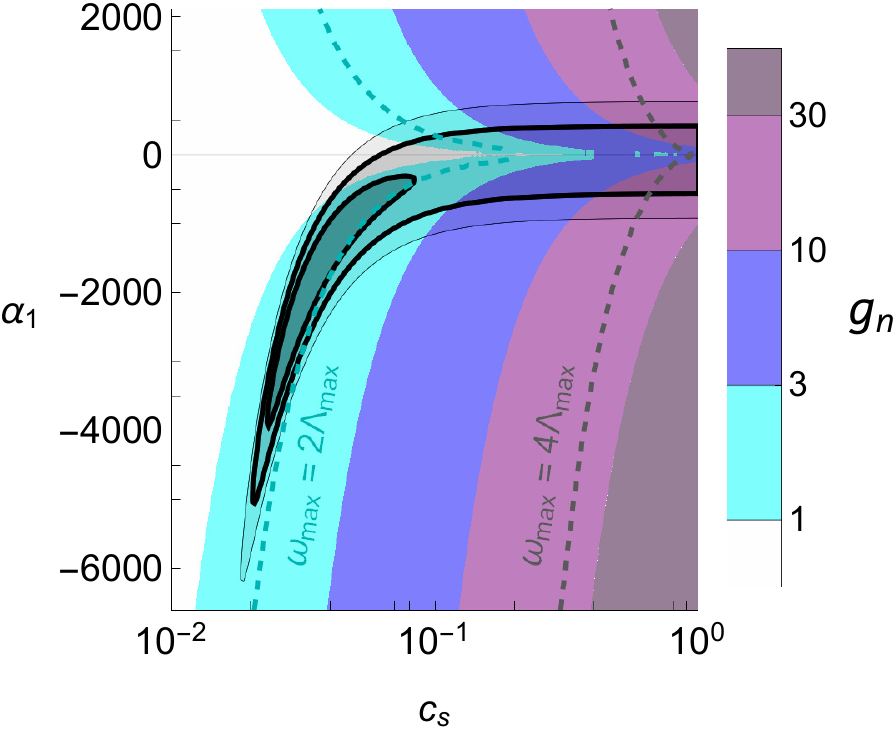}
\caption{
Left panel shows the highest EFT cutoff compatible with the radiative stability of \eqref{eqn:newEFT2}, $\Lambda_{\rm max}^4 = 16 \pi^2 c_s^4 f_\pi^2 / g_n^2$ (where the minimum $g_n$ is given in \eqref{eqn:gnmin2}), with dashed lines indicating where $\Lambda_{\rm max} = 3H$, $10H$ and $30H$ respectively (though note that these are order-of-magnitude estimates of $\Lambda_{\rm max}$ and may be subject to order unity corrections). 
Right panel shows this minimum $g_n$ (again subject to order unity corrections).  
The maximum energy identified in section~\ref{sec:3} from unitary $2\to 2$ scattering, $\omega_{\rm max}^4 / f_\pi^4 = 480 \pi c_s^4 / (1- c_s^2 )$, is never larger than this $\Lambda_{\rm max}$ estimate by more than an $\mathcal{O}(1)$ factor, so it possible to neglect higher order corrections in a radiatively stable way.
%Dashed lines show features which will be relevant in section~\ref{sec:3}, namely that for the majority of the parameter space we will consider $\Lambda_{\rm max}$ can be larger than $H$, and also $\omega_{\rm max} < \Lambda_{\rm max}$.  
}
\label{fig:radstab}
\end{figure}

Furthermore, although the EFT cutoff $\Lambda$ and symmetry breaking parameter parameter $g_n$ which suppress higher-order corrections are determined by the underlying UV physics, we have shown that they can be bounded in terms of the leading-order coefficients $c_s$, $\alpha_1$ and $\beta_1$ using radiative stability (assuming the power counting \eqref{eqn:newEFT}). In particular, 
%If the higher order corrections have the form \eqref{eqn:newEFT2}, then the symmetry breaking is,
\begin{align}
1 \gtrsim g_n^2 \gtrsim   \text{max} \left(  1 - c_s^2   \; , \; \left| \alpha_1 c_s^2 \right|^{1/2} \; , \;  \left| \beta_1 c_s^4 \right|^{1/3}  \right)
\label{eqn:gnmin2}
\end{align}
determines the maximum allowed cutoff, $\Lambda_{\rm max}^4 = 16 \pi^2c_s^4  f_\pi^4/ g_n^2$. 
In Figure~\ref{fig:radstab}, we plot the largest possible $\Lambda_{\rm max}$ as a function of $c_s$ and $\alpha_1$, and compare with the 2018 constraints from Planck on the (equilateral and orthogonal) bispectrum.  
For illustration, if we take the ``best fit'' values\footnote{
Since $f_{NL}^{\rm equil} = -26 \pm 47$ and $f_{NL}^{\rm orth} = -38 \pm 24$, the uncertainty in these values is order one. %However they are useful for illustration, because  
},
$ c_s \approx - 0.031$  and $\alpha_1 \approx  -2100 $, then the strength of the symmetry breaking is at least $g_n^2  \sim 1.4$ and the maximum cutoff is $\Lambda_{\rm max} \sim 0.093 f_\pi = 5.4 H$. Since these dimensional estimates are subject to order one corrections (from the explicit loop integrals, which we have not performed), one should not conclude from this that radiative stability has been violated---however it does motivate a more careful analysis of these higher order corrections in future, since observational constraints currently include regions of parameter space in which $g_n \sim \mathcal{O} (1)$ and $\Lambda_{\rm max} \sim \mathcal{O} (H)$.  
 
The central conclusion from this section is that the maximum energy $\omega_{\rm max}$ which we identify in section~\ref{sec:3} from one-loop corrections to tree-level $2\to 2$ scattering is always below (or approximately) the $\Lambda_{\rm max}$ estimated dimensionally from $L$-loop diagrams with $V$ vertices, as shown in Figure~\ref{fig:radstab}. 
This is important because the unitarity bounds we develop below are only numerically precise \emph{up to higher order corrections from $S_{\rm HO}$}, and we have now established that it is always possible for these corrections to be small in a radiatively stable way.    
Schematically, when $\epsilon_*$ is tuned as low as possible \eqref{eqn:epradstab} (but allowing $g_n$ and $\Lambda$ to be arbitrary), we can compare the expected size of the $2 \to 2$ amplitude from $S_{\rm LO}$ \eqref{eq:EFT_action_decoupling_intro} with that from $S_{\rm HO}$ (whose interactions start at $\partial^4 (\partial \pi)^n$), 
\begin{align}
\mathcal{A}_{2 \to 2}^{ (S_{\rm LO} )}  &\sim  ( 4 \pi )^4 \frac{\omega^4}{ \Lambda_{\rm max}^4 }  \left[ \frac{c_2}{g_n^2}  + g_n^2   \frac{c_2^2+c_3}{g_n^4}  + g_n^4   \frac{c_2^3 +c_2 c_3 + c_4}{g_n^6}  + g_n^6   
\frac{c_2^2 c_3 + c_3^2}{g_n^8}    \right]   \\
 \mathcal{A}_{2 \to 2}^{(S_{\rm HO})}  &\sim  (4 \pi )^2  \frac{\omega^8}{ \Lambda_{\rm max}^8 }    \times \text{max} \left( \frac{ c_n}{ g_n^{2n-2} }   \right)   \left[  1  +  \mathcal{O} \left( \frac{\omega}{\Lambda} \right)   \right]    \, .
\end{align}
%in a radiatively stable way (when the $\epsilon_*$ bound is saturated). 
Although very softly breaking boosts by making $g_n$ as low as possible will raise $\Lambda_{\rm max}$ and suppress the corrections from $S_{\rm HO}$, it will also suppress the relative contributions of $c_3$ and $c_4$ from $S_{LO}$---for instance, if $g_n^{2n-2} < \frac{1}{16 \pi^2} \omega^4 / \Lambda_{\rm max}^4$ then the $c_n$ contribution from $S_{\rm LO}$ can no longer be reliably included. Instead, taking $g_n = 1$ and breaking boosts strongly ensures that all corrections are suppressed by $( \omega / 4 \pi f_\pi )^4$. Note that the cutoff $\Lambda$, although it must be $> \omega$, does not need to be significantly larger if $\omega \ll \Lambda_{\rm max}$. 
%We will derive from perturbative unitarity a scale $\omega_{\rm max}$ at which perturbation theory breaks down, and providing $\omega_{\rm max}^4 < 16 \pi^2 \Lambda_{\rm max}^4$ then is a radiatively stable regime in which $S_{\rm HO}$ could be ignored. 
We will now study the amplitudes from $S_{\rm LO}$ in more detail, neglecting $S_{\rm HO}$ hereafter.

%%%%%%%%%%%%%%%%%%%%%%%%%%%%%%%%%%%%%%%%%%%%%%%%%%%%%%%%%%%%%%%%%%%%%%%%%%%%%%%%%%%%%%%%%%%%%%%%
%%%%%%%%%%%%%%%%%%%%%%%%%%%%%%%%%%%%%%%%%%%%%%%%%%%%%%%%%%%%%%%%%%%%%%%%%%%%%%%%%%%%%%%%%%%%%%%%

% 								IR CONSISTENCY

%%%%%%%%%%%%%%%%%%%%%%%%%%%%%%%%%%%%%%%%%%%%%%%%%%%%%%%%%%%%%%%%%%%%%%%%%%%%%%%%%%%%%%%%%%%%%%%%
%%%%%%%%%%%%%%%%%%%%%%%%%%%%%%%%%%%%%%%%%%%%%%%%%%%%%%%%%%%%%%%%%%%%%%%%%%%%%%%%%%%%%%%%%%%%%%%%

%%%%%%%%%%%%%%%%
\section{Unitarity Bounds}
\label{sec:3}
%%%%%%%%%%%%%%%%

The $\hat{S}$-matrix is a cornerstone of scattering theory (see for instance \cite{Chew, Eden}).
%Let us begin by recounting the constraint from unitarity (conservation of total probability) on $2 \to 2$ elastic scattering amplitudes. 
%Given a Lagrangian, 
The central idea is to construct an operator $\hat{S}$ which maps asymptotic in-states (in the far past) to asymptotic out-states (in the far future), such that $\langle n' | \hat{S} | n \rangle$ gives the probability of transitioning from an initial $n$-particle state, $| n \rangle$, into a final $n'$-particle state, $| n' \rangle$.
The scattering amplitude associated with this process, $\, \mathcal{A}_{n \to n'} \, $, is defined as the matrix element, 
$ \langle n' | \hat{T} | n \rangle = \mathcal{A}_{n \to n'} (2 \pi )^4 \delta^4 ( \sum_i p_i )$,
where $\hat{S} = \mathbbm{1} - i \hat{T}$ has been separated into free and interacting parts, and an overall momentum conserving $\delta$-function has been extracted. 

In order for states to remain properly normalised (necessary for consistent probabilities), $\hat{S}$ must be a unitary operator, $\hat{S}^\dagger \hat{S} = \mathbbm{1}$. 
This can be expressed as a condition on the scattering amplitudes by replacing $\hat{S}$ with $\hat{T}$ and inserting a complete set of states, 
%$\hat{S} = \mathbbm{1} - i \hat{T}$, 
\begin{equation}
- i  \langle n' | \hat{T} | n \rangle + i \langle n' | \hat{T}^\dagger | n \rangle   =  \sum_{j} \,\langle n' | \hat{T} | j \rangle \langle j | \hat{T}^\dagger | n \rangle    \,,
\label{eqn:Tunitary}
\end{equation}
which in terms of momentum-space amplitudes can be written as,
%\begin{equation}
%- i \left( \mathcal{A}_{n\to n'}  - \mathcal{A}_{n' \to n}^* \right)   =  \sum_{j} \, \mathcal{A}_{n \to j} \;  \mathcal{A}_{n' \to j}^*   \,,
%\label{eqn:Aunitary}
%\end{equation}
\begin{equation}
2\,  \text{Disc}_s \, \mathcal{A} ( p_1 ... p_n \to p'_1 ... p'_{n'}  )  =  \sum_{j} \int d \Pi_j (q) \;\; \mathcal{A} ( p_1 ... p_n \to q_1 ... q_j ) \mathcal{A}^* ( p_1' ... p_{n'}' \to q_1 ... q_j )\,,
\label{eqn:Aunitary}
\end{equation}
%\begin{equation}
%2 \text{Disc}_s \, \mathcal{A} ( p_1 p_2 \to p_3 p_4 )  =  \sum_{n} \int d \Pi_n \;\; \mathcal{A} ( p_1 p_2 \to q_1 ... q_n ) \mathcal{A}^* ( p_3 p_4 \to q_1 ... q_n )\,,
%\label{eqn:Aunitary}
%\end{equation}
where $\int d \Pi_j (q)$ represents the phase space of each $j$-particle state (with momenta $q_1, ... q_j$), and the $s$-channel\footnote{
We will refer to the $n \to n'$ process as the $s$-channel. The amplitude $\mathcal{A}_{n \to n'}$ is related via crossing to other channels in which some outgoing and ingoing particles have been exchanged---however these will not be needed since we will work entirely within the $s$-channel process.  
} discontinuity is given by,
\begin{align}
\text{Disc}_s \, \mathcal{A}_{n \to n'} = \frac{1}{2i} \left[ \mathcal{A}_{n \to n'} - \mathcal{A}^*_{n' \to n}  \right]
\end{align}
and is equal to the imaginary part $\text{Im} \, \mathcal{A}_{n \to n'}$ in theories with time-reversal invariance. 

The unitarity condition \eqref{eqn:Aunitary} will be the focus of this section. 
We will first review how the perturbative unitarity of scattering amplitudes constrains Lorentz-invariant EFTs and then extend this to EFTs in which boosts are broken, finding new bounds for every centre-of-mass velocity, and arrive at constraints on the EFT of Inflation's Wilson coefficients $\{ \cpi , \alpha_1 , \, \beta_1 \}$.

Throughout this section, we will denote the momentum of the $j^{\rm th}$ particle by $p_{j \, \mu} = ( \omega_j,  \mathbf{p}_j  )$, and make use of the usual Mandelstam variables $s_{ij} = - Z^{\mu\nu}  ( p_i + p_j )_\mu  ( p_i + p_j )_\nu$ (where the kinetic matix $Z^{\mu\nu}$ is used so that the propagator at momentum $ p_i + p_j$ is simply $\sim 1/s_{ij}$). We also define $p_{s \, \mu} = ( \omega_s , \mathbf{p}_s )$ as the total incoming momentum and $\rho_s = c_\pi | \mathbf{p}_s | / \omega_s$ as the velocity of the centre-of-mass (i.e. $\rho_s = v_{\rm CoM} / c_\pi$, where $v_{\rm CoM}$ is the speed measured with respect to\footnote{
If one instead used the metric $g^{\mu\nu}$ to define a $v_{\rm CoM}'$, then $\rho_s = c_s  v_{\rm CoM}' / c_T$. Consequently, $p_{s\, \mu}$ satisfies the $\pi$ dispersion relation when $\rho_s \to 1$ ($c_\pi | \mathbf{p}_s| \to \omega_s$), and satisfies the tensor dispersion relation when $\rho_s \to c_s$ ($c_T | \mathbf{p}_s | \to \omega_s$). Since we work in the decoupling limit, there is nothing special about the point $\rho_s = c_s$ in our (purely scalar) amplitudes. 
%We will discuss corrections to the decoupling limit at the end of this section.
} the kinetic matrix $Z^{\mu\nu}$).

%%%%%%%%%%%%%%%%%%%%%%%%%%%%%%%%%%%%%%%%%%%%%%%%%%%%%%%%%%%%%%%%%%%%%%%%%%%%%%%%%%%%%%%%%%%%%%%%
%							Lorentz Invariant Unitarity
%%%%%%%%%%%%%%%%%%%%%%%%%%%%%%%%%%%%%%%%%%%%%%%%%%%%%%%%%%%%%%%%%%%%%%%%%%%%%%%%%%%%%%%%%%%%%%%%

%%%%%%%%%%%%%%%%
\subsection{EFTs with Lorentz Invariance}
\label{sec:31}
%%%%%%%%%%%%%%%%

In Lorentz-invariant theories, the amplitude can be written solely in terms of relativistic invariants (e.g. $\mathcal{A}_{2 \to 2} ( p_1 p_2 \to p_3 p_4 )  = \mathcal{A}_{2 \to 2}  (s_{12}, s_{13}, s_{23} )$)
%for example,
%\begin{align}
%\mathcal{A}_{2 \to 2} ( p_1 p_2 \to p_3 p_4 )  = \mathcal{A}_{2 \to 2}  (s_{12}, s_{13}, s_{23} ) \; , 
% \end{align}
% (where momentum conservation removes $p_4$ and there is an additional kinematic relation $s_{12} + s_{13} + s_{23} = 4m_\pi^2$), 
and the boost symmetry can be used to replace arbitrary kinematics with $\mathbf{p}_s = 0$ (transforming to the ``centre-of-mass frame''). These two features allow simple unitarity constraints to be placed on $\mathcal{A}_{n \to n'}$ from \eqref{eqn:Aunitary}, which we now demonstrate. These results are well-known, but will highlight what needs to be changed when we develop new bounds for EFTs with broken boosts in section~\ref{sec:32}. 
First we sketch schematic bounds on $n \to n$ scattering (in the foward limit) from a simple dimensional analysis, and then give numerically precise bounds on $2 \to 2$ scattering using an expansion in partial waves.

%%%%%%%%
\subsubsection*{Elastic $n \to n$ Scattering}
%%%%%%%%

An elastic process (in which the particles in $| n' \rangle$ are the same as those in $| n \rangle$) has a forward limit in which the outgoing momenta are equal to the ingoing momenta, $p_i' = p_i$ for all $i$. In this limit, the initial and final states are identical, and the unitarity condition \eqref{eqn:Aunitary} becomes, 
\begin{align}
2 \,  \text{Im} \, \mathcal{A} ( p_1 ... p_n \to p_1 ... p_n )  \geq  \int d \Pi_n (q) \, | \mathcal{A} ( p_1 ... p_n \to q_1 ... q_n )  |^2 \, .
 \label{eqn:Aelastic}
\end{align} 
Since $2 | \mathcal{A}_{n \to n}| \geq 2 \text{Im} \, \mathcal{A}_{n\to n} \geq \int d \Pi_n | \mathcal{A}_{n\to n} |^2$, by estimating the size of $d \Pi_n$ one can use unitarity to place an upper bound on $| \mathcal{A}_{n \to n} |$.
%By estimating the size of $d \Pi_n$, one can place an upper bound on $|\mathcal{A}_{n\to n}|$ (since $| \mathcal{A}_{n \to n}| \geq \text{Im} \, \mathcal{A}_{n\to n}$). 
%
The $n$-particle phase space element is given explicitly by integrating over all future-pointing, on-shell 4-momenta which conserve the total ingoing momenta, 
\begin{align}
d \Pi_n (q) &= (2\pi)^4  \delta^4 \left( p_s - \sum_i^n q_i   \right) \;  \prod_{i=1}^n  \frac{d^4 q_i}{(2\pi)^4} \;\; 2 \pi \; \delta \left( q^2_i + m^2  \right) \Theta ( q^0_i ) \,  ,
\end{align}
and is Lorentz invariant. On purely dimensional grounds, the phase space volume scales with the total energy as $\int d \Pi_n \sim (4 \pi )^{-1}  \left( s / 16 \pi^2 \right)^{n-2}$, 
%and\footnote{
%The factors of $4\pi$ are inferred from each spherical volume element, $d^3 \mathbf{p}_i \sim 4 \pi$, and each denominator $1/(2\pi)^3$ after the on-shell $\delta$-functions have been imposed. 
%} $\int d \Pi_j  \sim  1/4\pi ( \omega_s^2/ 16\pi^2 )^{n-2}$. 
and so \eqref{eqn:Aelastic} becomes a simple bound,
\begin{align}
| \mathcal{A}_{n \to n } |  \; \lesssim  \; 4 \pi   \left(  \frac{16 \pi^2}{s} \right)^{n-2} \, , 
\label{eqn:AnnLI}
\end{align}
where $\sim$ denotes the various order unity numerical factors which we have neglected. 
%Since $\mathcal{A}_{n \to n}$ can depend only on the Lorentz-invariant $s$, writing $\omega_s^2 = s + | \mathbf{p}_s |^2$ we see that this bound is always strongest when $|\mathbf{p}_s | =0$ (in the centre-of-mass frame).  

%Although \eqref{eqn:Aelastic} (and the resulting \eqref{eqn:AnnLI}) apply non-perturbatively, we will use them to constrain the amplitude $\mathcal{A}_{n \to n}^{\rm tree}$ at leading order in a perturbative loop expansion. In perturbation theory, 
When applied to a perturbative loop expansion of $\mathcal{A}_{n \to n}$,
unitarity \eqref{eqn:Aelastic} provides a lower bound on the size of the non-analytic (imaginary) part of the amplitude generated at one-loop in terms of the tree-level amplitude---this is shown graphically in figure~\ref{fig:3} for $n=2$.  A violation of the bound \eqref{eqn:AnnLI} applied to $\mathcal{A}_{n \to n}^{\rm tree}$ signals that the one-loop contribution to the amplitude has exceeded the tree-level contribution, i.e. the loop expansion has broken down\footnote{
This does not necessarily correspond to new physics since a non-perturbative (all-loop) calculation of $\mathcal{A}_{n \to n}$ may still satisfy the unitarity condition \eqref{eqn:Aunitary}. 
%On the other hand, the EFT cutoff $\Lambda$ discussed in section~\ref{sec:2} does correspond to new physics. 
}.  

For example, for the interaction $C_{ab}  \partial^{2a} \phi^b / f_\phi^{2a+b-4}$, the tree level scattering amplitude with $V$ vertices is,
\begin{align}
 \mathcal{A}_{n \to n}^{\rm tree}  \sim  \left(  \frac{ C_{ab} s^{a} }{ f_\phi^{2a+b-4} }   \right)^V  \frac{1}{s^{V-1} }
\end{align}
and the number of external legs is $2n = (b-2) V + 2$. 
This tree-level amplitude only satisfies the inequality \eqref{eqn:AnnLI} providing,
\begin{align}
 s^{a - 2 + b/2}   \lesssim   ( 4 \pi )^{ b -2 -  1/V  }  \;  f_{\phi}^{2a -4 + b} / | C_{ab} | \, ,
\label{eqn:CabUboundLI}
\end{align}
which defines a maximum energy, $s_{\rm max}$, above which the theory 
%Beyond the threshold $s_{\rm max}$, the theory 
is strongly coupled and loops must be resummed (or new physics beyond the EFT must be included) if one is to restore unitarity. In terms of naive factors of $4\pi$, the lowest $s_{\rm max}$ comes from scattering with the fewest vertices---i.e. the contact diagram with $V=1$ if $b=2n$ is even, or the single-exchange diagram with $V=2$ if $b = 2n + 1$ is odd---but in reality the unitarity bound which gives the lowest numerical $s_{\rm max}$ will depend on the details of the phase space integration. 
%The lowest $s_{\rm max}$ comes from scattering with $n=2$. 
 
When there is a scale of interest in the problem which the EFT must resolve, such as a light mass (e.g. $m_\phi$) or background scale (e.g. $H$ when we discuss inflation), then demanding that $s_{\rm max} \gtrsim$ this scale allows \eqref{eqn:CabUboundLI} to be written as an upper bound on the size of $|C_{ab}|$. For example, the EFT is only consistent if $s_{\rm max} > 4m_\phi^2$, and consequently $| C_{ab} | <  (4 \pi )^{b-3} ( f_\phi / 2m_\phi )^{2a -b -4}$ if $b$ is even or $| C_{ab} | <  (4 \pi )^{b-3/2} ( f_\phi / 2m_\phi )^{2a -b -4}$ if $b$ is odd.
 
It is only possible to trust such unitarity bounds on $|C_{ab}|$ providing this $s_{\rm max}$ scale at which we are computing the amplitude is $\lesssim \Lambda$, the EFT cutoff.
Otherwise, higher-order EFT corrections will also contribute to $\mathcal{A}_{n \to n}$ and give large corrections to the bounds. 
From our discussion of radiative stability in section~\ref{sec:2}, in the presence of $C_{ab} \partial^{2a} \phi^b / f_\phi^{2a+b-4}$ the EFT cutoff can be at most $\Lambda_{\rm max}^{2a+4b-4} = g_\pi^{b-2} f_{\pi}^{2a+4b-4}/ |C_{ab}|$, with $g_\pi \lesssim 4\pi$. This shows that there is at least one radiatively stable tuning of higher-order corrections (namely \eqref{eqn:Spi} with $g_\phi \sim 4 \pi$) which ensures that they are no more than an $\mathcal{O}(1)$ correction\footnote{
Radiative stability requires that the scale of the higher-order corrections, $1/\Lambda_{\rm max}$, may not be tuned smaller than the floor set by loops of the $C_{ab}$ interaction. Since $s_{\rm max}$ is the scale at which the loop contributions from $C_{ab}$ exceed its tree-level contribution to the amplitude, 
%and the higher derivative corrections at this scale cannot be made significantly smaller than loops, 
one might imagine that $s_{\rm max} \sim \Lambda_{\rm max}^2$ is inevitable. 
However, since it is only the real part of the one-loop amplitude which determines the running and the radiative stability floor, the imaginary part of the amplitude could be much larger and violate unitarity at a scale $s_{\rm max} \ll \Lambda_{\rm max}$. In this simple example, there is an additional factor of $4 \pi$ in our estimate of the imaginary part (in practice this would come from the taking the discontinuity of the polylogarithmic branch cut in the $L$-loop amplitude), which for instance allows for $s_{\rm max}^2$ as low as $\Lambda_{\rm max}^4/4 \pi$ for the interaction $(\partial \phi )^4$.   
%We shall see that this is not the case, since $\Lambda$ is only the scale suppressing arbitrarily high-derivative terms, and it is quite possible for finite-order terms to be suppressed by a higher scale---for instance $s_{\rm max} / \Lambda_{\rm max}^2$ can be as small as $\sim 1/4\pi$ for low $n$ scattering. 
} to the $\mathcal{A}_{n \to n}$ bound \eqref{eqn:AnnLI} for any $n$.  

%The higher-order corrections may have tree-level couplings that are much larger than the loop contributions from the leading-order action (in which case $g_{\pi} \ll 4 \pi$ and loops are less important than tree-level insertions of higher-order operators), but the lowest their couplings may be tuned in a radiatively stable way is the size of those loops (in which case $g_{\pi} \sim 4 \pi$ and loops and higher derivative corrections are equally important).  
%Since perturbative unitarity bound is really an estimate of when the one-loop correction from $C_{ab}$ becomes large, it makes sense that only when $g_\pi \sim 4\pi$ are higher derivative corrections roughly the same (not larger) than the $s_{\rm max}$. 

%Demanding perturbative unitarity for all energies up to $f_\phi$ then requires that $| C_{ab} | \lesssim (4 \pi )^{b-2}$. 
%This is consistent with our previous discussion of radiative stability, since setting $C_{ab} = \epsilon^2 g_\pi^{b-2} \hat{C}_{ab}$ with $\epsilon^2 \lesssim 1$ and $g_\pi \lesssim 4 \pi$ only allowed for $\hat{C}_{ab}$ as large as $\sim \mathcal{O} (1)$. 
%
%We stress that the two conditions, radiative stability and perturbative unitarity, are logically independent requirements. The first requires that the correction to higher derivative operators generated by loops is small, while the second requires that the correction to 
%any tree-level computation of any $\mathcal{A}_{n\to n}^{\rm tree}$ amplitude from loops is small.

The bound \eqref{eqn:AnnLI} is only schematic---
%---while useful for estimating orders of magnitude and 
%%getting a feel for 
%what range of interaction coefficients will be relevant for a particular problem,
%%is only schematic---in particular, 
there is a (roughly order one) numerical prefactor from the phase space integral which cannot be obtained using simple dimensional arguments. We will now demonstrate how this precise factor can be calculated by focussing on the unitarity bound from $\mathcal{A}_{2 \to 2}$ scattering.

%%%%%%%%
\subsubsection*{Elastic $2 \to 2$ Scattering}
%%%%%%%%

%%%%%
%\paragraph{Kinematics:}
%%%%%
Consider the transition between two-particle states, $| \mathbf{p}_1, \mathbf{p}_2 \rangle$ and $| \mathbf{p}_3, \mathbf{p}_4 \rangle$. 
In Lorentz-invariant theories, since the amplitude does not depend on the choice of inertial frame, we can without loss of generality use three Lorentz boosts to set ${\bf p}_1 + {\bf p}_2 = 0$. Two spatial rotations can be used to orient the $z$-axis along ${\bf p}_1$, and the remaining azimuthal rotation ensures the amplitude can depend only on $\cos \theta = \hat{\bf p}_1 \cdot \hat{\bf p}_3$. Only one of $|\mathbf{p}_1|$ and $|\mathbf{p}_3|$ is independent (thanks to the mass-shell condition of $p_4^\mu = -p_1^\mu-p_2^\mu-p_3^\mu$), and can be written in terms of the total incoming energy $\omega_s = \omega_1 + \omega_2$.  
The amplitude for a general scattering process in a Lorentz-invariant theory can therefore be expressed in terms of just two variables, $\mathcal{A} ( \omega_s, \theta)$, for instance by taking,
\begin{align}
 \omega_1 = \omega_2 = -\omega_3 = -\omega_4 =  \frac{\omega_s}{2} \;\; , \;\;\;\; s_{12} = \omega_s^2 \;\; , \;\;\;\; s_{13} = - \frac{\omega_s^2}{2} \left( 1 -  \cos \theta \right) \, . 
 \label{eqn:kine_CoM}
\end{align}

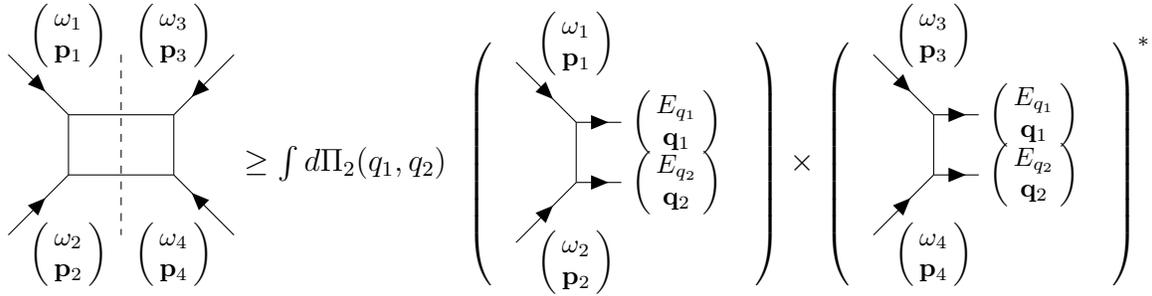
\begin{figure}
\centering
 		\begin{tikzpicture}[baseline=-2.35cm]
			\begin{feynman}

				\vertex (a0);
				\vertex [below=0.8cm of a0] (a1);
				\vertex [below=0.8cm of a1] (b1);
				\vertex [below=0.8cm of b1] (c1);
				\vertex [below=0.8cm of c1] (d1);

				\vertex [right=0.8cm of a1] (a2);
				\vertex [right=0.8cm of b1] (b2);
				\vertex [right=0.8cm of c1] (c2);
				\vertex [right=0.8cm of d1] (d2);

				\vertex [right=0.7cm of a2] (a3i);
				\vertex [right=0.7cm of b2] (b3i);
				\vertex [right=0.7cm of c2] (c3i);
				\vertex [right=0.7cm of d2] (d3i);

				\vertex [right=1.4cm of a2] (a3);
				\vertex [right=1.4cm of b2] (b3);
				\vertex [right=1.4cm of c2] (c3);
				\vertex [right=1.4cm of d2] (d3);

				\vertex [right=0.8cm of a3] (a4) ;
				\vertex [right=0.8cm of b3] (b4);
				\vertex [right=0.8cm of c3] (c4);
				\vertex [right=0.8cm of d3] (d4);

				\vertex [above=0.5cm of b2] (ta2) {\small $\left( \begin{array}{c} \omega_1 \\ \mathbf{p}_1 \end{array} \right)$};
				\vertex [below=0.5cm of c2] (td2) {\small $\left( \begin{array}{c} \omega_2 \\ \mathbf{p}_2 \end{array} \right)$};
				\vertex [above=0.5cm of b3] (ta3) {\small $\left( \begin{array}{c} \omega_3 \\ \mathbf{p}_3 \end{array} \right)$};
				\vertex [below=0.5cm of c3] (td3) {\small $\left( \begin{array}{c} \omega_4 \\ \mathbf{p}_4 \end{array} \right)$};

				\diagram*{
				(a1) --  [fermion] (b2) -- (b3) -- [anti fermion] (a4),  
				(d1)-- [fermion] (c2) -- (c3) --[anti fermion] (d4),
				(c2) -- (b2), (c3)--(b3),
				(a3i) -- [scalar] (d3i)
				};
				
			\end{feynman}	
		\end{tikzpicture}
 {\large $\geq \int d \Pi_2 ( q_1, q_2 )$ }
$\left( \begin{array}{c} \\ \\ \\ \\ \\\\  \\ \end{array} \right. $
		\begin{tikzpicture}[baseline=-2.25cm]
			\begin{feynman}

				\vertex (a0);
				\vertex [below=0.8cm of a0] (a1);
				\vertex [below=0.8cm of a1] (b1);
				\vertex [below=0.8cm of b1] (c1);
				\vertex [below=0.8cm of c1] (d1);

				\vertex [right=0.8cm of a1] (a2);
				\vertex [right=0.8cm of b1] (b2);
				\vertex [right=0.8cm of c1] (c2);
				\vertex [right=0.8cm of d1] (d2);

				\vertex [right=0.6cm of a2] (a3);
				\vertex [right=0.6cm of b2] (b3);
				\vertex [right=0.6cm of c2] (c3);
				\vertex [right=0.6cm of d2] (d3);

				\vertex [above=0.5cm of b2] (ta2) {\small $\left( \begin{array}{c} \omega_1 \\ \mathbf{p}_1 \end{array} \right)$};
				\vertex [below=0.5cm of c2] (td2) {\small $\left( \begin{array}{c} \omega_2 \\ \mathbf{p}_2 \end{array} \right)$};
				\vertex [right=0.0cm of b3] (tb3) {\small $\left( \begin{array}{c} E_{q_1} \\ \mathbf{q}_1 \end{array} \right)$};
				\vertex [right=0.0cm of c3] (tc3) {\small $\left( \begin{array}{c} E_{q_2} \\ \mathbf{q}_2 \end{array} \right)$};
 				
				\diagram*{
				(a1) --  [fermion] (b2) -- [fermion] (b3),  
				(d1)-- [fermion] (c2) -- [fermion] (c3),
				(c2) -- (b2)
				};
				
			\end{feynman}	
		\end{tikzpicture}
		$\left. \begin{array}{c} \\ \\ \\ \\ \\\\  \\ \end{array} \right) $
		{\large $\times$} $\left( \begin{array}{c} \\ \\ \\ \\ \\\\  \\ \end{array} \right. $
		\begin{tikzpicture}[baseline=-2.35cm]
			\begin{feynman}

				\vertex (a0);
				\vertex [below=0.8cm of a0] (a1);
				\vertex [below=0.8cm of a1] (b1);
				\vertex [below=0.8cm of b1] (c1);
				\vertex [below=0.8cm of c1] (d1);

				\vertex [right=0.8cm of a1] (a2);
				\vertex [right=0.8cm of b1] (b2);
				\vertex [right=0.8cm of c1] (c2);
				\vertex [right=0.8cm of d1] (d2);

				\vertex [right=0.6cm of a2] (a3);
				\vertex [right=0.6cm of b2] (b3);
				\vertex [right=0.6cm of c2] (c3);
				\vertex [right=0.6cm of d2] (d3);

				\vertex [above=0.5cm of b2] (ta2) {\small $\left( \begin{array}{c} \omega_3 \\ \mathbf{p}_3 \end{array} \right)$};
				\vertex [below=0.5cm of c2] (td2) {\small $\left( \begin{array}{c} \omega_4 \\ \mathbf{p}_4 \end{array} \right)$};
				\vertex [right=0.0cm of b3] (tb3) {\small $\left( \begin{array}{c} E_{q_1} \\ \mathbf{q}_1 \end{array} \right)$};
				\vertex [right=0.0cm of c3] (tc3) {\small $\left( \begin{array}{c} E_{q_2} \\ \mathbf{q}_2 \end{array} \right)$};
 				
				\diagram*{
				(a1) --  [fermion] (b2) -- [fermion] (b3),  
				(d1)-- [fermion] (c2) -- [fermion] (c3),
				(c2) -- (b2)
				};
				
			\end{feynman}	
		\end{tikzpicture}
		$ \left. \begin{array}{c} \\ \\ \\ \\ \\\\  \\ \end{array} \right)^{*} $
\caption{While the non-linear unitarity relation \eqref{eqn:Aunitary} holds non-perturbatively, it can be applied in perturbation theory to relate the one-loop discontinuity (shown left) to the square of the tree-level amplitude (shown right), where the inequality holds whenever the kinematic states of particles 1 and 2 coincide with those of particles 3 and 4. Perturbative unitarity breaks down if the $| \mathcal{A}_{2\to 2}^{\rm loop}|$ required by this (perturbative) form of the unitarity condition is larger than the $| \mathcal{A}_{2 \to 2}^{\rm tree}|$ contribution.}
\label{fig:3}
\end{figure}

For two-particle intermediate states, the phase space integral $d \Pi_2$ can be performed exactly, yielding unitarity bounds with precise numerical coefficients. 
%First, we use the invariance under Lorentz boosts to transform to the centre-of-mass frame, in which $\mathbf{p}_1 + \mathbf{p}_2 = 0$. Then 
Performing the integrals over the six $\delta$-functions in $d \Pi_2$ leaves an integral over just two angular variables, which we choose to be the solid angle $d^2 \hat{\bf q}_1$. In the CoM frame, this becomes, 
\begin{align}
\int d \Pi_2  &=  \int \frac{d^2 \hat{\bf q}_1 }{ (4 \pi )^2 }  \frac{ | \mathbf{q}_1 | }{ \omega_s } =: \int \frac{ d^2 \hat{\bf q}_1 }{4\pi} \; \mathcal{N}_2  \, 
\end{align}
where $| {\bf q}_1 | = \sqrt{ (\omega_s/2)^2  - m^2 }$ is fixed by momentum conservation, and we define the constant factor arising from this phase space volume as $\mathcal{N}_2 = | \mathbf{q}_1 |/ 4\pi \omega_s$. 
%We will focus initially on the $n=2$ term in \eqref{eqn:Aunitary}, since for $4m^2 \leq s < 9m^2$ it is only the non-zero term (and will return to the other terms at the end of this section). 
The unitarity condition \eqref{eqn:Aunitary}, which contains this angular integral, therefore relates $\mathcal{A}$ at different kinematics (different scattering angles), making it a functional relation which is difficult to analyse. 

However, while we have accounted for translational invariance (factoring out an overall $\delta^4 ( \sum_j p_j )$) and Lorentz boosts (setting $\mathbf{p}_s = 0$), we have yet to exploit rotational invariance. This is achieved by transforming the angle $\theta$ to an angular momentum $\ell$ (which is conserved by the interactions in $\hat{T}$), resulting in the the well-known \emph{partial wave expansion},
%, which we will now review in the context of Lorentz-invariant theories, before turning to a new extension of this formalism in section~\ref{sec:32} which incorporates preferred-frame effects.   
%
%The partial wave expansion transforms the angle $\theta$ to an angular momentum $\ell$, 
\begin{align}
 \mathcal{A} ( \omega_s , \theta ) =   \mathcal{N}_a \, \sum_{\ell} \, a_{\ell} ( \omega_s ) f_{\ell} ( \theta ) 
\label{eqn:pw_1}
\end{align}  
where $\mathcal{N}_a$ is an overall normalisation, and the mode functions $f_{\ell} ( \theta ) = (2\ell + 1 ) \, P_{\ell} ( \cos \theta)$ are the longitudinal analogue of Fourier modes\footnote{
The mode functions $f_{\ell} (\theta)$ form a complete orthonormal basis,
\begin{align}
\frac{1}{2} \int_{0}^\pi d \theta \,  \sin \theta \; f_{\ell_1}^* ( \theta ) f_{\ell_2} ( \theta ) = (2 \ell_1 + 1 ) \delta_{\ell_1 \ell_2} \, ,
\label{eqn:florth}
\end{align} 
and satisfy an addition formula,
\begin{align}
 \int \frac{d^2 \hat{\bf q} }{4 \pi } \; f_{\ell_1} ( \hat{\bf p}_1 \cdot \hat{\bf q}  )  \;  f_{\ell_2}^* (  \hat{\bf p}_3 \cdot \hat{\bf q}  )  =  \delta_{\ell_1 \ell_2}  \;  f_{\ell_1} (  \hat{\bf p}_1 \cdot \hat{\bf p}_3 ) \, .  
 \label{eqn:fladd}
\end{align}
}.
Using \eqref{eqn:pw_1} in \eqref{eqn:Aunitary} gives,
\begin{align}
2 \, \text{Im} \, a_{\ell} ( \omega_s ) =  \mathcal{N}_a \mathcal{N}_2 \, | a_{\ell} ( \omega_s ) |^2  \; .
\end{align}
%Note that since $\omega_s$ is the same for $p_1p_2$ and $p_3 p_4$, the discontinuity $\text{Disc}_s$ of a function of $\omega_s$ only is simply its imaginary part. 
%This transformation to angular momentum $\ell$ has simplified the angular integrals appearing in \eqref{eqn:Aunitary}, and is now a non-linear condition on $a_{\ell} (\omega_s)$ at fixed $\ell$.
%Since interactions are rotationally invariant, the $\hat{S}$-matrix is diagonal when written in $\ell$-space, which greatly simplifies the unitarity condition \eqref{eqn:Aunitary}.
Thanks to rotational invariance, the $\hat{S}$-matrix is diagonal when written in $\ell$-space, and this has removed the angular integral appearing in the unitarity condition \eqref{eqn:Aunitary}.
Choosing the overall normalisation to be $\mathcal{N}_a = 2/\mathcal{N}_2$ gives a particularly simple condition on the $a_{\ell}$ coefficients: since $| a_{\ell} | \geq \text{Im} \, a_{\ell}$,
\begin{equation}
 | a_{\ell} ( \omega_s ) | \leq 1 \;\; \Rightarrow \;\;  |\text{Re} \, a_{\ell} (\omega_s) | \leq 1/2 \;\;  , 
 \label{eqn:al_U_CoM}
\end{equation}
and so every partial wave coefficient is bounded by unity. 

For example, for the interaction $C_{24} \,  (\partial \phi )^4 / f_\phi^{4}$, only two partial wave coefficients are non-zero at tree-level,
\begin{align}
a_0 (\omega_s )  =  \frac{ 5 C_{24} }{ 6 }   \frac{ \omega_s^4 }{ 4 \pi f_\phi^4 }        \;\;\;\; , \;\;\;\; 
a_2 (\omega_s )  =  \frac{ C_{24} }{ 30 }   \frac{  \omega_s^4 }{ 4 \pi f_\phi^4 }   \, . 
\end{align}
The unitarity condition \eqref{eqn:al_U_CoM} applied to these tree-level partial waves defines a maximum energy above which the perturbative unitarity is lost (when the loop shown in Figure~\ref{fig:3} becomes large). 
Although we have worked in a particular frame, with $| \mathbf{p}_s | = 0$, Lorentz symmetry allows us to replace $\omega_{s}^2 \to s$ and translate these bounds to a general frame,
\begin{align}
s^2  \lesssim  \frac{3}{5} \; 4 \pi \,  f_\phi^4 / | C_{24}|  \, . 
\end{align}
This is the numerically precise version of \eqref{eqn:CabUboundLI} when $n=2$. 

In summary, in a Lorentz-invariant theory one is free to work in the centre-of-mass frame with $\mathbf{p}_s=0$, in which the partial wave expansion,  
\begin{align}
 \mathcal{A}(\omega_s,\theta) =  \frac{16 \pi \, \omega_s }{ \sqrt{\omega_s^2 - 4m^2} } \sum_{\ell} a_{\ell} ( \omega_s ) (2\ell + 1 ) P_{\ell} ( \cos \theta ) \, , 
 \label{eqn:pw_flat}
\end{align}
contains coefficients $a_{\ell} (\omega_s)$ which are bounded by unitarity \eqref{eqn:al_U_CoM}. Using \eqref{eqn:florth} we can extract each of these coefficients, for instance when the masses are negligible,
\begin{equation}
  a_{\ell} ( \omega_s) = \frac{1}{32 \pi}  \int_{-1}^1 d \cos \theta  \; P_{\ell} (\cos \theta ) \; \mathcal{A} ( \omega_s, \theta )   \, ,
  \label{eqn:al_from_A}
\end{equation}
and this allows one to place unitarity constraints on parameters appearing in the EFT.

%%%%%%%%%%%%%%%%%%%%%%%%%%%%%%%%%%%%%%%%%%%%%%%%%%%%%%%%%%%%%%%%%%%%%%%%%%%%%%%%%%%%%%%%%%%%%%%%
%							Lorentz Breaking Unitarity
%%%%%%%%%%%%%%%%%%%%%%%%%%%%%%%%%%%%%%%%%%%%%%%%%%%%%%%%%%%%%%%%%%%%%%%%%%%%%%%%%%%%%%%%%%%%%%%%

%%%%%%%%%%%%%%%%
\subsection{EFTs with Broken Boosts}
\label{sec:32}
%%%%%%%%%%%%%%%%

In this section we discuss how to impose unitarity in a scalar field theory with broken boosts. 
Due to the lack of boost symmetry, the $n\to n$ scattering amplitude now depends on the center-of-mass velocity $\rho_s$, and in general there is both an $s_{\rm max}$ and a $| \mathbf{p}_s|_{\rm max}$ above which perturbative unitarity is lost. 
In particular, since boosts are broken by some preferred time-like direction $n^\mu$, the total momentum $\mathbf{p}_s$ can no longer be freely set to zero---this introduces a preferred spatial direction, as shown in figure~\ref{fig:cosmic_motion}, and for $2\to 2$ a more general partial wave expansion is required since the angular momentum of individual particles is no longer conserved. 

For computation, it is often most convenient to express amplitudes as an overall power of $\omega_s$ (the total energy) multiplying a dimensionless function of $\rho_s$ (the centre-of-mass velocity) or $\gamma_s = 1/\sqrt{1-\rho_s^2}$ (the centre-of-mass Lorentz factor). However, since it is not particularly transparent physically what a ``maximum velocity'' means for an EFT\footnote{
One might argue that the natural maximum for $\rho_s$ should be $c_s$, since an intermediate state with $\rho_s > c_s$ can be transformed using a $c_T$-preserving Lorentz boost to one with a negative invariant mass squared (which should be unstable). However, since the $c_T$-preserving Lorentz boosts are only a symmetry of theory in the UV, where $\pi$ is no longer the correct degree of freedom, it seems premature to conclude that there is any issue with having $\rho_s > c_s$ in the EFT. We will come back to this at the end of section \ref{sec:33}.
}, we will quote our final bounds in terms of $s$ (the centre-of-mass-frame energy) and $|\mathbf{p}_s|$ (the centre-of-mass momentum), where
\begin{align}
 \omega_s^2 = s + c_\pi^2 | \mathbf{p}_s |^2 = \gamma_s^2 s \, . 
\end{align}
$s$ is the internal energy that the system has (in its ``rest frame'', when $\mathbf{p}_s = 0$), while $
c_\pi | \mathbf{p}_s |$ is the energy the system has a result of its overall motion.
%\footnote{
%This is analogous to decomposing the total angular momentum into a rest frame contribution (i.e. the spin) plus an additional contribution (i.e. the orbital angular momentum). 
%}. 
Unitarity will place maximum thresholds on both $s$ and $|\mathbf{p}_s|$, which are readily interpreted as a bound on how strongly the particles can interact with each other (their relative energy, $s$) and a bound on how strongly the particles can interact with the background (their $|\mathbf{p}_s|$ relative to $n^\mu$).

%%%%%%%%
\subsubsection*{Elastic $n \to n$ Scattering}
%%%%%%%%

The $n$-particle phase space is again given by integrating over all future-pointing, on-shell 4-momenta which conserve the total ingoing momenta\footnote{
The constant factors of $\sqrt{-Z}$ arise from the Fourier transform, $f (p) = \int d^4 x \, \sqrt{-Z} \, e^{i p_\mu x^\mu} f (x)$
%\begin{align}
%f (p) = \int d^4 x \, \sqrt{-Z} \, e^{i p_\mu x^\mu} f (x)\, \quad \text{with inverse} \quad \quad f(x) = \int \frac{ d^4 p }{ (2\pi)^4 \sqrt{-Z}} \,  e^{-i p_\mu x^\mu} f (x)\,. 
%\end{align}
}, 
\begin{align}
d \Pi_n &= (2\pi)^4 \sqrt{-Z} \delta^4 ( p_s - \sum_i^n q_i   ) \;  \prod_{i=1}^n  \frac{d^4 q_i}{(2\pi)^4 \sqrt{-Z}} \;\; 2 \pi \; \delta \left(  \mathcal{E} ( q_i )  \right) \Theta ( q^0_i ) \,  
\end{align}
where $\mathcal{E} ( q ) =   Z^{\mu\nu} q_\mu q_\nu + m^2 $ is the classical equation of motion. 
%Below we will perform this integral explicitly for the case $n=2$, but before specialising we can already use \eqref{eqn:} to provide schematic unitarity bounds. 
Since this is invariant under rotations of $q_\mu$ which preserve the kinetic matrix $Z^{\mu\nu}$, we can again estimate the phase space volume, $\int d\Pi_n  \sim (4\pi)^{-1} \left( s / 16 \pi^2 \right)^{n-2}$, where now $s = - Z^{\mu\nu} p_{s\, \mu} p_{s\, \nu}$.
% since the overall momentum-conserving $\delta$-function can be used to fix $q_n$, leaving the on-shell condition $\delta ( \epsilon (q_n) )$ to fix $|\mathbf{q}_{n-1} |$---this leaves an integral over $(n-2)$ of the 3-momenta (each of which $\sim \omega_s^2 / 16 \pi^2$), plus an integral over the angular part of $\mathbf{q}_{n-1}$ (which $\sim 4\pi / 16\pi^2$).
Then the unitarity condition \eqref{eqn:Aunitary} implies the same upper bound \eqref{eqn:AnnLI} on the amplitude, providing $\mathcal{A}_{2 \to 2}$ is computed using $Z^{\mu\nu}$ as an effective metric.  
%\begin{align}
%$ \big| \mathcal{A}_{n \to n} \big|  \lesssim 4\pi  \left(  \frac{16\pi^2}{ \omega_s^2 } \right)^{n-2}   . $
%\end{align}

%There are two important kinematic differences: firstly, $\mathcal{A}_{n \to n}$ should now be computed in momentum space where $Z^{\mu\nu}$ is treated as an effective metric\footnote{
%See \cite{deRham:2017aoj} for how factors of $\cpi$ otherwise enter the bound if $\mathcal{A} (p_i)$ is computed using $g^{\mu\nu}$. 
%}, and secondly it is no longer true that the strongest constraint will come from setting $| \mathbf{p}_s |=0$ because the amplitude may depend explicitly on $| \mathbf{p}_s |$.

%For the interaction $c_{abc} ( n^\mu \partial_\mu )^c ( Z^{\mu\nu} \partial_\mu \partial_\nu  )^{a} \phi^b / f_\phi^{2a+b+c-4}$,
For the general interaction $C_{abc} ( n^\mu \partial_\mu )^c ( Z^{\mu\nu} \partial_\mu \partial_\nu  )^{a} \pi^b / f_\phi^{2a+b+c-4}$, the coefficient $C_{abc}$ can be constrained by estimating the tree-level amplitude from $V$ such vertices, 
\begin{align}
| \mathcal{A}_{n \to n}^{\rm tree} |  \; \sim \;  \left(  \frac{C_{abc} \; \omega_s^c s^a }{ f_\pi^{2a+b+c-4}}   \right)^{ V } \frac{1}{s^{V-1} }  
\end{align} 
where the number of external legs is $2n = (b-2)V + 2$ and we have neglected masses since we are interested in high energy ($s \gg 4m_\pi^2$) behaviour of this amplitude. The schematic unitarity bound \eqref{eqn:AnnLI} then requires that,
\begin{align}
s^{ ( 2a +b + c - 4 )/2 } \left(  1 + \frac{ | \mathbf{p}_s |^2 }{s}   \right)^{c/2}   \; \lesssim \; (4 \pi )^{b-2 - 1/V} f_\pi^{2a+b+c-4 } / | C_{abc} |  \; . 
 \label{eqn:CabUbound}
\end{align}
When the number of time derivatives $c \neq 0$ the range of allowed interaction energies $s$ depends on the centre-of-mass motion $| \mathbf{p}_s |$. 
The largest possible $s$ is achieved when $| \mathbf{p}_s | =0$, and coincides with the $s_{\rm max}$ result expected from the Lorentz invariant case (because $n^\mu \partial_\mu$ is now also $\propto \sqrt{s}$ in this frame). Now, at lower values of $s$, there is always a maximum $| \mathbf{p}_s |$ at which unitarity is violated: there is a separate strong coupling scale for the centre-of-mass motion (which can also be viewed as a maximum velocity) above which perturbative unitarity is violated. 

%The breaking of Lorentz boosts can be viewed as introducing a fixed background direction, $n_\mu = ( 1,0,0,0)$. 
A complementary way to view this scattering is to change to coordinates in which $\mathbf{p}_s = 0$, but since the system is no longer Lorentz-invariant this change affects the interaction strengths. In particular, the transformation to centre-of-mass coordinates results in a change in $n_\mu$,
\begin{align}
p_{s \, \mu} =\left(  \begin{array}{c}
\omega_s \\
\mathbf{p}_s  \end{array} \right)  \to p_{s \,\mu}' = \left( \begin{array}{c}
  \sqrt{ s }  \\
  \mathbf{0}
\end{array}\right)    \;\;\;\; \Rightarrow \;\;\;\; n_\mu = \left( \begin{array}{c} 
1 \\
\mathbf{0}  \end{array} \right) \to  n'_\mu = \left( \begin{array}{c} 
\omega_s / \sqrt{s}  \\
- \mathbf{p}_s / \sqrt{s}  \end{array} \right) \, ,
\end{align}
shown in Figure~\ref{fig:cosmic_motion}. 
Intuitively, the faster the whole system is moving with respect to the background, the larger the couplings induced by the background (since $|\mathbf{p}_s| \gg s$ corresponds to $n_\mu$ having large components). The system cannot move at arbitrarily high centre-of-mass velocities because the interaction strengths cannot be arbitrarily large without violating unitarity. 

Whenever $2a+b \geq 4$, the largest possible $|\mathbf{p}_s|$ is achieved when $s \sim 4m_\pi^2$ is a minimum,
\begin{align}
\left(  | \mathbf{p}_s |_{\rm max} \right)^{c} \sim  f_\pi^c \,  \frac{ (4 \pi )^{b-2-1/V} }{ |C_{abc}| } \, \left( \frac{ f_{\pi} }{ m_{\pi}} \right)^{2a+b-4} \; , 
\end{align}
%or when $s \sim m_\pi |\mathbf{p}_s|$
%\begin{align}
%| \mathbf{p}_s |^{(2a +b +c -4)/2}  | \mathbf{p}_s |^c \sim f_{\pi}^{(2a+b+c-4)/2} f_\pi^c \left( \frac{m_\pi}{f_\pi} \right)^{c/2} \,  \frac{ (4 \pi )^{b-2-1/V} }{ |C_{abc}| } \, \left( \frac{ f_{\pi} }{ m_{\pi}} \right)^{(2a+b-4)/2} \; , 
%\end{align}
with $V=1$ when $b$ is even and $V=2$ when $b$ is odd, and where typically $f_\pi / m_\pi \gg 1$ (since $f_\pi$ controls the EFT cutoff $\Lambda_{\rm max}$). 
However, something interesting happens when $a=0$, $b=3$: then unitarity demands that $| \mathbf{p}_s|$ be small at both large and small $s$,  reaching a maximum only at an intermediate value $m_\pi^2 \ll s \ll f_\pi^2$ (this is shown in Figure~\ref{fig:a1b1}).
For interactions of the form $(n^\mu \partial_\mu )^c \pi^3$, at a fixed $|\mathbf{p}_s|$ 
%exchange diagrams with $V = 2$ vertices lead to 
there is always a \emph{minimum} $s$ at which unitarity is violated, given by \eqref{eqn:CabUbound}---for example, when $|\mathbf{p}_s | \lesssim f_\pi$ this is approximately, 
%\begin{align}
%s^{-1 +2c }    \lesssim   m_\pi^{2c}  ( 16  \pi^2 )^{1-1/V} \; f_\pi^{2c-2} / | C_{03c} |^2 \;\;\;\; \Rightarrow  \;\;\;\;  \gamma_s^{2c} \gamma_s^{2c-2} \lesssim  4 \pi \; ( f_\pi / m_\pi )^{2c-2} / | C_{03c} |^2
%\end{align}
%when $s$ is fixed and $| \mathbf{p}_s |$ is increased to its kinematic maximum. 
%So cutoff when $c=3$ is $( m_\pi^{3} f_\pi^2 )^{1/5}$, so $\Lambda_{3/2}$ in massive gravity language.
%Conversely, if $|\mathbf{p}_s |$ is fixed and $s$ is decreased to it's minimum, $s \sim m | \mathbf{p}_s|$, then,
%\begin{align}
% ....
%\end{align}
%\begin{align}
% \gamma_s^{2c} s^{c-1} \lesssim  4 \pi \; f_\pi^{2c-2} / | C_{03c} |^2 \;\;\;\; \Rightarrow  \;\;\;\;  \gamma_s^{2c} \gamma_s^{2c-2} \lesssim  4 \pi \; ( f_\pi / m_\pi )^{2c-2} / | C_{03c} |^2
%\end{align}
%\begin{align}
%\frac{\left( s + | \mathbf{p}_s |^2 \right)^c }{s} \lesssim  4 \pi \; f_\pi^{2c-2} / | C_{03c} |^2
%\end{align}
%and cannot be satisfied for fixed $|\mathbf{p}_s|$ and $| C_{03c}|$ when $s \ll |\mathbf{p}_s|^2$. 
\begin{align}
s \gtrsim f_\pi^2 \left( \frac{ | \mathbf{p}_s | }{f_\pi}  \right)^{2c} \, \frac{ | C_{03c} |^2 }{4\pi}\, .
\label{eqn:smin}
\end{align}
Unless all $| C_{03c} |  \lesssim  m_\pi / f_\pi$ are suppressed (bringing this minimum $s$ below $4m_\pi^2$), scattering with low $s$ will violate perturbative unitarity.  
Physically, this is because a large $| \mathbf{p}_s |$ generates a large $\pi^3$ vertex, which allows a propagating $\pi$ particle to emit a single off-shell $\pi$. As $s$ is decreased this emission becomes longer lived and contributes more to the scattering amplitude, until eventually (perturbative) unitarity is lost. 
The reason this effect is not observed for higher-point interactions is that their $n>2$ particle phase space volume shrinks as $s$ decreases, and so the two competing effects (longer-lived emissions versus smaller available phase space) cancel out and unitarity is not violated. However, we stress that we have only considered forward limit scattering in which there are no hierarchies between the incoming momenta, i.e. all $n^\mu \partial_\mu \sim \omega_s$ and $Z^{\mu\nu} \partial_\mu \partial_\nu \sim s$. A more detailed study of higher $n$-point unitarity bounds, in which say the total $s$ is held fixed (so the phase space volume is fixed) but one of the channels $s_{ij} \ll | \mathbf{p}_s |^2$ (allowing for production of a long-lived intermediate), may lead to a similar IR cutoff on $s$ for any interaction of the form $(n^\mu \partial_\mu )^c \pi^b$. We will not pursue that further here, and instead focus on the $2 \to 2$ unitarity bound and its consequences for a $\dot \pi^3$ interaction.

%%%%%%%%
\subsubsection*{Elastic $2 \to 2$ Scattering}
%%%%%%%%

A scattering process between four on-shell momenta with $\sum_j p_{j\, \mu} = 0$ in a theory with broken boosts can depend on at most five independent kinematic variables: the two Lorentz invariant variables $\{ s, t \}$, plus now we have $n^\mu$ with which to extract three energies, $\{ n^\mu p_{1\mu} , n^\mu p_{2\mu} , n^\mu p_{3\mu} \}$.  
The analogue of the $(\omega_s, \theta)$ variables from the Lorentz-invariant problem are $(
\omega_s  ,   \rho_s  ,  \theta_1   ,  \theta_3  ,  \phi_{13}  
)$
where $\cos \theta_j$ is the angle between $\mathbf{p}_j$ and $\mathbf{p}_s$, and $\phi_{13} = \phi_1 - \phi_3$ where $\phi_j$ is the azimuthal angle about $\mathbf{p}_s$.  
To convert an amplitude $\mathcal{A} ( p_{1\mu} , p_{2\mu} , p_{3\mu}, p_{4\mu} )$ to these variables, first use momentum conservation to set $p_2^\mu = p_s^\mu - p_1^\mu$ and $p_4^\mu = -p_s^\mu - p_3^\mu$, then the remaining momenta (using two spatial rotations to align the $\hat{ \mathbf{z}}$ axis along $\mathbf{p}_s$) are,
\begin{align}
{\bf p}_s &= | \mathbf{p}_s |   \left( \begin{array}{c} 
0  \\
0  \\
1 \end{array}
\right) \;\; , \;\;\;\;  
&{\bf p}_1 &= | \mathbf{p}_1 |  \left(   \begin{array}{c} 
\sin \theta_1  \cos \phi_1    \\
\sin \theta_1   \sin \phi_1  \\
\cos \theta_1 \end{array}
\right) \;\; , \;\;\;\;  
&  {\bf p}_3 &= | \mathbf{p}_3 |  \left( \begin{array}{c} 
\sin \theta_3  \cos \phi_3    \\
\sin \theta_3   \sin \phi_3  \\
\cos \theta_3  \end{array}
\right) \;\; , \;\;\;\;  
\end{align}
where the energies are fixed by the four mass-shell conditions as\footnote{
We have neglected the small mass in \eqref{eqn:wj}. With finite mass, $\omega_j = \sqrt{ c_\pi^2 |\mathbf{p}_j |^2 + m_\pi^2}$, with
\begin{align}
c_\pi | \mathbf{p}_j | &= \frac{\omega_s}{2} \frac{ 1 -\rho_s^2 }{1  - \rho_s \cos \theta_j} \frac{  \sqrt{ 1 - \frac{4m^2}{\omega_s^2} \frac{ 1- \rho_s^2 \cos^2 \theta_j }{(1 - \rho_s^2 )^2}   } + \rho_s \cos \theta_j  }{   1  + \rho_s \cos \theta_j  }  \, . 
\label{eqn:wj_m}
 \end{align}
 The mass is unimportant providing the scattering energy is sufficiently large, $s > 4 \gamma_s^2 m_\pi^2$, where $\gamma_s = 1/\sqrt{1- \rho_s^2}$ (else the mass forbids large angles, where the square root becomes imaginary). 
%This is only real providing $\rho_s^2 \cos^2 \theta_j  > - \frac{ s }{ 4 \gamma_s^2 m_\pi^2 } + 1 $, i.e. need $s > 4 \gamma_s^2 m_\pi^2$ else the mass starts to become important at large angles. This places a maximum value on the speed, $\gamma_s^2 < \frac{s}{4 m_\pi^2}$. 
We will always work with $s \gg m_\pi^2$, and hereafter any $\gamma_s \gg 1$ limit is always taken with the understanding that it remains smaller than $s/4m_\pi^2$, i.e. $1 + |\mathbf{p}_s |^2/s  < s/4m_\pi^2  \gg 1$. This was not needed for our discussion of $n \to n$ scattering because that focussed solely on $\theta_j \approx 0$ (so that $\omega_j \sim \omega_s$).
},
\begin{align}
\omega_j =  c_s| \mathbf{p}_j | = \frac{\omega_s}{2} \frac{1 - \rho_s^2}{ 1 - \rho_s \cos \theta_j } ,
\label{eqn:wj}
\end{align}
and finally use the remaining spatial rotation (around $\mathbf{p}_s$) to set $\phi_1 + \phi_3 = 0$. 
Note that in the CoM frame, $\rho_s \to 0$, all particles have the same energy,
$ c_\pi | \mathbf{p}_j | =  \omega_s/2$, and in this limit the three angles can appear only in the combination,
$ \hat{ \bf p}_1 \cdot \hat{ \bf p}_3 = \cos \theta_1 \cos \theta_3 + \sin \theta_1 \sin \theta_3 \sin \phi _3 \equiv \cos \theta ,$ 
which is the usual relativistic scattering angle appearing in the Mandlestam invariants,
\begin{align}
 s_{12} = \omega_s^2 ( 1 - \rho_s^2 ) \;\;\;\; , \;\;\;\; s_{13} =  2 \omega_1 \omega_3 ( 1 - \cos \theta ) \, .
\end{align}

%\paragraph{Unitarity condition:}
Performing integrals over the six $\delta$-functions in $d \Pi_2$ leaves an integral over just two angular variables, which we choose to be the solid angle $d^2 \hat{\bf q}_1$ as before\footnote{
We note for completeness that for general masses the phase space element in a frame with $\rho_s \neq 0$ is given by,
\begin{align}
d \Pi_2 &=   \frac{d^2 \hat{\bf q}_1 }{ (4 \pi )^2 }  \frac{ c_\pi | \mathbf{q}_1 | }{ \omega_s } \left(  1 - \frac{ \rho_s E_{q_1} }{  c_\pi | \mathbf{q}_1 | }  \,   \hat{\bf p}_s \cdot \hat{\bf q}_1  \right)^{-1}    =: \frac{ d^2 \hat{\bf q}_1 }{4\pi} \; \mathcal{N}_2 ( \hat{\bf q_1 } ) \,  , 
\label{eqn:dPi2_m}
\end{align}
where $E_{q_1} = \sqrt{  c_\pi^2 | {\bf q}_1 |^2  - m^2}$ and $| {\bf q}_1 |$ is fixed as in \eqref{eqn:wj_m} in terms of the angle $\hat{\bf p}_s \cdot \hat{\bf q}_1$.
}, 
\begin{align}
 d \Pi_2 
 =  \frac{d^2 \hat{\bf q}_1 }{ (4 \pi )^2 } \;  \frac{2 E_{q_1}^2  }{ s }   
 =: \frac{ d^2 \hat{\bf q}_1 }{4\pi} \; \mathcal{N}_2 ( \hat{\bf q_1 } ) 
  \, ,
  \label{eqn:dPi2}
\end{align} 
where the energy $E_{q_1}$ is fixed as in \eqref{eqn:wj} in terms of the the angle between $\hat{\bf q}_1$ and $\mathbf{p}_s$, and we have defined the angle-dependent factor $\mathcal{N}_2 ( \hat{\mathbf{q}_1} ) = 2 E_{q_1}^2 / 4 \pi s$. 
%We will focus initially on the $n=2$ term in \eqref{eqn:Aunitary}, since for $4m^2 \leq s < 9m^2$ it is only the non-zero term (and will return to the other terms at the end of this section). 

We now seek an angular momentum expansion of the non-relativistic scattering amplitude analogous to the partial-wave expansion (\ref{eqn:pw_flat}) adapted to a general ${\bf p}_1 + {\bf p}_2 \neq 0$ frame. Since the amplitude now depends on both longitudinal and azimuthal angles the expansion should naturally contain both the total angular momentum $L^2$ (eigenvalues $\ell_j (\ell_j +1)$) as well as its projection $L_z$ (eigenvalues $m_j$) for each particle, 
\begin{align}
 \mathcal{A} ( p_1 p_2 \to p_3 p_4 )  =  \mathcal{N}_a \sum_{ \substack{ \ell_1, \ell_3 \\ m_1 m_3}} a_{\ell_1 \ell_3}^{m_1 m_3} ( \omega_s , \rho_s ) Y_{\ell_1}^{m_1} ( \hat{\bf p}_1 )  Y_{\ell_3}^{m_3*} (  \hat{\bf p}_3  ) 
 \label{eqn:all_exp}
\end{align}
where $\mathcal{N}_a$ is again an overall normalisation, and the spherical harmonics  $Y_\ell^m$ are a complete set of orthonormal mode functions on the sphere,
\begin{align}
\int  d^2 \hat{\bf q}  \; Y_{\ell_1}^{m_1} (  \hat{\bf q} )  Y_{\ell_2}^{m_2 *} (  \hat{\bf q} )   =  \delta_{\ell_1 \ell_2} \delta_{m_1 m_2} \, .
\label{eqn:fllorth}
\end{align}
As shown in Figure~\ref{fig:cosmic_motion}, even once $n^\mu$ is fixed (which uses all three boosts) and the direction of $\mathbf{p}_s$ is fixed (which uses two rotations), there is still one rotational freedom left -- this imposes a selection rule on the spherical wave coefficients, $a^{m_1m_3}_{\ell_1 \ell_3} \sim \delta_{m_1 m_3}$ (since angular momentum along $\mathbf{p}_s$ is conserved). 
We will write,
\begin{equation}
	\label{eq:selection_rule}
 a^{m_1 m_3}_{\ell_1 \ell_3} = \delta_{m_1 m_3} \; ( \mathfrak{a}^{m_1} )_{\ell_1 \ell_3}  
\end{equation}
and refer to $\mathfrak{a}^{m_1}$ as a matrix with indices $\{ \ell_1 , \ell_3 \}$ running from $m_1$ onwards.
Finally, our spherical wave coefficients $a^{m_1 m_3}_{\ell_1 \ell_3} ( \omega_s , \rho_s) $ in a general frame are related to the usual centre-of-mass coefficients $a_{\ell} ( \omega_s)$ by\footnote{
This follows from the Legendre addition formula \eqref{eqn:fladd}, which can be written as: 
$$ P_{\ell} ( \hat{\bf p}_1 \cdot \hat{\bf p}_3 ) = \frac{4 \pi}{2 \ell + 1 } \sum_m Y^{m}_{\ell} ( \hat{\bf p}_1 ) Y^{m*}_{\ell} ( \hat{\bf p}_3 )\,. $$
},
\begin{align}
 a_{\ell} ( \omega_s ) = \frac{1}{2 \ell + 1} \sum_{m=-\ell}^{+\ell}  \mathfrak{a}^{m}_{\ell \ell} ( \omega_s, 0 )  \, , 
\label{eqn:all_to_al}
\end{align}
which is effectively averaging over all possible $L_z$ projections when particles 1 and 3 have fixed total angular momentum $\ell (\ell + 1)$. 

Now that we are equipped with the spherical-wave expansion \eqref{eqn:all_exp}, 
we can apply the unitarity condition \eqref{eqn:Aunitary} to scattering amplitudes with no boost-invariance. 
Explicit substitution of \eqref{eqn:all_exp} into \eqref{eqn:Aunitary} yields the condition,
\begin{align}
\text{Disc}_s \, a^{m_1 m_3}_{\ell_1 \ell_3 } ( \omega_s,  \rho_s ) \;\; \geq  \;\; \sum_{ k, j } a^{m_1 k}_{\ell_1 j} ( \omega_s,  \rho_s ) a^{m_3 k *}_{\ell_3 j} ( \omega_s,  \rho_s )  \; , 
\label{eqn:330}
\end{align}
where we have taken care to fixe the overall normalisation, $\mathcal{N}_a = 8 \pi / \sqrt{ \mathcal{N}_2 ( \hat{\mathbf{p}}_1 ) \mathcal{N}_2 ( \hat{\mathbf{p}}_3 ) }$, such that the angular integral from $\int d \Pi_2$ becomes simply \eqref{eqn:fllorth}. 
In terms of the matrix $\mathfrak{a}^{m}$, 
\begin{align}
\frac{1}{2 i} \, \left(  \mathfrak{a}^{m}  -    \mathfrak{a}^{m \dagger}   \right)  \geq \mathfrak{a}^{m} \mathfrak{a}^{m \, \dagger}  \; , 
\end{align}
and our unitarity condition \eqref{eqn:330} can be seen as taking a particular matrix element of \eqref{eqn:Tunitary} (in a basis of 2-particle spherical wave states rather than a 2-particle plane wave state \cite{Richman:1984gh}).
Focussing on a particular diagonal element of $\mathfrak{a}^m$, we have,
\begin{equation}
 \text{Im} \, \mathfrak{a}^m_{\ell \ell}  \geq \sum_j | \mathfrak{a}_{\ell j}^m |^2   \geq  |  \mathfrak{a}^{m}_{\ell \ell} |^2 
 \label{eqn:all_U_diag}
\end{equation}
and hence an analogous boundedness to \eqref{eqn:al_U_CoM}, namely $| \text{Re} \,  \mathfrak{a}^m_{\ell \ell} (\omega_s, \rho_s) | < 1/2$, which now holds at \emph{any} centre-of-mass velocity. 
In fact, we show in Appendix \ref{sub:submatrix_proof} that this extends to any minor of $\mathfrak{a}^m$,
\begin{align}
 | \det \, \mathfrak{a}^m_{\rm sub}  |  < 1 
 \label{eqn:all_U}
\end{align}
where $\mathfrak{a}^m_{\rm sub}$ is any (finite) submatrix of $\mathfrak{a}^m$. We will focus on the simplest bound \eqref{eqn:all_U_diag}, and leave further exploration of the (infinite number of) unitarity bounds \eqref{eqn:all_U} for the future.

In Appendix~\ref{app:pw}, we show how to systematically evaluate the angular integrals in \eqref{eqn:all_from_A} for any $\mathcal{A}_{2 \to 2}^{\rm tree}$ (which is analytic up to simple poles in $s, t$ and $u$). There are three particular features which seem to apply quite generally,
\begin{itemize}

\item At small $\rho_s$, the strongest bound is from low $\ell$ modes. 

\item At large $\rho_s \to 1$, the strongest bound is from high $\ell$ modes. 

\item There is an apparent (at most dilogarithmic) branch cut at $\rho_s = \pm 1$, which arises from an (anti)collinear singularity and would be resolved by finite mass effects. Since we assume that $s \gg 4 m_\pi^2$, we never approach this cut. 

\end{itemize}
 
\begin{figure}
\includegraphics[height=4.5cm]{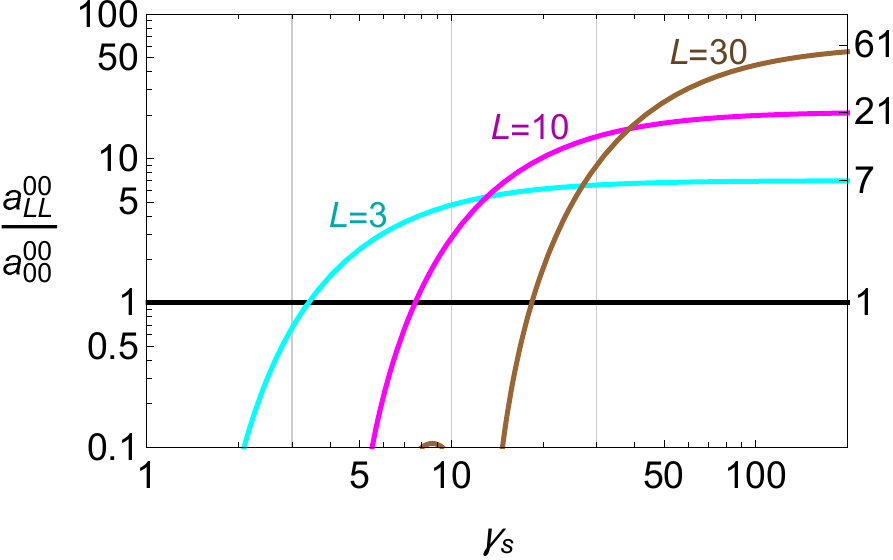}
\includegraphics[height=4.4cm]{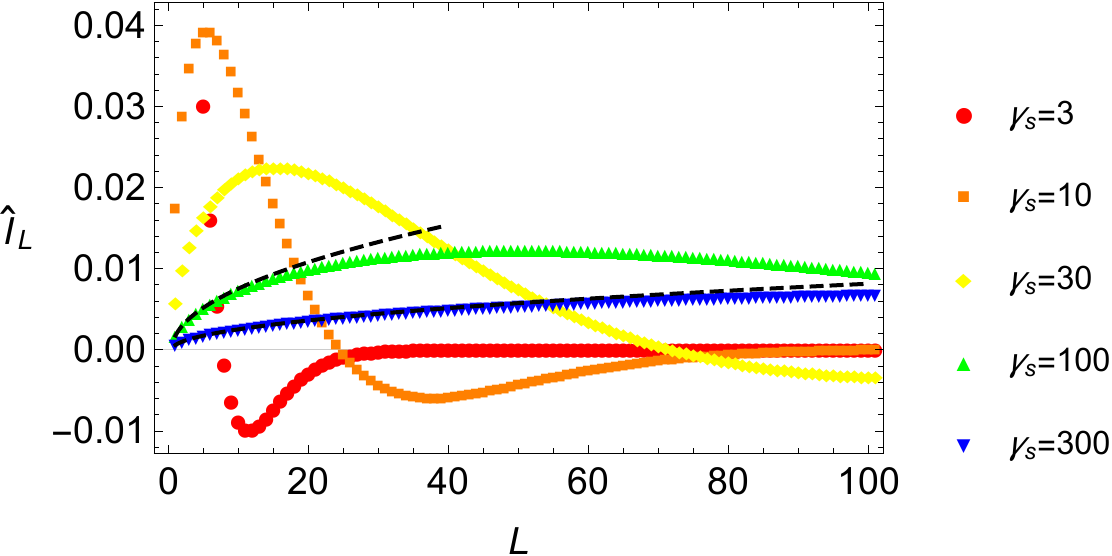}
\caption{Properties of the $\dot \pi^4$ spherical-wave amplitudes. 
Left: 
The ratio $a^{00}_{\ell \ell} / a^{00}_{00}$ is plotted against $\gamma_s = 1/\sqrt{1-\rho_s^2}$ for $L=3$, $10$ and $30$, and gray grid lines show $\gamma_s = 3, 10$ and $30$.
At small $\gamma_s$ it is $a^{00}_{00}$ which dominates, while at large $\gamma_s$ the $a^{00}_{\ell \ell}$ amplitude is larger by a factor of $2\ell +1$. The $\gamma_s$ at which $a^{00}_{\ell \ell}$ first exceeds $a^{00}_{00}$ scales $\propto \ell$, so terms with $\ell \gg \gamma_s$ can always be neglected from the sum in \eqref{eqn:Ub1}. Right: The angular integral $\hat{I}_{\ell}$ is plotted against $\ell$ at fixed $\gamma_s = 3, 10, 30, 100$ and $300$. In general they display a large positive maximum at a finite $\ell$ near to $\gamma_s$, which is followed by a smaller negative minimum before they approach zero from below as $\ell \to \infty$ (sufficiently fast for the spherical wave expansion \eqref{eqn:pw_n} to converge). Black dashed lines show the approximation \eqref{eqn:Ismallrsb1} for $\gamma_s = 100$ and $\gamma_s = 300$, agreeing well with $\hat{I}_{\ell}$ at sufficiently small $\ell < \gamma_s/\sqrt{2}$.}
\label{fig:Ib1}
\end{figure}

\begin{figure}
\includegraphics[width=0.5\textwidth]{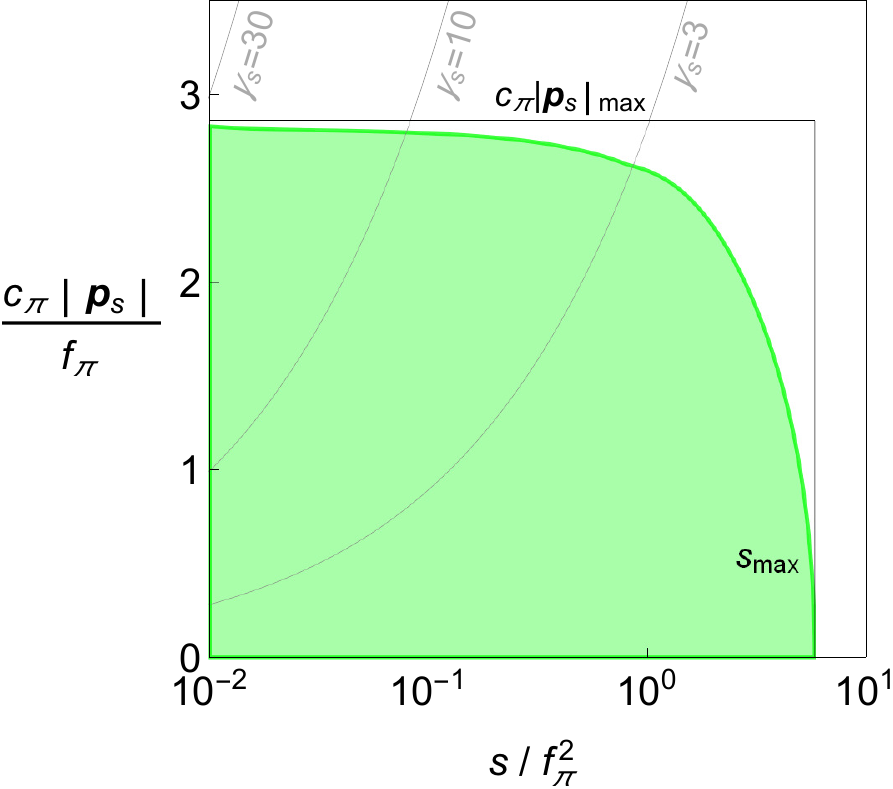}
\includegraphics[width=0.5\textwidth]{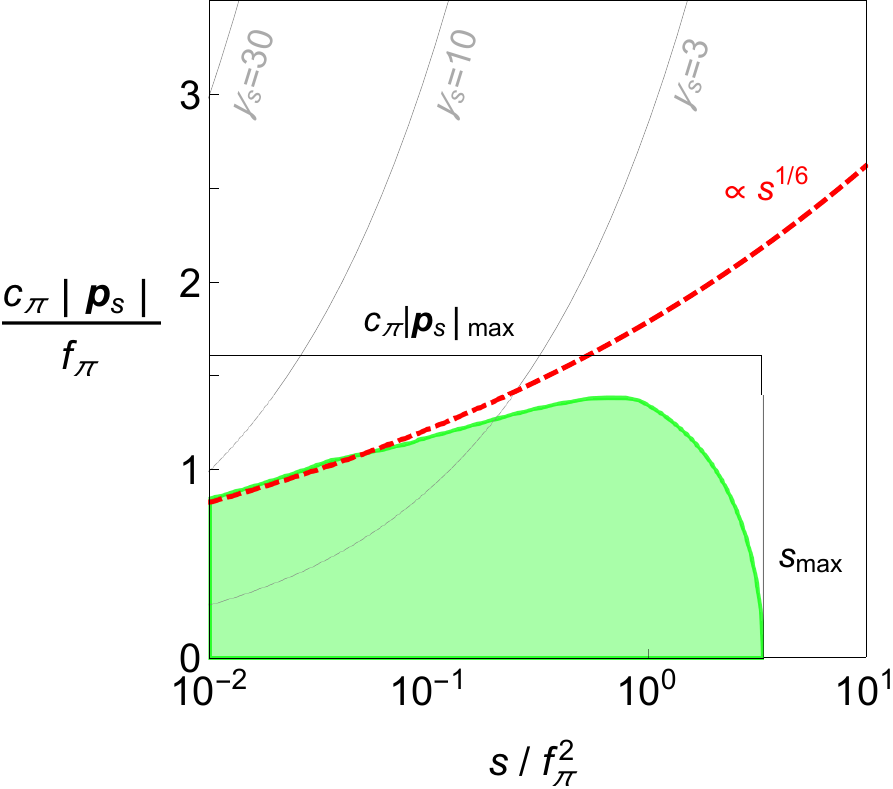}
\caption{The values of $s$ and $|\mathbf{p}_s|$ consistent with perturbative unitarity for a $\dot \pi^4/f_\pi^4$ interaction (left) and a $\dot \pi^3/f_\pi^2$ interaction (right). 
Gray grid lines show $\gamma_s = 3, 10$ and $30$---the spherical waves with $\ell\approx \gamma_s$ are most responsible for the constraints.  
The cubic vertex has the special feature that at fixed $|\mathbf{p}_s|$ there is a minimum $s$, and the black dashed line shows the $s^{1/6}$ scaling expected from \eqref{eqn:smina1}.
The maximum $s_{\rm max}$ and $|\mathbf{p}_s|_{\rm max}$ are given in \eqref{eqn:smaxpmaxb1} with $\alpha_1 =1$ and in \eqref{eqn:smaxpmaxa1} with $\beta_1 =1$ respectively---note that the actual maximum $|\mathbf{p}_s|$ for the cubic vertex is set by $\ell \approx 1$ spherical-waves and is lower than the upper bound in \eqref{eqn:smaxpmaxa1} from the $\ell = 0$ wave. 
}
\label{fig:a1b1}
\end{figure}

For example, consider the single interaction $\beta_1 \dot \pi^4 / f_\pi^4$. The corresponding scattering amplitude is $\mathcal{A} = 24 \beta \omega_1 \omega_2 \omega_3 \omega_4 / f_\pi^4$, and since this does not depend on $\phi_{13}$ the only non-zero spherical-wave amplitudes are those with $m_1=m_3=0$. Furthermore, since $\mathcal{A}$ can be written in the form $f (\omega_1) f (\omega_3)$, it has the convenient property that the spherical-wave amplitudes factorise into a product of two angular integrals,
\begin{align}
 a^{00}_{\ell_1 \ell_3} ( \omega_s , \rho_s ) = \beta_1 \frac{\omega_s^4}{ f_\pi^4 } \hat{I}_{\ell_1} ( \rho_s ) \hat{I}_{\ell_3}^* (\rho_s) 
 \label{eqn:IIb1}
\end{align}
which are given in \eqref{eqn:Ihatb1} and plotted in Figure~\ref{fig:Ib1}. Perturbative unitarity then requires that,
\begin{align}
 \beta_1 \frac{ \omega_s^4}{f_\pi^4} \; \sum_{\ell} | \hat{I}_{\ell} ( \rho_s ) |^2  \; \leq  \; 1  \; . 
 \label{eqn:Ub1}
\end{align}
%which comes from truncating the sum in \eqref{eqn:all_U_diag} at $\ell$. 
%The sum on the left-hand side is plotted in Figure~\ref{fig:a1b1} up to various maximum $\ell$, and exhibits the feature that 
At small $\rho_s$ the constraining power comes mostly from $\ell =0$, but as $\rho_s$ is increased the higher $\ell$ modes begin to dominate the sum. 
In particular, when $\rho_s \to 0$ only the first spherical-wave coefficient,
\begin{align}
 a^{00}_{00} ( \omega_s  , \rho_s )  =  \frac{3 \beta_1}{32 \pi} \,  \frac{\omega_s^4}{f_\pi^4}   ( 1 - \rho_s^2 ) 
\label{eqn:a00b1}
\end{align}
is non-zero, as required by \eqref{eqn:all_to_al}. But as $\rho_s$ approaches $1$ the higher order spherical-waves become important and the integrals take the form,
\begin{align}
\hat{I}_{\ell} (\rho_s ) = \sqrt{ \frac{ 2 \ell + 1 }{4\pi} } \frac{ \sqrt{3} }{2 \gamma_s }  \left[ 1  +   \frac{ 2 \ell^2 }{\gamma_s^2} \log \left(  \frac{ \ell }{2 \gamma_s}  \right)   + \mathcal{O} \left( \frac{\ell}{\gamma_s^2} ,  \frac{\ell^2}{\gamma_s^2}   \right)   \right]  \, , 
 \label{eqn:Ismallrsb1}
\end{align}
where $\gamma_s = 1/\sqrt{1-\rho_s^2}$ and we have assumed that $1 \ll \ell <  \gamma_s$ in order to perform an expansion of \eqref{eqn:IalF}. By summing only up to $\ell \approx \gamma_s/\sqrt{2}$, we can discard all but the first term of \eqref{eqn:Ismallrsb1} and write a simple approximation for the sum when $\gamma_s$ is large, $\sum_{\ell} | \hat{I}_{\ell}  |^2 \approx  3 /64 \pi$.
Unitarity \eqref{eqn:Ub1} is therefore violated if $s$ exceeds the $s_{\rm max}$ threshold defined by \eqref{eqn:a00b1} at small $|\mathbf{p}_s|$, or if $| \mathbf{p}_s |$ exceeds the $|\mathbf{p}_s |_{\rm max}$ threshold defined using \eqref{eqn:Ismallrsb1} at small $s$, 
\begin{align}
 s_{\rm max}  = 4 f_\pi^2 \left(  \frac{2 \pi }{ 3 \beta_1}   \right)^{1/2}   \;\;\;\; c_\pi | \mathbf{p}_s |_{\rm max} =  2 f_\pi \left(  \frac{4 \pi }{3 \beta_1 }  \right)^{1/4}  \; .
 \label{eqn:smaxpmaxb1}
\end{align}
The full region of $\{s , |\mathbf{p}_s | \}$ compatible with perturbative unitarity is shown in Figure~\ref{fig:a1b1}.

Now consider the cubic interaction, $\alpha_1 \dot \pi^3/ f_\pi^4$. The $t$- and $u$-channel exchange poles in $\mathcal{A}$ lead to more complicated expressions for the $a^{m_1 m_3}_{\ell_1 \ell_3}$---in particular, they are non-zero for all $m_1=m_3$ and do not factorise like \eqref{eqn:IIb1}---however they still exhibit the same three features as the $\beta_1$ amplitude above (and can be found in the Appendix). In particular, in the low $\rho_s$ regime only the first spherical-wave coefficient,
\begin{align}
 a^{00}_{00} ( \omega_s , \rho_s )  = - \frac{3 \alpha_1^2 }{64 \pi} \frac{\omega_s^4}{f_\pi^4}  (3 + 2 \rho_s^2 ) 
 \label{eqn:a00a1}
\end{align}
is non-zero, again consistent with \eqref{eqn:all_to_al}.
In the high $\rho_s$ regime,
\begin{align}
 a_{\ell_1 \ell_3}^{m_1 m_3} (\omega_s , \rho_s ) = -  \frac{15 \alpha_1^2 }{64 \pi}  \frac{\omega_s^4}{f_\pi^4 }  \sqrt{2 \ell_1 + 1} \sqrt{2 \ell_3 +1 }  \left(  \delta^{m_1}_0  \delta^{m_3}_0  +  \mathcal{O} \left(  \frac{\ell_j}{ \gamma_s} \, , \, \frac{\ell_j^2}{\gamma_s^2}   \right) \right) \, . 
 \label{eqn:smallrsa1}
\end{align} 
When $\ell_1 = \ell_3 $, the first subleading term is $\frac{12}{5} \, \ell^2 / \gamma_s^2  \log \left( \ell / \gamma_s \right)$, so once again we can approximate \eqref{eqn:330} by truncating the sum at $\ell^2 \approx \frac{5}{12} \gamma_s^2 $ and using only the leading term in \eqref{eqn:smallrsa1}, which gives $\left( \sum_{j} a_{\ell j} a_{\ell j}^* \right) /  a_{\ell \ell}  \approx  - \frac{ 15 \alpha_1^2}{64 \pi} \frac{\omega_s^4}{f_\pi^4}  \frac{5 \gamma_s^2}{12}$ when $\gamma_s$ is large. 
Unlike for the quartic interaction, as $s$ is made small at fixed $|\mathbf{p}_s|$ perturbative unitarity becomes a \emph{stronger} constraint: arbitrarily small $s$ are not allowed, since demanding $\frac{ 15 \alpha_1^2}{64 \pi} \frac{\omega_s^4}{f_\pi^4}  \frac{5 \gamma_s^2}{12} < 1 $ requires,
\begin{align}
s \;  \geq \;\frac{25}{64} \;  f_\pi^2 \left(  \frac{ |\mathbf{p}_s | }{ f_\pi }  \right)^6  \,  \frac{\alpha_1^2}{4 \pi}
\label{eqn:smina1}
\end{align}
when $\gamma_s \approx | \mathbf{p}_s|^2/s \gg 1$. This agrees with \eqref{eqn:smin}, anticipated earlier using dimensional analysis, now with a numerical prefactor which has been determined by performing the $d \Pi_2$ integral explicitly using the spherical-wave expansion.  
The full region of $\{s , |\mathbf{p}_s | \}$ compatible with perturbative unitarity is shown in Figure~\ref{fig:a1b1}. The maximum $|\mathbf{p}_s|$ is no longer reached at small $s$, but rather at some intermediate scale, and cannot be larger than that allowed by\footnote{
The $a^{00}_{00}$ mode can provide this upper bound on $|\mathbf{p}_s |_{\rm max}$ at small $s$ since it is a monotonic function of $\rho_s$, while the higher $\ell$ modes are not. 
} \eqref{eqn:a00a1}, which can be used to infer both the maximum allowed energy and a maximum possible momentum,
\begin{align}
 s_{\rm max}  = 4 f_\pi^2 \left(   \frac{ 2 \pi }{ 9 \alpha_1^2}   \right)^{1/2}   \;\;\;\; c_\pi | \mathbf{p}_s |_{\rm max} \leq  2 f_\pi \left(  \frac{ 2 \pi }{15 \alpha_1^2 }  \right)^{1/4}  \; .
 \label{eqn:smaxpmaxa1}
\end{align}

In summary, in a general frame in which ${\bf p}_s  \neq 0$, the spherical wave expansion\footnote{
\eqref{eqn:pw_n} has used the small mass expression for $\mathcal{N}_2$ given in \eqref{eqn:dPi2}, but more generally if \eqref{eqn:dPi2_m} is used in \eqref{eqn:all_exp} with $\mathcal{N}_a = 8\pi / \sqrt{\mathcal{N}_2 (\hat{\bf p}_1 )\mathcal{N}_2 (\hat{\bf p}_3 )  }$ then the unitarity bound \eqref{eqn:330} holds for any finite mass. 
},
\begin{align}
\mathcal{A}(\omega_s,\rho_s,\theta_1,\theta_3,\phi_{13}) = \frac{ 16 \pi^2 \omega_s^2 (1 - \rho_s^2 )  }{  \omega_1 \omega_3 }   \sum_{ \substack{ \ell_1, \ell_3  }} \sum_{\substack{-\ell_1 < m_1 < \ell_1 \\ -\ell_3 < m_3 < \ell_3}}  a_{\ell_1\ell_3}^{m_1m_3}(\omega_s,\rho_s) Y^{m_1 *}_{\ell_1}(\hat{\vec p}_1) Y^{m_3}_{\ell_3}(\hat{\vec p}_3) \; , 
\label{eqn:pw_n}
\end{align}
contains coefficients $a^{m_1 m_3}_{\ell_1 \ell_3} = \delta_{m_1 m_2} ( \mathfrak{a}^m )_{\ell_1 \ell_3}$ in which all minors of $\mathfrak{a}^m$ are bounded as in \eqref{eqn:all_U}. Using \eqref{eqn:fllorth}, we can extract each of these coefficients,  
\begin{equation}
 a^{m_1 m_3}_{\ell_1 \ell_3} (  \omega_s,  \rho_s  ) =  \frac{ 1}{ 1- \rho_s^2} \int \frac{d^2 {\bf p}_1}{4 \pi} \frac{d^2 {\bf p}_3}{4 \pi} \, Y^{m_1}_{\ell_1} ( \hat{\bf p}_1 ) Y^{m_3*}_{\ell_3} ( \hat{\bf p}_3  ) \;  \frac{\omega_1}{\omega_s}  \, \frac{\omega_3}{\omega_s} \mathcal{A} ( p_1 p_2 \to p_3 p_4 ) \, ,
 \label{eqn:all_from_A}
\end{equation}
and thus translate the unitarity condition \eqref{eqn:all_U} into bounds on EFT coefficients. 
In Appendix~\ref{app:pw}, we show how to systematically evaluate the angular integrals in \eqref{eqn:all_from_A} for any $\mathcal{A}_{2 \to 2}^{\rm tree}$ (which is analytic up to simple poles in $s, t$ and $u$). 
We will now apply these new spherical-wave unitarity bounds to the EFT of Inflation. 

%\begin{figure}
%\includegraphics[width=0.475\textwidth]{fig4a.eps}
%\includegraphics[width=0.525\textwidth]{fig4b.eps}
%\caption{
%For fixed parameters $c_s = 0.032$ and $\alpha_1 = -250$, the left panel shows that at a fixed center-of-mass energy ($s = 10^{-2} f_\pi$) there is a critical $| \mathbf{p}_s |$ above which the minimum $\beta_1$ exceeds the maximum $\beta_1$ allowed by perturbative unitarity, and the right panel shows how this critical $|\mathbf{p}_s|$ changes as $s$ is varied. The red region under the curve represents the kinematics allowed by perturbative unitarity. At very small $|\mathbf{p}_s|$, there is a maximum center-of-mass energy. As $s$ is decreased, more of the system's energy may go into the center-of-mass motion $|\mathbf{p}_s|$ without violating unitarity, however once $s$ becomes too small the exchange contributions becomes large (this is eventually limited by the particle masses, which we have neglected). 
%}
%\end{figure}

%%%%
\subsection{EFT of Inflation}
\label{sec:33}
%%%%

Since in the decoupling limit the EFT of Inflation becomes a simple theory of a single scalar field with a preferred direction, $n^\mu$, we can use the unitarity constraints developed above to constrain the Wilson coefficients $\alpha_1$, $\beta_1$ of \eqref{eq:EFT_action_decoupling_intro}.
In particular, 
%going to momentum space, and writing each particle's momenta as $( p_i )_\mu = ( \omega_i , {\bf p}_i )$, we have,
the momentum space scattering amplitude mediated by $S[\pi]$ is,
\begin{align}
f_\pi^4 \mathcal{A} ( p_1 p_2 \to p_3 p_4 ) &=   24  \beta_1 \omega_1 \omega_2 \omega_3 \omega_4   
+ 2  \beta_2 \left(   \;   \omega_{12}^{34}  s_{12}   +  \omega_{13}^{24}  s_{13} + \omega_{14}^{23}  s_{23}     \right)  
+2 \beta_3 \left(  s_{12}^2 + s_{13}^2 + s_{23}^2   \right)    \nonumber \\ 
&+ \frac{ 1 }{s_{12}} \left( 6  \alpha_1 \omega_1 \omega_2 \omega_{12}   +  2 \alpha_2  \omega_{12}  s_{12}   \right) \left( 6  \alpha_1 \omega_3 \omega_4 \omega_{34}   +   2\alpha_2  \omega_{34}  s_{34}    \right)   \nonumber \\
&+  \frac{ 1 }{s_{13}  } \left( 6 \alpha_1 \omega_1 \omega_3 \omega_{13}   +  2 \alpha_2   \omega_{13}  s_{13}  \right) \left( 6  \alpha_1 \omega_2 \omega_4 \omega_{24}   +  2 \alpha_2   \omega_{24}  s_{24}    \right)   \nonumber \\
&+  \frac{ 1 }{s_{23}} \left( 6 \alpha_1 \omega_1 \omega_4 \omega_{14}   +  2 \alpha_2   \omega_{14}  s_{14}   \right) \left( 6 \alpha_1 \omega_2 \omega_3 \omega_{23}   +  2 \alpha_2   \omega_{23}  s_{23}    \right)   
\label{eqn:Acomp}
\end{align}
where $s_{ij} = - Z^{\mu\nu} ( p_i + p_j )_\mu (p_i + p_j )_\nu $, $\omega_{ij} = \omega_i + \omega_j$ and $\omega_{ij}^{kl}= \omega_i \omega_j + \omega_k \omega_l$, and we have used momentum conservation and mass-shell conditions (neglecting all terms in $m_\pi$ to be consistent with neglecting $\eta_*$ terms in the effective action).

%%%%
\subsubsection*{Scattering in the Inflationary Rest Frame}
%%%%

%Before we conclude this discussion of Lorentz invariant perturbative unitarity we recall how this condition was applied to the EFT of Inflation in \cite{Baumann:2015nta}. 
Let us recall how this amplitude was used in \cite{Baumann:2011su, Baumann:2014cja}. 
By enforcing the restriction $\mathbf{p}_s = 0$ by hand (i.e. focussing on a particular choice of kinetmatics), then one may use the partial wave expansion \eqref{eqn:pw_flat}. 
Taking the amplitude \eqref{eqn:Acomp}  in the frame \eqref{eqn:kine_CoM}, one finds that the only non-zero partial waves are:
\begin{align}
 a_0 ( \omega_s ) &= \frac{1}{192 \pi} \left( -3 ( 3 \alpha_1 + 4 \alpha_2 )^2 + 18 \beta_1 + 24 \beta_2 + 40 \beta_3   \right)   \frac{\omega_s^4}{ f_\pi^4}   \\ 
 a_2 ( \omega_s ) &=  \frac{ \beta_3 }{120 \pi} \; \frac{\omega_s^4}{ f_\pi^4} \, . 
\end{align}
Since perturbative unitarity is lost when $| \text{Re} \, a_{\ell} (\omega_s )|$ first reaches $1/2$, which by definition does not happen at energies below the strong coupling scale, $\omega_s \lesssim 2 \Lambda_{\mathcal{O}}$, each of these partial waves gives an upper bound on $\Lambda_{\mathcal{O}}$. 
Since $a_2$ depends only on $\cpi$, it gives a bound which is independent of the bispectrum and trispectrum,
\begin{align}
s_{\rm max}^2  &=  \; f_\pi^4 \times  \frac{480\pi \cpi^4}{ 1 - \cpi^2 }   \, .
\label{eqn:CoM_U_cs}
\end{align}
%and is only corrected by loops and higher derivative terms. 
As described in \cite{Baumann:2014cja}, since the unitarity bound from $a_0$ mixes $\cpi$ and $\alpha_1$ with $\beta_1$ (which is far less constrained observationally \cite{Bartolo:2010di, Smith:2015uia, Akrami:2019izv}), it is essentially always possible to tune $\beta_1$ such that the bound is satisfied. For example, when $\beta_1 \gg \alpha_1 \gg 1 - \cpi^2$, we can focus on the contributions from $\alpha_1^2$ and $\beta_1$ only, 
\begin{align}
 a_0 ( \omega_s ) = \frac{3}{64 \pi} \; \frac{\omega_s^4}{ f_\pi^4} \left( - 3 \alpha_1^2 + 2 \beta_1 \right)  \,, 
 \label{eqn:lim0}
\end{align}
and so if $2 \beta_1 \approx 3 \alpha_1^2$ then this bound does not lower $\Lambda_{\mathcal{O}}$ even if $\alpha_1$ is very large. 
 
However, since the amplitude \eqref{eqn:Acomp} is not Lorentz invariant, focussing on the CoM frame restricts the possible kinematics of the scattering particles unnecessarily. 
%Specifically, one is restricting their attention to processes in which the total incoming momentum, $p_1^\mu + p_2^\mu$, coincides with the cosmic rest frame, $n^\mu$. 
Satisfying \eqref{eqn:al_U_CoM} is therefore necessary but not sufficient for the $2\to2$ scattering processes to respect perturbative unitarity, since we must also demand the unitarity condition \eqref{eqn:Aunitary} in other frames. One immediate consequence of this is that although a tuning (such as $3 \alpha_1^2 \approx 2 \beta_1$) always seemed possible to guarantee perturbative unitarity, this is an artefact of CoM scattering -- we will now use scattering in a more general frame to show that $\alpha_1$ and $\beta_1$ are separately bounded.  %(see Figure~\ref{fig:U}).   

%%%%
\subsubsection*{Beyond the Inflationary Rest Frame}
%%%%

Using the spherical wave expansion \eqref{eqn:pw_n} at an arbitrary centre-of-mass velocity, there is an upper bound on the strong coupling scale for every $\mathfrak{a}^{m}_{\ell \ell} ( \omega_s, \rho_s)$ -- for every choice of $m$ and $\ell$, there is a whole function of $\rho_s$ which must be bounded. A systematic approach for computing these spherical wave coefficients is described in Appendix~\ref{app:pw}. 
%Although the $t$ and $u$ channel exchange diagrams contribute to all possible $\ell$ and $m$, here we will focus on $\ell \leq 4$ only. 
For example, when $\beta_1 \gg \alpha_1 \gg 1 - \cpi^2$, the first spherical wave coefficient is simply,
\begin{align}
   \mathfrak{a}^{0}_{00} ( \omega_s , \rho_s )  =  \frac{3}{ 64 \pi}  \frac{\omega_s^4}{ f_\pi^4}  \left[  ( - 3 + 2 \rho_s^2 ) \alpha_1^2  + ( 2 - 2 \rho_s^2 ) \beta_1 \right]  \, . 
 \label{eqn:a000}
\end{align}
Now, a tuning like $2 \beta_1 \approx 3 \alpha_1^2$ is no longer enough to guarantee perturbative unitarity of the $2 \to 2$ scattering amplitude for all centre-of-mass velocities. 
%The unitarity bounds from $\ell_1 =\ell_3 = 1$ result in three further bounds ($m=-1, 0, +1$), and those from $\ell_1 = \ell_3 = 2$ result in five further bounds ($m =-2,-1,0,1,2$)\footnote{
%In fact, thanks to the $t$ and $u$ channel exchange contributions, there is a non-zero $\mathfrak{a}^m_{\ell_1 \ell_3}$ for \emph{every} possible $\ell_1$ and $\ell_3$. 
%}. 
%Together with \eqref{eqn:a000}, 
In fact, since each spherical-wave amplitude gives a separate bound of the form,
\begin{align}
| \mathfrak{a}^m_{\ell \ell} ( \omega_s , \rho_s) | = \frac{1}{4 \pi} \, \frac{\omega_s^4}{ f_\pi^4} \left| F^m_{\ell} ( \rho_s ) \alpha_1^2 + G^m_{\ell} (\rho_s ) \beta_1 \right| < \frac{1}{2} \, ,
\label{eqn:lim1}
\end{align}
where $F^m_{\ell}$ and $G^m_{\ell}$ are known functions of $\rho_s$,  
unitarity requires that both $\alpha_1$ and $\beta_1$ be separately bounded (since there is no tuning for $\beta_1$ which simultaneously satisfies all bounds. 

Since the observational constraints on $\beta_1$ are much weaker than those on $\alpha_1$ and $c_s$, our goal is now to remove $\beta_1$ from these unitarity bounds.
We do this by noting that, for a fixed $\{ \alpha_1, c_s  \}$, each spherical wave amplitude implies there is a maximum $\beta_1^{\rm max}$ (at which $\mathfrak{a}^m_{\ell \ell} = 1/2)$ and a minimum $\beta_1^{\rm min}$ (at which $\mathfrak{a}^m_{\ell \ell} = - 1/2$). For example, from \eqref{eqn:a000}, these maximum and minimum values occur at,
\begin{align}
\beta_1 =  \frac{1}{2 (1- \rho_s^2)} \left[  \pm  \frac{32 \pi}{3} \frac{f_\pi^4}{ \omega_s^4 }  + ( 3- 2 \rho_s^2) \alpha_1^2   \right]   \, . 
\label{eqn:lim2}
\end{align}
When $\rho_s = 0$, this restricts $\beta_1$ to the range $2 \beta_1 = 3 \alpha_1^2  \pm 32 \pi f_\pi^4 /3 \omega_s^4$. As $\rho_s$ is increased (at fixed $\omega_s$), this allowed range of $\beta_1$ shifts, until eventually there are no real values of $\beta_1$ which can simultaneously satisfy both bounds. This happens when,
\begin{align}
\beta_1^{\rm min} (\omega_s, \rho_s )  \geq \beta_1^{\rm max} ( \omega_s,  0 )  \;\;\;\; \Rightarrow \;\;\;\; 
\omega_s^4 \;  \frac{ \rho_s^2 }{ 2 - \rho_s^2}   \geq f_\pi^4 \,  \frac{32 \pi}{15 \alpha_1^2}    
\label{eqn:lim3}
\end{align}
irrespective of $\beta_1$. In particular, when $s$ is made very small this implies that $| \mathbf{p}_s |_{\rm max} \leq 2 f_\pi^4 \left(  2 \pi / 15 \alpha_1^2  \right)^{1/4}$, which agrees with \eqref{eqn:smaxpmaxa1} since at small $s$ it is the $\alpha_1^2$ interaction which dominates the spherical-wave amplitudes. Comparing $\beta_1^{\rm max}$ with $\beta_1^{\rm max}$ in this way for every $\mathfrak{a}^m_{\ell \ell}$ then provides a set of constraints on $\alpha_1$ independently of $\beta_1$. 
We also mention in passing that another way to remove $\beta_1$ is to look for bounds \eqref{eqn:lim1} where its coefficient $G^m_{\ell} (\rho_s)$ vanishes. As discussed in section~\ref{sec:22}, this happens whenever $m\neq 0$, and also when $m=0$ for particular values\footnote{
For $\ell > 2$ there is always at least one zero,
e.g. for $\ell = 3$ this is at $\rho_s = 0.715931...$, and at $\ell = 4$ this is at $\rho_s = 0.842835...$.
In fact Figure~\ref{fig:Ib1} shows that for any fixed $\rho_s$ there is always exactly one value of $\ell$ (not necessarily an integer) for which $I_{\ell}$ vanishes. 
} of $\rho_s$ (which depend on $\ell$)---both of these have the potential to constrain $\alpha_1$ independently of $\beta_1$, and could be explored further in future.
  
\begin{figure}
\includegraphics[width=0.475\textwidth]{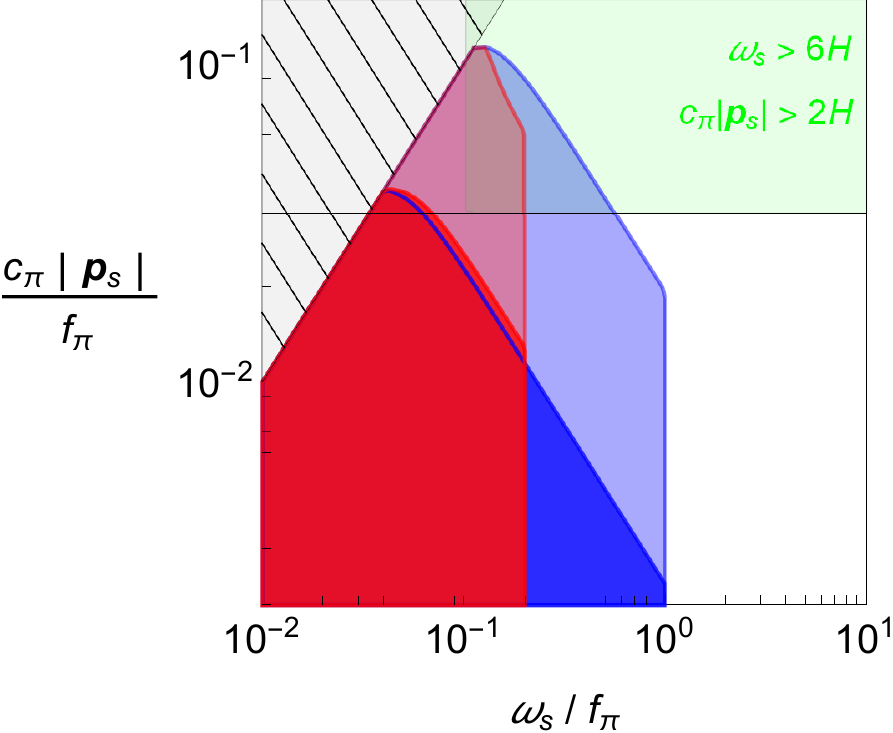}
\includegraphics[width=0.525\textwidth]{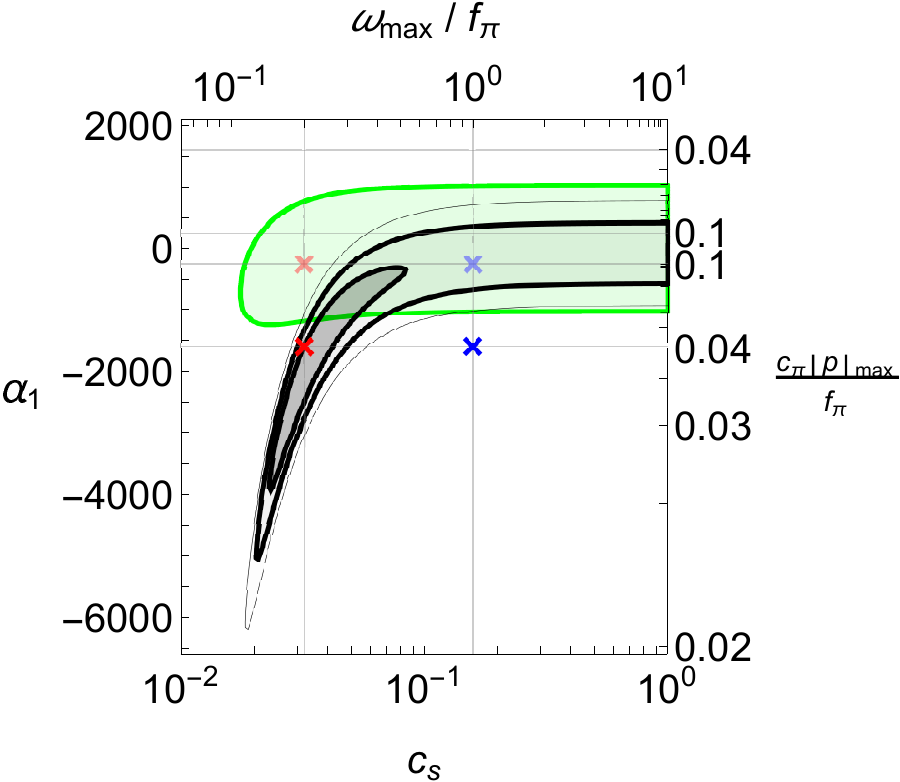}
\caption{
Left panel shows the unitary values of $\omega_s$ and $|\mathbf{p}_s|$ for four different values of $(\cpi, \alpha_1)$---namely $(0.16,-250)$ in light blue, $(0.16, -1600)$ in dark blue, $(0.032,-250)$ in light red, $(0.032,-1600)$ in dark red. The hatched region $c_\pi | \mathbf{p}_s| > \omega_s$ cannot be accessed kinematically (no real time-like $p_1$ and $p_2$ give such a $p_s$). The maximum energy scale, $\omega_{\rm max}$, is independent of $\alpha_1$, while the maximum spatial scale, $|\mathbf{p}|_{\rm max}$, is independent of $c_s$. 
Right panel shows the $(c_s, \alpha_1 )$ parameter space, with corresponding values of $(\omega_{\rm max} , | \mathbf{p} |_{\rm max} )$. Grid lines indicate $\alpha_1 = \pm 250 , \pm 1600$ and $c_s = 0.032, 0.16$ (crosses correspond to the points depicted on the left). The gray contours show the Planck 2018 $68\%$, $95\%$ and $99.7 \%$ confidence intervals from $f_{NL}^{\rm equil}$ and $f_{NL}^{\rm orth}$. Finally, the region in which there is at least some $\omega_s  > 6H$ for which $| \mathbf{p}_s|$ can be $> 2H$ is shown is green. Although when $\alpha_1 = -1600$ both $\omega_{\rm max} > 6H$ and $c_\pi | \mathbf{p} |_{\rm max} > 2 H$, they cannot both be subhorizon.
}
\label{fig:5}
\end{figure}

While \eqref{eqn:lim2} and \eqref{eqn:lim3} are useful illustrations, it is not necessary to assume that $\beta_1 \gg \alpha_1 \gg 1-c_s^2$. 
In general, each spherical-wave amplitude is quite lengthy, and has contributions from $\{ \alpha_1^2 , \alpha_1 \alpha_2 , \alpha_2^2, \beta_1, \beta_2, \beta_3  \}$. But our strategy is the same: from each $\mathfrak{a}^m_{\ell \ell}$, first identify the maximum/minimum $\beta_1$ at $\rho_s =0$, and then increase $ \rho_s $ at fixed $\omega_s$ until there is no longer any value of $\beta_1$ which can all simultaneously satisfy all bounds. 
The resulting range of momenta and energies which are consistent with perturbative unitarity is shown in Figure~\ref{fig:5} for a few fixed values of $\{ c_s, \alpha_1 \}$.
In general there are maxima at,
\begin{align}
 s_{\rm max}^2 = 16 f_\pi^4 \,  \frac{ 30 \pi c_s^4 }{1- c_s^2} \;\;\;\; \text{and} \;\;\;\; c_\pi | \mathbf{p}_s |_{\rm max} \leq 2 f_\pi \left(  \frac{2\pi}{15 \alpha_1^2}  \right)^{1/4} \; , 
\end{align}
as described in \eqref{eqn:CoM_U_cs} and \eqref{eqn:lim3}. 
 
\begin{figure}[h!]
\includegraphics[width=0.5\textwidth]{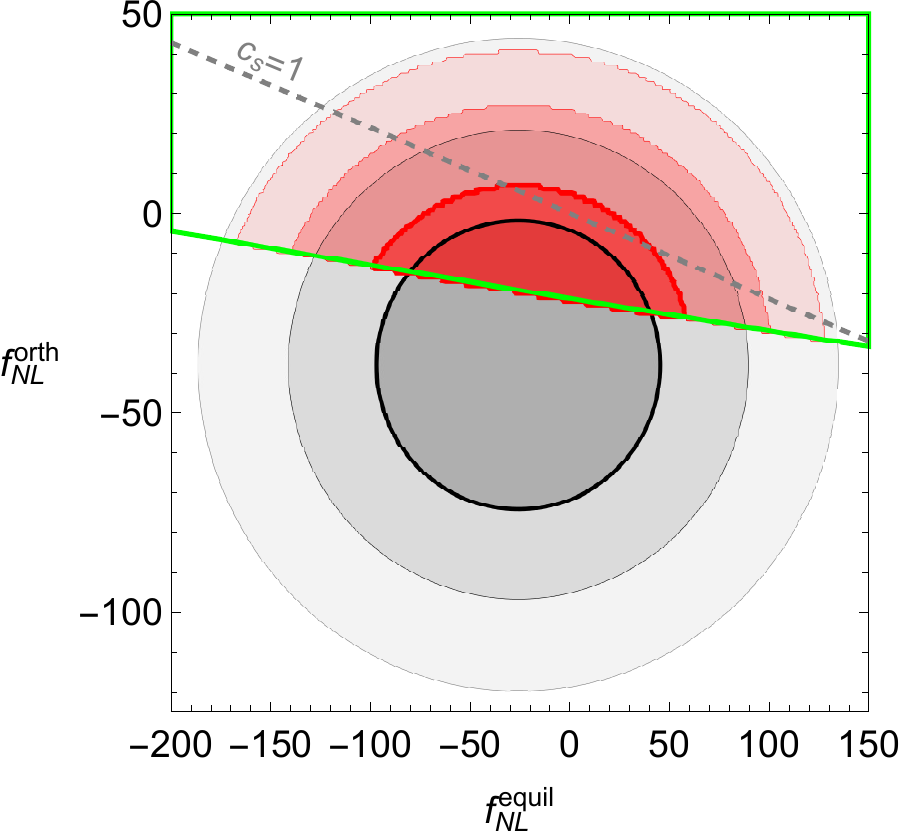}
\includegraphics[width=0.5\textwidth]{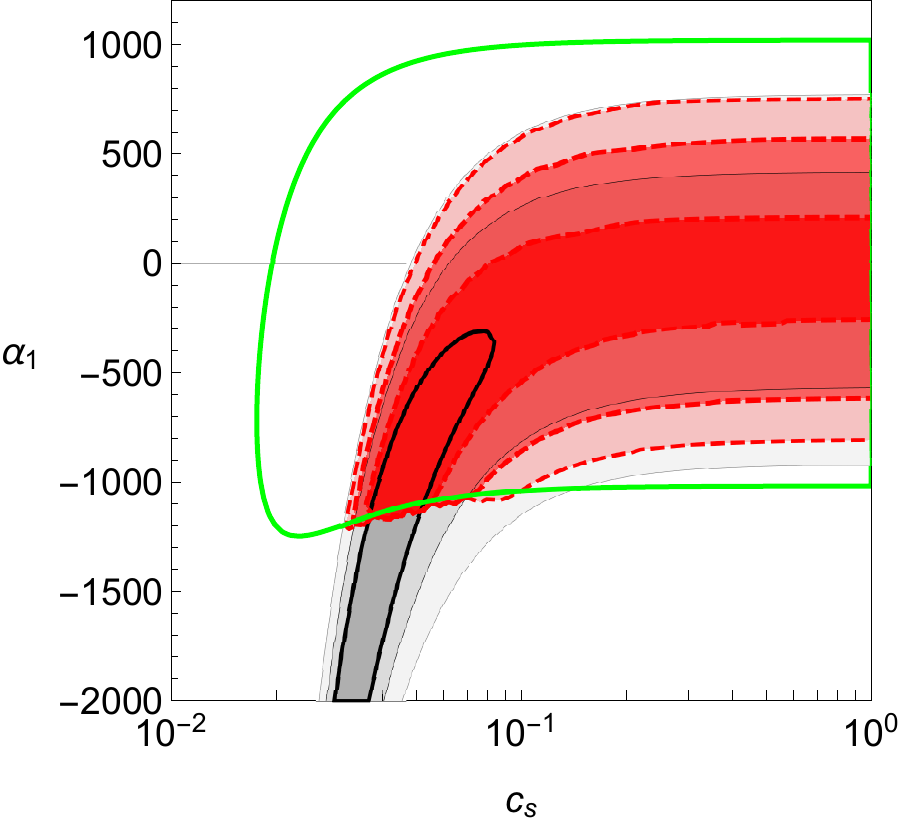}
\caption{
Gray contours show the $68\%$, $95 \%$ and $99.7\%$ confidence intervals for equilateral ($f_{NL}^{\rm equil}$) and orthogonal ($f_{NL}^{\rm orth}$) shapes in the bispectrum \eqref{eq:bispectrum} from Planck 2018 \cite{Akrami:2019izv}. The red contours are the same Planck likelihood, but with the additional prior of perturbative unitarity of $\mathcal{A}_{2\to 2}^{\rm tree}$ for at least some range of $\omega_s$ and $|\mathbf{p}_s|$ which qualifies as subhorizon \eqref{eqn:subhorizon}, as shown in Figure~\ref{fig:5}. The $68\%$ confidence interval in $\{ f_{NL}^{\rm equil}, f_{NL}^{\rm orth} \}$ is a factor of $\approx 3$ smaller with the additional assumption that subhorizon physics is approximately single-field and weakly coupled. 
\label{fig:bispectrum}
}
\end{figure}

Using these spherical-wave bounds, we can now ask precise questions about which regions in $\{ c_s , \alpha_1 \}$ parameter space can describe subhorizon physics \eqref{eqn:subhorizon} (i.e. when does the tree-level scattering mediated by those interactions respect perturbative unitarity). 
For instance, if we find that the scattering of on subhorizon scales (at $c_\pi | \mathbf{p}_s | / H  \gg  | \nu^2 - \frac{1}{4} | \approx 2$) is only unitary if $c_\pi | \mathbf{p}_s | < 2 H$, then we can conclude that it is \emph{never} possible for the EFT to describe this regime\footnote{
This is the most conservative possible requirement, since in practice if unitarity required $c_\pi | \mathbf{p}_s | < 2H + \delta$ for some small $\delta$ then one might still conclude that the subhorizon regime \eqref{eqn:subhorizon} is never unitarity---this simply leads to even stronger bounds than those we present in Figure~\ref{fig:bispectrum}.  
}. Analogously, since both incoming particles have $\omega_1 > 2H$ and $\omega_2 > 2H$ (and we must allow for any relative scattering angle), then the total energy $\omega_s$ must be larger than $6H$ to be in the subhorizon regime \eqref{eqn:subhorizon}.
  
This can be translated into a theoretical prior: demanding that inflation was approximately single-field (and weakly coupled) on subhorizon scales \eqref{eqn:subhorizon} restricts us to the green region shown in Figure~\ref{fig:bispectrum}.
%Since for each point in $\{ c_s, \alpha_1 \}$ parameter space we can now infer 
We can then translate our allowed region of $\{ \cpi , \alpha_1 \}$ into an allowed region for bispectrum shapes $\{ f_{NL}^{\rm equil} , f_{NL}^{\rm orth} \}$, and multiply a simple Gaussian likelihood (using Planck's mean and variance) by a prior distribution which is uniform for the allowed $\{ f_{NL}^{\rm equil} , f_{NL}^{\rm orth} \}$ (and zero otherwise)---this is described in more detail in Appendix~\ref{app:priors}. 
%The resulting posterior is used to generate contours within which $68 \%$, $95 \%$ and $99.7\%$ of the total probability lie (preferentially filling the most likely bins first). 
This produces the posterior contours shown in Figure~\ref{fig:bispectrum}.

Finally, we note that there are several other ways to investigate perturbative unitarity. We have focussed on the most conservative possible application, identifying the range of parameter space in which subhorizon scattering (approximated by plane waves) could possibly be unitary in perturbation theory. One could further demand that there are no new states (or non-perturbative effects) until at least some scale, say $10 H$, above the background---
%---i.e. that inflation is approximately single-field (and weakly coupled), described by \eqref{eq:EFT_action_decoupling_intro}---
this would give an even stronger theoretical prior. Also, rather than focussing on energy/momentum cutoffs relative to the background scale $H$, one could instead compare the maximum velocity cutoff $\rho_s^{\rm max}$ with the speed of tensors modes: in fact, if the theory is to be perturbatively unitary at all centre-of-mass speeds up to $c_T$ (which corresponds to $\rho_s = c_s$), then, 
\begin{align}
\frac{  \alpha_1^2 \omega_s^4 }{4 \pi f_\pi^4 }  \leq    \frac{ 64 }{25}  \left( 1 - c_s^2 \right)  \;\;\;\; \text{when} \;\;\;\; 1  - c_s^2 \ll 1 \, . 
\label{eqn:last}
\end{align} 
Interestingly, taking $c_s \to 1$ and demanding unitarity up to the speed $c_T$ forces $\alpha_1$ to vanish at least as fast as $1 - c_s^2$.

Going beyond the CoM frame has therefore furnished us with a number of new unitarity bounds with interesting consequences for inflation.
%, which we show in Figure \ref{fig:US}. 
These give tighter constraints on the Wilson coefficients $\alpha_3$ and $\beta_1$, and are just the tip of the unitarity iceberg---there are further bounds from higher $\ell$ coefficients, as well as non-linear constraints coming from the determinant condition \eqref{eqn:all_U}, which we leave for future investigation.

%%%
%%%

%%%%%%%%%%%%%%%%
\section{Discussion}
\label{sec:disc}
%%%%%%%%%%%%%%%%

Demanding consistent scattering amplitudes has proven to be a valuable tool in constraining low-energy Effective Field Theories (EFTs).
However, due to their reliance on Lorentz invariance, to date existing techniques have not been fully exploited in cosmology (where the background spacetime spontaneously breaks Lorentz invariance). 
Here, we have taken the first steps towards extending EFT constraints from radiative stability and unitarity to theories in which boosts are broken. This has allowed us to constrain the dynamics of subhorizon modes produced during inflation, and identify the region of parameter space in which these subhorizon scales are approximately single-field and weakly coupled. 

We began by showing that the higher-order EFT corrections can be parameterised according to a simple power counting scheme in terms of a single heavy scale $\Lambda$, a field coupling $g_\pi$, and an order parameter $g_n$ which controls the breaking of boosts. Radiative stability, the requirement that quantum corrections are at most order unity, then places bounds on these power counting parameters in terms of the leading-order interaction coefficients (which are not renormalised on dimensional grounds and so we allow them to take independent values). 
In particular, when boosts are broken spontaneously the field coupling is fixed in terms of $f_\pi / \Lambda$ (the ratio of the decay constant to the EFT cutoff), and then radiative stability can place constraints directly on the leading-order coefficients. 
When applied to the EFT of Inflation, this shows that the leading-order interactions $S_{\rm LO} [ \delta g^{\mu\nu}]$ in \eqref{eqn:SLO1} can only be accompanied by higher-order corrections of the form \eqref{eqn:newEFT} in a radiatively stable way providing $M_n^4 \lesssim f_\pi^4 c_s^{1-2n}$ (or $\alpha_j \lesssim 1/c_s^2$ and $\beta_j \lesssim 1/c_s^4$). 
The the EFT cutoff in $Z^{\mu\nu} \partial_\mu \partial_\nu$ is $\Lambda^2 = 4 \pi c_s^2 f_\pi^2 / g_n$, while the cutoff in time derivatives $n^\mu \partial$ is $\Lambda / g_n$, and the lowest value that $g_n$ can take is given by whichever $\left( M_n^4/ f_\pi^4 c_s^{1-2n} \right)^{1/(2n-2)}$ is largest. In particular, $g_n$ may not be lower than $\sqrt{1-c_s^2}$.
 
We then focussed on the scattering amplitude mediated by $S_{\rm LO}$. By extending the usual partial wave expansion of the $2 \to 2$ scattering amplitude to include preferred-frame effects (a breaking of boosts due to a constant time-like $n^\mu$), 
\begin{align}
\mathcal{A} ( \omega_s, \rho_s , \hat{\bf p}_1 , \hat{\bf p}_3  ) = \frac{ 16 \pi^2 \omega_s^2 (1 - \rho_s^2 )  }{  \omega_1 \omega_3 }   \sum_{ \substack{ \ell_1, \ell_3 \\ m_1 m_3  }}   \mathfrak{a}_{\ell_1\ell_3}^{m_1}  \delta_{m_1 m_3}  \; Y^{m_1 *}_{\ell_1} ( \hat{ \bf p}_1   ) Y^{m_3}_{\ell_3} ( \hat{\bf p}_3  ) \; ,
\end{align}
we have shown that unitarity requires $| \det \mathfrak{a}^m_{\rm sub} (\omega_s, \rho_s) | < 1$ for any finite submatrix of $\mathfrak{a}^m$ and for any centre-of-mass velocity, $\rho_s$. 
Applying these new constraints to the EFT of Inflation, we have shown how the leading order Wilson coefficients $\{ \cpi, \alpha_1, \beta_1 \}$ are related to the strong coupling scales (at which perturbation theory breaks down) for $s$ (the internal interaction energy) and $c_\pi | \mathbf{p}_s|$ (the energy in the centre-of-mass motion). 
Combining Planck observational constraints on the equilateral and orthogonal bispectrum shapes with the prior that inflation is approximately single-field and weakly coupled on subhorizon scales (i.e. the regime \eqref{eqn:subhorizon} in which mode functions behave as plane waves) results in an improvement of the $68 \% $ confidence interval by a factor of $\approx 3$. 
For comparison, Simons Observatory (due to begin taking data in early 2020s) has a goal sensitivity \cite{Ade:2018sbj} which would improve this interval by a factor of $\approx 3.6$, while the later CMB-S4
%(late 2020s at least)
experiment \cite{Abazajian:2016yjj, Abazajian:2019eic} forecasts an improvement by a factor of $\approx 6.0$.
 
% Connection to $c_s =1$ conjecture
One of our main observations is that the EFT cutoffs in energy and momentum need not be the same in EFTs with broken boosts. In fact, we have shown that they must necessarily differ in the presence of Lorentz-violating cubic interactions (like $\dot \pi^3$). It is worth pointing out that, if the new physics which UV completes the EFT \eqref{eq:EFT_action_decoupling_intro} was Lorentz-invariant, one might expect that the energy and momentum cutoffs should be the same, particularly in the limit $c_s \to 1$ (since then at the level of particle propagation there is no Lorentz-violation either from $\pi$ or from the heavy physics). We certainly have not proved this. But it might be pointing in the same direction as the $c_s =1$ conjecture made in \cite{Baumann:2015nta} that the existence of a Lorentz-invariant UV completion requires $\alpha_1 \to 0$ as $c_s \to 1$.

%Combining our radiative stability constraints and unitarity constraints, we find that if scattering at a centre-of-mass velocity $\rho_s$ is to be unitarity for all $s$ up to the EFT cutoff $\Lambda$, then the symmetry-breaking parameter must lie within,
%\begin{align}
% 1 - \rho_s^2  \gtrsim  g_n^2  \gtrsim 1 - c_s^2 \, . 
%\end{align}
%This suggests that, unless higher derivative terms are very finely tuned, they will inevitably become important when $\rho_s$ approaches $c_s$, i.e. when the $\pi$ centre-of-mass is boosted to velocities above $c_T$. One may wonder if this is perhaps related to a Cerenkov effect. 

Looking ahead, there are many other avenues for future work. 
% Bootstrapping higher order correlators
For instance, here we have deliberately removed the $\beta_1$ dependence from the spherical-wave unitarity bounds in order to provide robust constraints on the bispectrum alone, but in future, as bispectrum measurements improve, there is the possibility of using these unitarity bounds to instead constrain $\beta_1$. Exploiting this unitarity connection between bispectrum and trispectrum may allow us to use future measurements of $f_{NL}$ to bootstrap information about higher order non-Gaussianities. 

There are also other systems to which our unitarity bounds may be applied. The most closely related is dark energy fluctuations in the late Universe, which can be described by a functionally similar EFT \cite{Gubitosi:2012hu}. Further afield are condensed matter systems (with approximately linear dispersion relations) which share the same symmetry breaking pattern, for instance those studied in \cite{Nicolis:2015sra} (and the upcoming \cite{Grall:2020ibl}).

Furthermore, there is the potential to constrain the EFT of Inflation even further by extending other amplitude techniques to finite centre-of-mass velocities. 
One example is the positivity bounds put forward in \cite{Adams:2006sv} and applied to the EFT of Inflation in \cite{Baumann:2015nta}, which we have also extended to $\rho_s \neq 0$ and will present in a following work \cite{Grall2}.
 
Also, since we have worked entirely within the decoupling limit, our amplitudes are not sensitive to the propagation of any degrees of freedom at $c_T$ (rather $c_T$ enters only as the invariant-speed preserved by the non-linear boosts). This means that we do not encounter any particular effects when $\rho_s$ crosses $c_s$ (i.e. when the $\pi$ centre-of-mass speed exceeds $c_T$), in particular there is no Cerenkov radiation in the decoupling limit.
In fact, the graviton exchange contribution to $\mathcal{A}_{2 \to 2}^{\rm tree}$ can be estimated by assuming that the propagator receives $\mathcal{O} (H \omega)$ corrections as it is taken on-shell, giving $\sim \mathcal{O} \left( \omega^3 /M_P^2 H \right)$, which is indeed negligible. It would be interesting to go beyond the decoupling limit to explore the consistency of the theory at speeds $\rho_s > c_s$ further in future.

% Connecting flat space estimates to in-in estimates of errors
And finally, developing a more thorough understanding of how radiative stability and unitarity manifest directly in the in-in correlators (the natural observables from inflation), removing the need for our restriction to strictly subhorizon scales, is particularly important. One possible in-road may be the recent connection uncovered in \cite{Maldacena:2011nz,Raju:2012zr,Arkani-Hamed:2015bza,Arkani-Hamed:2018kmz} that cosmological correlators contain the corresponding flat space scattering amplitudes  as the residue of the total momentum pole. 
 
To sum up, we have considered how EFTs with broken boosts can be constrained using radiative stability and perturbative unitarity and used these to improve constraints on the primordial bispectrum produced during inflation. 
%\TG{
%%Of course the main limitation of our analysis is that we could only probe perturbative unitarity of the EFT in its flat-space, decoupling limit. In particular this prevented us from probing the EFT closer to the Hubble scale during inflation and directly at the level of correlators.
%While it is currently an open %(and challenging!) 
%problem to probe unitarity of cosmological bulk evolution (away from the limit \eqref{eqn:DL}) directly in cosmological correlators. 
%A resolution of this problem would certainly allow us to probe unitarity of the EFT of Inflation away from the limit \eqref{eqn:DL} and place more robust constraints on Non-Gaussianities.\\
In this work we have made a step forward in importing powerful Lorentz-invariant EFT techniques to cosmology, a programme which is particularly important in light of next generation experiments.

% \vspace{-0.1in}
%\subsubsection*{Acknowledgements}
%\vspace{-0.1in}
\acknowledgments
We thank Sadra Jazayeri, Daan Meerburg, Johannes Noller and David Stefanyszyn for useful discussions, and in particular Enrico Pajer for comments on a draft. TG is supported by the Cambridge Trust. SM is supported by an Emmanuel College Research Fellowship. 
This work is supported in part by STFC under grants ST/L000385, ST/L000636 and
ST/P000681/1.

\appendix
%%%%%%%%
\section{The Spherical Wave Expansion}
\label{app:A}
%%%%%%%%

%%%%%%%%
\subsection{Unitarity Bound on the Minors of $\mathfrak{a}^m$}
\label{sub:submatrix_proof}
%%%%%%%%

Strictly speaking, $(\mathfrak{a}^m)_{\ell_1 \ell_3}$ is an infinite dimensional matrix, and operations like $\det ( \cdot )$ must be handled with care. However, if we focus on any (finite) submatrix, $\mathfrak{a}^m_{\rm sub}$, in which both $\ell_i$ indices run over a (finite) set of integers $N$, then we can write the unitarity condition as,
\begin{equation}
\text{Im} \,  ( \mathfrak{a}^m_{\rm sub} )_{\ell_1 \ell_2} = \sum_{\ell \in N} ( \mathfrak{a}^{m}_{\rm sub} )_{\ell_1 \ell}  ( \mathfrak{a}^{m \dagger}_{\rm sub} )_{\ell \ell_2} + \sum_{L \notin N} a^m_{\ell_1 L} a^{m*}_{\ell_2 L} =: A_{\ell_1 \ell_2} + B_{\ell_1\ell_2}
\end{equation}
where $A$ and $B$ are both positive-definite matrices.

Since $A$ and $B$ are both positive-definite,  then for any vector $v$,
\begin{equation}
 v^T ( A +B ) v  \geq  v^T A v  \geq  0 
\end{equation}
This implies $\det (A + B) \geq \det (A)$, since we can use,
\begin{equation}
 e^{- v^T ( A + B ) v} \leq  e^{- v^T A v }
\end{equation}
to show that,
\begin{align}
\int d v \;\; e^{- v^T ( A + B ) v} &\leq  \int d v \;\; e^{- v^T A v }  \nonumber  \\
\Rightarrow \;\;\;\; \frac{1}{\sqrt{\det (A+B) } } &\leq  \frac{1}{\sqrt{\det ( A) }} \, . 
\end{align}
The unitarity condition can therefore be written as,
\begin{equation}
 \det \left( \text{Im} \, \mathfrak{a}^m_{\rm sub} \right) \geq \det \, \left(   \mathfrak{a}^m_{\rm sub} \mathfrak{a}^{m \dagger}_{\rm sub}    \right) = |  \det ( \mathfrak{a}^m_{\rm sub} )  |^2 \, . 
\end{equation}

Finally, note that if we perform the Cartesian decomposition, $\mathfrak{a}^m = \text{Re} \, \mathfrak{a}^m + i \text{Im}\, \mathfrak{a}^m$, and $\text{Im} \, \mathfrak{a}^m$ is positive-definite, then
%\begin{align}
$ \det ( \mathfrak{a}^m ) \geq \det \left(  \text{Im} \, \mathfrak{a}^m \right)  \, $. 
%\end{align}
This allows us to write the following unitarity condition for \emph{any} finite submatrix $\mathfrak{a}^m_{\rm sub}$ of $\mathfrak{a}^m_{\ell_1 \ell_2}$, 
\begin{align}
 \det ( \mathfrak{a}^m_{\rm sub} )  \leq 1 \, . 
\end{align}
as quoted in the main text.

%%%%%%%%
\subsection{Evaluating the Spherical Wave Coefficients}
\label{app:pw}
%%%%%%%%

In the main text we focused only on $\ell_1 = \ell_3 \leq 2$, but here we provide general formulae which can be used to generate any desired $a^{m_1 m_3}_{\ell_1 \ell_3}$.  
For convenience, we split the partial wave coefficient into three pieces,
\begin{align}
a^{mm}_{\ell_1 \ell_3}  = [ a^{mm}_{\ell_1 \ell_3}  ]_s + [ a^{mm}_{\ell_1 \ell_3}  ]_t + [a^{mm}_{\ell_1 \ell_3} ]_u
\end{align}
which correspond to the parts of $\mathcal{A}$ with no angular poles (i.e. the contact and $s$-channel exchange diagrams, as well as the analytic parts of the $t$- and $u$-channel exchange diagrams), a pole in $t$ (from the $t$-channel exchange diagram) and a pole in $u$ (from the $u$-channel exchange diagram). We will discuss the computation of each piece in turn. 

We will use our single remaining rotation (around $\mathbf{p}_s$) to fix $\phi_1 = \pi/2$, and focus on the remaining integral over $\phi_3$, $\theta_1$ and $\theta_3$. 
The overall strategy is to first fix an $m$ and do the $\phi_3$ integral explicitly, and then fix $\ell_1$ and $\ell_3$ and do the $\theta$ integrals. In the limit $\rho_s \to 1$, it is possible to write down simple closed form expressions for $a^{mm}_{\ell_1 \ell_3}$ for any $m$ and $\ell$.

%%%%
\subsubsection{The $s$-channel pole}
%%%%

\subsubsection*{The $\phi$ integrals}
The integral over $\phi$ can be done immediately, since there are no angular poles in $[a^{mm}_{\ell_1 \ell_3}]_s$ by construction. For any positive integer $n$, we have that\footnote{
One simple way to derive \eqref{eqn:tint} is to expand $t^n$ as a power series in $\sin \phi_3$, do the integration term by term, and then resum into a standard hypergeometric form. 
},  
\begin{align}
 \int_0^{2\pi} \frac{d\phi_3}{2\pi} \, e^{-i m \phi_3} \frac{ t^n }{\omega_1^n \omega_3^n} =&  \frac{ n! i^m 2^{n-|m|} }{ |m|! (n-|m|)! }  
  (1- x y)^{n-|m|}  ( 1-x^2 )^{|m|/2} (1-y^2)^{|m|/2}      \nonumber \\
&\quad \times {}_2 F_1 \left(  \frac{ |m| - n }{2},  \frac{ 1 + |m| - n}{2} ;   1 + |m| ;  \frac{ (1 - x^2) (1 - y^2 ) }{ (1 - x y)^2 }  \right) , 
\label{eqn:tint}
\end{align}
where we have used that $t = 2 \omega_1 \omega_3 ( 1 - \cos \theta_1 \cos \theta_3 - \sin \theta_1 \sin \theta_3 \sin \phi_3 )$ when $\phi_1 = \pi/2$ is fixed, and introduced $x= \cos \theta_1, y=\cos \theta_3$. Note that the integral vanishes unless $n \geq |m|$, and so when combined with the $(1-x^2)^{|m|/2}$ from the spherical harmonic this integral always produces an analytic polynomial in $x^p y^q$. 
Using $u= 4m^2 -s-t$, all of the integrals appearing in $[ a^{mm}_{\ell_1 \ell_3} ]_s$ can be written in this form (multiplied by $\phi$-independent factors). 

For the EFT amplitude \eqref{eqn:Acomp} and up to $\ell = 2$, we require only the first few,
\begin{align}
 \int_0^{2\pi} \frac{d\phi_3}{2\pi} \,  t &=  2 \omega_1 \omega_3 (1- x y)    \nonumber \\
 \int_0^{2\pi} \frac{d\phi_3}{2\pi} \,  t^2 &=  2 \omega_1^2 \omega_3^2 \left( 2 (1 - x y)^2 +  (1 - x^2) (1 - y^2 )  \right)  \nonumber \\
 \int_0^{2\pi} \frac{d\phi_3}{2\pi} \, e^{\mp i \phi_3} t &= \pm i  \omega_1 \omega_3  \sqrt{1-x^2} \sqrt{1-y^2}    \nonumber \\
 \int_0^{2\pi} \frac{d\phi_3}{2\pi} \, e^{\mp i \phi_3} t^2 &= \pm 4 i \omega_1^2 \omega_3^2 (1- x y ) \sqrt{1-x^2} \sqrt{1-y^2}   \nonumber  \\
  \int_0^{2\pi} \frac{d\phi_3}{2\pi} \, e^{\mp i 2\phi } t^2 &= - \omega_1^2 \omega_3^2 (1-x^2)(1-y^2)   \, .  
\end{align}

\subsubsection*{The $\theta$ integrals}
After carrying out the $\phi$ integral, this leaves a sum over terms in $\omega_1^a \omega_3^b x^p y^q$ with integer powers. 
We can then replace every $x$ and $y$ with inverse powers of $\omega_i$, 
\begin{align}
\rho_s  x = 1 -  \frac{1-\rho_s^2}{2} \frac{\omega_s}{\omega_1}  \;\;\;\; , \;\;\;\; \rho_s y =  1  +  \frac{1-\rho_s^2}{2} \frac{\omega_s}{\omega_3}
 \label{eqn:xyTow}
\end{align}
and express $\left[ a^{m m}_{\ell_1 \ell_3} \right]_s$ as a sum over integrals of $\omega_1^a \omega_3^b$. We then write this as a sum over $I^a I^b$, where $I^a$ is a set of elementary master integrals,
\begin{align}
 I^a ( \rho_s ) = \frac{1}{2} \int_{-1}^1 dx  \; ( 1- \rho_s x )^a \; , 
\end{align}
and the index $a$ runs from $-\max_1$ to $\ell_1-\min_1$, where $\min_1$ ($\max_1$) is related to the minimum (maximum) power of $\omega_1$ appearing in $\left[ \mathcal{A} \right]_s$ (and similarly $b$ runs from $-\max_3$ to $\ell_3 -\min_3$). For the EFT amplitude \eqref{eqn:Acomp}, the ranges are $-3 \leq a \leq \ell_1 - 1$ and $-3 \leq b \leq \ell_3 - 1$ for the sum over $I^a I^b$.
These integrals can be evaluated straightforwardly (e.g. by changing the integration variable to $u = 1-\rho_s x$), 
\begin{align}
I^a ( \rho_s) = \begin{cases} 
\frac{1}{2 \rho_s} \log \left(  \rho_+ / \rho_-  \right)  &\text{if} \;\; a = - 1 \, ,  \\[10pt]
\frac{  \rho_+^{a+1} - \rho_-^{a+1}  }{  2 \rho_s (a+1)  }   &\text{otherwise} \, ,
\end{cases}
\end{align}
where $\rho_- = 1 - \rho_s$ and $\rho_+ = 1 + \rho_s$. 

We also note that Rodriguez's formula for the Legendre polynomial can be used to provide a closed form expression directly for, 
\begin{align}
I^a_{\ell} ( \rho_s ) &:= \frac{1}{2} \int_{-1}^1 dx \, \left( \frac{ \omega_1}{\omega_s} \right)^a \; P_{\ell} ( x )
\end{align}
in terms of Gauss hypergeometric functions,
\begin{align}
 I^a_{\ell} ( \rho_s ) =  \frac{ \sqrt{\pi } \, \Gamma ( \ell + 1 ) \, \Gamma (a+ \ell ) }{ \ell! 2^{a+ \ell +1 } \Gamma (a)   \Gamma \left( \ell + \frac{3}{2}\right) }  
  \left(1- \rho_s^2 \right)^a \rho_s^{\ell}  \;\;
   _2F_1\left(\frac{a+\ell}{2} \; , \; \frac{ a+\ell+1 }{2}  \; ; \;  \ell + \frac{3}{2} \; ;\;  \rho_s^2 \right)  \, ,
   \label{eqn:IalF} 
\end{align}
which can speed up numerical implementation.

\subsubsection*{Small $1-\rho_s$}
When $\rho_s \to 1$, $\omega_1$ (and $\omega_3$) has support only at $x = 1$ (and $y = 1$), and can be expressed in terms of Dirac $\delta$ functions and their derivatives. For instance,
\begin{align}
 \frac{1 - \rho_s }{1 - \rho_s x } \approx  \rho_- \log \left(  \frac{\rho_-}{2}  \right) \delta ( x - 1 ) 
 +
 \rho_- \sum_{k=1}^\infty \frac{2^k}{k!} \delta^{(k)} (x-1 )
 \end{align} 
where $\rho_- = 1 - \rho_s$, and we have used $\approx$ to indicate that this is how the right-hand side behaves when integrated against a regular function of $x$. For finite $\rho_-$, the infinite sum indicates finite support in $x$, but when truncated to leading order $\omega_1$ behaves as a single insertion of $\delta ( x - 1)$.
For arbitrary powers,
\begin{align}
\left(   \frac{1 - \rho_s }{1 - \rho_s x }   \right)^n \approx \sum_{j=1}^{n-1} \rho_-^{j} \;   \frac{ \delta^{(j-1)} (x-1)   }{ (n-1)_j } 
 - \rho_-^n \log \left(   \frac{\rho_-}{2}  \right) \delta^{(n-1)} (x-1)
  + \mathcal{O} ( \rho_-^{n} )
  \label{eqn:w1series}
 \end{align}
where $(n-1)_j = (n-1) (n-2) ... (n-1 - j +1 )$ is the falling Pochhammer symbol, and the $\mathcal{O} (\rho_-^n )$ correction is generally an infinite sum over $\delta^{(k)} (x-1)$ (and so has finite support in $x$).   

Since derivatives of the Legendre polynomials obey $P_{\ell}^{(n)} (1) = (\ell + n)_{2n} / 2^n / n!$, where $a_n = a (a-1) ... (a-n+1)$ is the falling Pochhammer symbol, we have that $\delta^{(k)} (x-1) P_{\ell} (x) \approx \ell^{2k}$, and therefore the series expansion \eqref{eqn:w1series} of $\omega_1$ can be viewed as an expansion in powers of $(1-\rho_s) \ell^2$.  

So when $1 - \rho_s \ll \ell^{-2}$, we can replace the $\theta$ integrals over $P_{\ell}$ with simply\footnote{
Note that the whole series \eqref{eqn:w1series} can be resummed, yielding a hypergeometric expression which agrees with \eqref{eqn:IIb1}. 
},
\begin{align}
I^a_{\ell} ( \rho_s ) 
%&:= \frac{1}{2} \int_{-1}^1 dx \, \omega_1^a \; P_{\ell} ( x )   \nonumber  \\
&= \rho_- \frac{ P_{\ell} (1) }{a-1}   +  \mathcal{O} ( \rho_-^2 \ell^2 )    \, .  
\end{align}
%It is this relation which we have used to compute the $[ a^{00}_{\ell_1 \ell_3}]_s$ as $\rho_s \to 1$ given in the main text. 
Note that when $m \neq 0$, the $\phi$ integral vanishes when $x \to 1$ and $y\to 1$, and therefore in the $\rho_s \to 1$ limit the dominant partial waves are the $a^{00}_{\ell_1 \ell_3}$ modes. 

%We mention in passing that the whole series in $\rho_- \ell^2$ can even be resummed, e.g. using \eqref{eqn:w1series}, 
%\begin{align}
%I^{1}_{\ell} (\rho_s) = \frac{1 - \rho_s^2}{2} \, \ell (\ell + 1 ) \; {}_p F_q \left(  \begin{array}{c}
% 1,1,1, -\ell, 2+\ell\\
% 2,2,2
%\end{array}      \;, \; -1 \right)
%\end{align}
%which is now valid for any $\rho_s$ and $\ell$, and which gives an alternative (albeit more complicated) route to computing the $a^{mm}_{\ell\ell} (\omega_s, \rho_s)$.

%%%%
\subsubsection{The $t$-channel pole}
%%%%

\subsubsection*{The $\phi$ integral}
First, notice that for a scalar field the residue of the $t$-channel pole is always independent of $s$ and $u$. The $\phi$ integral over the $t$-channel pole therefore always takes the form,
\begin{align}
\mathcal{I}_m (c) := \int_0^{2\pi} \frac{d \phi}{2\pi} \, \frac{ e^{- i m \phi } }{1 - c \, \sin \phi } = \int_{-\infty}^{+\infty} \frac{dz}{\pi} \, \frac{1}{1 + z^2 - 2 c z} \, \left( \frac{i + z}{ i - z }  \right)^m
%\label{eqn:Imc}
\end{align}
where we have used the substitution $z = \tan (\phi/2)$. 
For $m>0$, this has two poles in the upper half-plane (at $z=i$ and $z=c+ i \sqrt{1-c^2}$) and one pole in the lower half-plane (at $z = c - i \sqrt{1-c^2}$). For $m<0$, there are two poles in the lower half-plane (now at $z=-i$) and one in the upper half-plane. Closing the contour in either direction gives the same answer, 
\begin{align}
\mathcal{I}_m (c) =   \frac{ i^m}{\sqrt{1-c^2}} \left(  \frac{-1 + \sqrt{1-c^2}}{c}  \right)^{|m|}   \label{eqn:Imc}
%\begin{cases}
%  \frac{ i^m}{\sqrt{1-c^2}} \left(  \frac{-1 + \sqrt{1-c^2}}{c}  \right)^m  \;\;\;\; &\text{when } m > 0 \, , \\[10pt]
%  \frac{ i^m}{\sqrt{1-c^2}} \left(  \frac{ -1 + \sqrt{1-c^2}}{c}  \right)^{-m}  \;\;\;\; &\text{when } m < 0 \, .
% \end{cases} 
\end{align} 
Consequently,
\begin{align}
\int_0^{2\pi} \frac{d \phi_3}{2 \pi} \, e^{-im\phi_3} \,  \frac{\omega_{13} \omega_s }{ s_{13} } &= \frac{\rho_s}{1-\rho_s^2} \;  \frac{ \cos \theta_1 - \cos \theta_3 }{1 - \cos \theta_1 \cos \theta_3} \; 
\mathcal{I}_m \left( \frac{ \sin \theta_1 \sin \theta_3 }{1 - \cos \theta_1 \cos \theta_3 }  \right)  \nonumber \\
&= \frac{  i^m \rho_s}{1-\rho_s^2}  \text{sgn} \left( \cos \theta_1 - \cos \theta_3   \right) \frac{ \left( -1 + \cos \theta_1 \cos \theta_3  + |\cos \theta_1  - \cos \theta_3 |   \right)^{|m|} }{ \left( \sin  \theta_1 \sin \theta_3 \right)^{|m|}}  
\end{align}
The way that even/odd $m$ contribute only an overall sign is also clear from that fact that $\phi_3 \to \pi - \phi_3$ leaves the integrand invariant. 
The first three such integrals can be written in terms of $x = \cos \theta_1$ and $y = \cos \theta_3$ as,
\begin{align}
\int_0^{2\pi} \frac{d \phi_3}{2 \pi} \,   \frac{\omega_{13} \omega_s }{ s_{13} }  &=  \frac{\rho_s}{1-\rho_s^2} \, \text{sgn} ( x - y  )   \label{eqn:Jeg} \\
\int_0^{2\pi} \frac{d \phi_3}{2 \pi} \,  e^{\pm i \phi } \frac{\omega_{13} \omega_s }{ s_{13} }  &= \mp  \frac{i \rho_s}{1-\rho_s^2} \; \frac{ 
\left(  x - y  \right)  +  \text{sgn} ( x - y  )   \left(  -1 + x y   \right)  
}{ \sqrt{1-x^2} \sqrt{1-y^2}  }    \nonumber  \\
\int_0^{2\pi} \frac{d \phi_3}{2 \pi} \,  e^{\pm 2 i \phi } \frac{\omega_{13} \omega_s }{ s_{13} }  &=    \frac{ - \rho_s}{1-\rho_s^2} \; \frac{ 
2 \left( - 1 + x y  \right) \left(  x - y  \right)  +  \text{sgn} ( x - y  )   \left(   1 + x^2  - 4 x y + y^2  + x^2 y^2   \right)  
}{ (1-x^2) (1-y^2 ) }  \, .   \nonumber  
\end{align}
Note that the denominators always cancel with the factor of $\left( 1-x^2 \right)^{|m|/2}$ appearing from the $P_{\ell}^m ( x )$ in the spherical harmonics, leaving an analytic polynomial in $x^p y^q$.

\subsubsection*{The $\theta$ integrals}
After carrying out the $\phi$ integral, this leaves a sum over terms in $\omega_1^a \omega_3^b x^p y^q$ with integer powers. 
We can then replace every $x$ and $y$ with inverse powers of $\omega_i$ using \eqref{eqn:xyTow}  and express $\left[ a^{m m}_{\ell_1 \ell_2} \right]_t$ as a sum over the elementary master integrals,
\begin{align}
 J_{ab} ( \rho_s )  &= \frac{1}{4} \int_{-1}^1 dx  \int_{-1}^1 dy \;  (1-\rho_s x )^a (1-\rho_s y )^b \; \text{sgn} ( x - y )    
\end{align}
where $a$ runs from  $-\max_1$ to $\ell_1-\min_1$, where $\min_1$ ($\max_1$) is related to the minimum (maximum) power of $\omega_1$ appearing in $\left[ \mathcal{A} \right]_t$---for the EFT amplitude \eqref{eqn:Acomp}, the ranges are $-4 \leq a \leq \ell_1 - 2$ and $-4 \leq b \leq \ell_3 - 2$.  

These integrals may be straightforwardly evaluated (for instance by changing integration variables, $u = 1- \rho_s x$ and $v = 1- \rho_s y$),
\begin{align}
 J_{ab} ( \rho_s ) = \begin{cases}
\frac{ \rho_-^{b+1} - \rho_+^{b+1}     }{2 \rho_s^2 (b+1)^2}  + \frac{ \rho_-^{b+1} + \rho_+^{b+1}   }{4 \rho_s^2 (b+1) } \log \left(  \frac{\rho_+}{\rho_-}  \right)    &\text{if} \;\; a=-1  \\[10pt]
 - \frac{1}{4 \rho_s^2 (a+1)^2}  \left(  \frac{\rho_+}{\rho_-}  \right)^{1+a} \left[  1 - \left( \frac{\rho_-}{\rho_+}  \right)^{2a+2} + 2 (a+1) \left( \frac{\rho_-}{\rho_+}  \right)^{a+1} \log \left( \frac{\rho_-}{\rho_+}   \right)    \right]   &\text{if} \;\; a+ b = -2  \\[10pt]
\frac{(b-a) \left(  \rho_+^{a+b+2}  -   \rho_-^{a+b+2}   \right)
+ (a+b+2) \left(  \rho_-^{b+1} \rho_+^{a+1}    - \rho_+^{b+1} \rho_-^{a+1} \right)}{4 \rho_s^2 (a+1) (b+1) (a+b+2) }     &\text{otherwise}
 \end{cases}
\end{align}
where $J_{ab} = - J_{ba}$ is antisymmetric.

%%%%%
%\subsubsection*{Small $\rho_s$}
%%%%%

%%%%
\subsubsection*{Small $1-\rho_s$}
%%%%
Since the $t$-channel pole contains a $1/|x-y|$ collinear divergence, the expression \eqref{eqn:w1series} is no longer valid. Nonetheless, the rationale that as $\rho_s \to 1$ quantities like $\omega_1^a \omega_3^b$ become infinitesimally supported is still useful, and indeed,
\begin{align}
 J^{ab}_{\ell_1 \ell_3}  (\rho_s)  &:= \frac{1}{4}  \int_{-1}^1 dx  \int_{-1}^1 dy \;  \left( \frac{\omega_1}{\omega_s} \right)^a \left( \frac{ \omega_3 }{ \omega_s} \right)^b \; P_{\ell_1} ( x ) P_{\ell_3} (y) \; \text{sgn} ( x - y )    \nonumber   \\
 &= \rho_-^2  \frac{ (a-b) P_{\ell_1} ( 1 ) P_{\ell_3} ( 1 ) }{ (a-1)(b-1) (a+b-2)}   + \mathcal{O} ( \rho_-^3 \ell^2 ) \; . 
\end{align}
Note that when $m \neq 0$, the $\phi$ integral vanishes when $x \to 1$ and $y\to 1$, and therefore in the $\rho_s \to 1$ limit the dominant partial waves are the $a^{00}_{\ell_1 \ell_3}$ modes.

%%%%
\subsubsection{The $u$-channel pole}
%%%%

\subsubsection*{The $\phi$ integral}
Since $\phi_4 = \pi + \phi_3$, we can also use the result \eqref{eqn:Imc} to write,
\begin{align}
\int_0^{2\pi} \frac{d \phi_3}{2 \pi} \, e^{-i m \phi_3}   \frac{\omega_{14} \omega_s }{ s_{14} }  &= 
\frac{  (-i)^m \rho_s}{1-\rho_s^2}  \text{sgn} \left( \cos \theta_1 - \cos \theta_4   \right) \frac{ \left( -1 + \cos \theta_1 \cos \theta_4  + |\cos \theta_1  - \cos \theta_4 |   \right)^{|m|} }{ \left( \sin  \theta_1 \sin \theta_4 \right)^{|m|}} 
\end{align}
and then use,
\begin{align}
\omega_3 \sin \theta_3 = \omega_4 \sin \theta_4 \;\;\;\; \text{and} \;\;\;\; \omega_3 \cos \theta_3 + \omega_4 \cos \theta_4  = - \rho_s \omega_s 
\end{align}
to write, 
\begin{align}
\int_0^{2\pi} \frac{d \phi_3}{2 \pi} \, e^{-i m \phi_3}   \frac{\omega_{14} \omega_s }{ s_{14} }  &= 
\frac{ (- i)^m \rho_s}{1-\rho_s^2}  \text{sgn} \left( \omega_{14}  \right) \frac{ \left(  - \omega_4 - x ( \rho_s \omega_s + \omega_3 y )  + \text{sgn} (\omega_{14} ) ( \omega_4 x   +  \omega_3 y + \rho_s \omega_s  )   \right)^{|m|} }{  \left( \omega_3 \,  \sqrt{1-x^2} \sqrt{1-y^2} \right)^{|m|}}  \, . 
\end{align}
Finally, use $\omega_4 = -\omega_s - \omega_3$ to express this integral in terms of $\omega_1, \omega_3, x$ and $y$. Note that $\text{sgn} ( \omega_{14} ) = \text{sgn} \left( \frac{x+y}{1+x y} - \frac{2 \rho_s}{1+\rho_s^2}   \right)$. 

\subsubsection*{The $\theta$ integral}
This leaves a sum over terms in $\omega_1^a \omega_3^b x^p y^q$ with integer powers.
We can then replace every $x$ and $y$ with inverse powers of $\omega_i$, and express $[ a^{m m}_{\ell_1 \ell_3} ]_u$ as a sum over the elementary integrals,
\begin{align}
 K^{ab} ( \rho_s ) := \frac{1}{4} \int_{-1}^1 dx \int_{-1}^1 dy \; (1-\rho_s x)^a (1-\rho_s y)^b \; \text{sgn} \left(
\frac{x+y}{1+x y} - \frac{2\rho_s}{1+\rho_s^2} 
    \right) 
\end{align}
where $a$ runs from $-\max_1$ to $\ell_1-\min_1$, where $\min_1$ is related to the minimum power of $\omega_1$ appearing in $[ \mathcal{A} ]_u$. For the EFT of Inflation amplitude, the $K^{ab}$ integrals range from $-4 \leq a \leq \ell_1 - 2$ and $-4 \leq b \leq \ell_3 - 2$. 

The $K^{ab}$ integrals can also be performed explicitly, for instance by substituting,
\begin{align}
 1 - \rho_s x = \frac{1-\rho_s^2}{ 1- \rho_s x' } \;\;\;\; , \;\;\;\;  1 - \rho_s y = \frac{1-\rho_s^2}{ 1- \rho_s y' }
\end{align}
which sends $\text{sgn} (\omega_{14})$ to $- \sgn ( x' + y' )$. This gives,
\begin{align}
 K^{ab} ( \rho_s ) = \begin{cases} 
\frac{1}{\rho_s^2} \left[   \frac{\pi^2}{12}    - \frac{1}{2} \log \left( \frac{\rho_-}{2} \right)^2 + \frac{1}{4} \log \left(  \frac{\rho_+}{\rho_-}   \right)^2 - \text{Li}_2 \left( \frac{\rho_-}{2} \right) 
\right]  \qquad\qquad \text{if} \;\;\;\; a = b = -1 , \;\; \text{else:} \\[10pt] 
- \frac{ (1-\rho_s^2 )^{a+b+2} }{4 \rho_s ( b+1 ) } \left[ 
\left( \rho_+^{-b-1} + \rho_-^{-b-1} \right) I^{-a-2} (\rho_s)  
- \frac{ 2  }{ \rho_s } \sum_{n} \left(  \begin{array}{c}
a - b + 1 \\ 2n
\end{array}  \right)   B_{\rho_s^2} \left( n+\tfrac{1}{2},  -b-1   \right) 
\right] 
% &\text{if} \;\;\;\; b \neq -1
 \end{cases}
 \label{eqn:Kab}
\end{align} 
where $\text{Li}_2 (z)$ is the dilogarithm and $B_z ( a, b)$ is the incomplete Beta function. Note that since $K^{ab} = K^{ba}$ is symmetric, one can always choose $b \neq -1$ unless $a=b=-1$. 
For small values of $a$ and $b$ (i.e. small values of $\ell_1$ and $\ell_3$), \eqref{eqn:Kab} produces relatively simple polynomials in $\rho_s$. 

%%%%%
%\subsubsection*{Small $\rho_s$}
%%%%%

%%%%
\subsubsection*{Small $1-\rho_s$}
%%%%
Since the $u$-channel pole also contains a collinear divergence, the expression \eqref{eqn:w1series} is no longer valid. Nonetheless, the rationale that as $\rho_s \to 1$ quantities like $\omega_1^a \omega_3^b$ become infinitesimally supported is still useful, and indeed,
\begin{align}
 K^{ab}_{\ell_1 \ell_3} (\rho_s)  &:= \frac{1}{4}  \int_{-1}^1 dx  \int_{-1}^1 dy \; \left( \frac{\omega_1}{\omega_s} \right)^a \left( \frac{\omega_3}{\omega_s} \right)^b \; P_{\ell_1} ( x ) P_{\ell_3} (y) \; \text{sgn} \left( \frac{x+y}{1+x y} - \frac{2\rho_s}{1+\rho_s^2}  \right)    \nonumber   \\
 &= \rho_-^2  \left[  \frac{1}{(a-1)(b-1)} - 2 \frac{(a-2)! (b-2)!}{ ( a+b-2 )! }   \right] P_{\ell_1} ( 1 ) P_{\ell_3} ( 1 ) + \mathcal{O} ( \rho_-^3 \ell^2 )  \, . 
\end{align}
Note that when $m \neq 0$, the $\phi$ integral vanishes when $x \to 1$ and $y\to 1$, and therefore in the $\rho_s \to 1$ limit the dominant partial waves are the $a^{00}_{\ell_1 \ell_3}$ modes. 

%%%%%%%%
\subsection{Some Explicit Examples}
%%%%%%%%

%%%%
\subsubsection*{$\dot \pi^4$ Interaction}
%%%%

For example, the simple $\dot \pi^4$ amplitude \eqref{eqn:a00b1} can be written in terms of \eqref{eqn:IalF},
\begin{align}
\hat{I}_{\ell} (\rho_s ) = \sqrt{ \frac{4! ( 2 \ell + 1  ) }{4\pi ( 1 - \rho_s^2 ) } }  \left( I^2_{\ell} (\rho_s) - I^3_{\ell} (\rho_s)   \right)  \, . 
\label{eqn:Ihatb1}
\end{align}
In particular, the first subleading term in the small $1-\rho_s$ expansion of \eqref{eqn:IIb1} is,
\begin{align}
 \hat{I}_\ell ( \rho_s )  = \sqrt{ \frac{3 (2 \ell +1 )}{16 \pi} } \sqrt{ 1 - \rho_s  }  & \Big[ 1   \nonumber \\
 & - \frac{ 1 - \rho_s}{4}  \left( 1 + 2 \ell (5 + \ell) - 
   4 \ell (1 + \ell ) ( H_{\ell/2}  + H_{(\ell+1)/2}  + \log (2 - 2 \rho_s ) )  \right)   \nonumber \\
  &+ \mathcal{O} \left(  (1 - \rho_s )^2  \right) \Big]
\end{align}
where $H_n$ is the $n^{\rm th}$ harmonic number, which grows logarithmically at large $n$. When $\ell \gg 1$, it is this logarithmic growth which dominates, and leads to the simple expression \eqref{eqn:Ismallrsb1} which we used to estimate at which $\ell$ the unitarity sum should be truncated.

%%%%
\subsubsection*{All leading order cubic/quartic interactions}
%%%%

Although \eqref{eq:EFT_action_decoupling_intro} contains five Wilson coefficients, the first spherical-wave coefficient depends on only four independent combinations of them, which we can denote,
\begin{align}
C_0 &=  - 27 \alpha_1^2 - 72 \alpha_1 \alpha_2 - 48 \alpha_2^2 + 18 \beta_1 + 24 \beta_2 + 40 \beta_3  \; ,  \\
 C_1  &= \frac{3}{2} \left( 
- 27 \alpha_1^2 + 108 \alpha_1 \alpha_2 + 112 \alpha_2^2 - 27 \beta_1 - 56 \beta_2 - 140 \beta_3 \right)   
 \; ,  \\ 
C_2 &=  \frac{ 2 \alpha_2^2 - \beta_2 - 4 \beta_3 }{300 }  \; , \;\;\;\;
C_3 =  \frac{ - 2 \alpha_2^2 + \beta_2 - 4 \beta_3 }{ 300 }  \,
\end{align}
so that,
\begin{align}
\mathfrak{a}^0_{00} = \frac{1}{192\pi } \left[  C_0 f_0 (\rho_s) + C_1 f_1 (\rho_s) + C_2 f_2 (\rho_s) + C_3 f_3 (\rho_s)   \right]
\end{align}
where the functions $f_j (\rho_s)$ have the convenient property that $f_j (\rho_s) = \rho_s^{2j} + ...$ at small $\rho_s$, and are given explicitly by,
\begin{align}
f_0 &= 1 \;\;\;\; , \;\;\;\;
f_1 = \rho_s^2 \; , \\
f_2 &= \rho_s^4 ( 3 F (1 - \rho_s^2)^2 + 
     5 (1 + \rho_s^2)) (3 F (1 - \rho_s^2)^2 (-3 + 7 \rho_s^2) + 
     5 (-91 + 4 \rho_s^2 + 7 \rho_s^4)) \; , \\
f_3 &= \rho_s^4 (-1 + \rho_s^2) (30 F (-1 + \rho_s^2) (9 + 20 \rho_s^2 + 7 \rho_s^4) + 
     25 (43 + 34 \rho_s^2 + 7 \rho_s^4) + 
     9 F^2 (3 - 2 \rho_s^4 - 8 \rho_s^6 + 7 \rho_s^8))
\end{align}
where $F = - \left( 5 (2 \rho_s (3 + \rho_s^2) + 3 \log ( \frac{1-\rho_s}{1+\rho_s} )  \right)/ 6 \rho_s^5$ has the limit $F (0) = 1$. 

The small $\rho_s$ expansion of the $\ell = 1$ and $\ell =2$ diagonal entries of $\mathfrak{a}^m$ are,
\begin{align}
%\mathfrak{a}^0_{00} &= \frac{-3 \left(3 \alpha _1+4 \alpha _2\right){}^2+18 \beta _1+24 \beta _2+40 \beta_3}{192 \pi }-\frac{\left(27 \alpha _1^2-108 \alpha _2 \alpha _1-112 \alpha _2^2+27   \beta _1+56 \beta _2+140 \beta _3\right) \rho _s^2}{288 \pi }\\
 \mathfrak{a}^{0}_{11} &= - \frac{7 \rho_s^2}{150 \pi} \beta_3 + \frac{\rho_s^2}{192 \pi} \left[ - (3\alpha_1 + 4 \alpha_2 )^2 + 6 \beta_1 + 8 \beta_2 +24 \beta_3    \right] +  ...  \\
 \mathfrak{a}^{\pm 1}_{11} &=  \frac{\rho_s^2}{150 \pi} \beta_3 + ...  \\
 \mathfrak{a}^{0}_{22} &= \frac{\beta_3}{120 \pi}  + \frac{ \rho_s^2 }{10 080 \pi} ( 135 \alpha_1^2 + 112 \alpha_2^2 - 56 \beta_2 - 452 \beta_3 )  + ...  \\
 \mathfrak{a}^{\pm 1}_{22} &= \frac{\beta_3}{120 \pi}  + \frac{ \rho_s^2 }{6720 \pi} ( 45 \alpha_1^2 + 56 \alpha_2^2 - 28 \beta_2 -  272 \beta_3 )    + ...  \\
 \mathfrak{a}^{\pm 2}_{22} &=  \frac{\beta_3}{120 \pi}  - \frac{ \rho_s^2 }{3360 \pi} ( 45 \alpha_1^2 + 92 \beta_3 )  + ... 
\end{align}
which indeed satisfy \eqref{eqn:all_to_al} and reduce to the $a_{\ell}$ of \cite{Baumann:2015nta} when $\rho_s \to 0$.

~\\
This concludes this section on evaluating the partial wave coefficients $a^{mm}_{\ell_1 \ell_3}$. We have successfully broken a general partial wave up into a sum over simple master integrals, $I^a I^b, J^{ab}$ and $K^{ab}$, and given explicit closed expressions for them. Furthermore, when $\rho_s \to 1$, the $\omega_1$ and $\omega_3$ act as $\delta (x-1)$ and $\delta (y-1)$ and their derivatives, allowing us to write particularly compact expressions for the Legendre-weighted integrals $I^a_{\ell_1} I^b_{\ell_3}, J^{ab}_{\ell_1 \ell_3}$ and $K^{ab}_{\ell_1 \ell_3}$ which make up $a^{00}_{\ell_1 \ell_3}$.

%%%%%%%%

%%%%%%%%

%%%%%%%%%%%%%%%%
\section{Unitarity Priors}
\label{app:priors}
%%%%%%%%%%%%%%%%

%%%%%
%\paragraph{CMB Correlations:}
%%%%%

In this short appendix we describe how our unitarity bounds were used to produce Figure~\ref{fig:bispectrum}.
In simple terms, we first translate our allowed region of $\{ \cpi , \alpha_1 \}$ into an allowed region for bispectrum shapes $\{ f_{NL}^{\rm equil} , f_{NL}^{\rm orth} \}$, and then multiply a simple Gaussian likelihood (using Planck's mean and variance) by a prior distribution which is uniform for the allowed $\{ f_{NL}^{\rm equil} , f_{NL}^{\rm orth} \}$ (and zero otherwise). The resulting posterior is used to generate contours within which $68 \%$, $95 \%$ and $99.7\%$ of the total probability lie (preferentially filling the most likely bins first). 
%This produces the contours shown in Figure~\ref{fig:bispectrum}.

The observed limits on the bispectrum reported by the Planck Collaboration \cite{Akrami:2019izv} on equilateral and orthogonal shapes are:
\begin{align}
 f^{\rm eqil }_{NL} =  -26 \pm 47\, , \;\;\;  f^{\rm orth}_{NL} = -38 \pm 24\,, 
\end{align}
from $T+E$ correlations once lensing is subtracted (uncertainties correspond to $68\%$ confidence).  From the EFT of Inflation's scalar sector \eqref{eq:EFT_action_decoupling_intro}, the expected non-Gaussianity with model parameters $\cpi$ and $\alpha_1$ are (using the Planck 2018 Fisher matrix),
\begin{align}
f^{\rm equil}_{NL} &= - 0.156  \alpha_1  - 0.353  \frac{ 1 - \cpi^2 }{\cpi^2}   \\
f^{\rm orth}_{NL} &= 0.0334 \alpha_1 - 0.0008 \frac{1- \cpi^2}{\cpi^2} +  0.0334 \frac{ (1-\cpi^2)^2 }{\cpi^2}  \, . 
\end{align}
We will assume that the uncertainties are Gaussian, such that the likelihood of measuring $(f^{\rm equil}_{NL} , f^{\rm orth}_{NL} )$ when the true values are $( -26, -38 )$ is simply, $L = e^{-\chi^2/2} / (2 \pi \sqrt{ \det \mathtt{C}} )$, where,
\begin{align}
 \chi^2 ( f_{NL}^{\rm equil} , f_{NL}^{\rm orth} )  = \left(   f^{\rm equil}_{NL} + 26 \;\; , \;\; 
  f^{\rm orth}_{NL} + 38
   \right) \mathtt{C}^{-1}  \left( \begin{array}{c}
  f^{\rm equil}_{NL} + 26 \\ 
  f^{\rm orth}_{NL} + 38 
 \end{array} \right)
\end{align}
coincides with the $\chi^2$ used in the analysis of \cite{Akrami:2019izv}, with a covariance matrix given by $\mathtt{C} \approx \text{diag} \left( 47^2  ,  24^2   \right)$. 
Contours of $\chi^2 ( f_{NL}^{\rm equil} , f_{NL}^{\rm orth} ) \leq 2.28,5.99,$ and $11.62$ correspond to an enclosed probability of $68\%$, $95\%$ and $99.7\%$ respectively, and are plotted in Figure~\ref{fig:bispectrum} as ``Planck 2018'', giving constraints on the parameters $(\cpi, \alpha_1)$ from Planck's observations alone. 

The advantage of assuming a simple likelihood, $L$, is that we can use Bayesian inference to incorporate our unitarity priors: once we identify the ``theoretically allowed'' region (in which, say, $\Lambda_{\mathcal{O}} \geq 10 H$), we use a uniform prior  $P( f_{NL}^{\rm equil} , f_{NL}^{\rm orth} )$ which is constant in the allowed region and zero elsewhere.
%\footnote{
%Using a uniform prior in $f_{NL}$ is numerically convenient for integrating the probability distribution, but one could also have chosen a uniform prior in $(\cpi, \alpha_1)$. 
%}. 
The analogous $68\%$, $95\%$ and $99.7\%$ contours of the posterior $\propto L \times \, P ( f_{NL}^{\rm equil} , f_{NL}^{\rm orth} )$ are then plotted in Figure~\ref{fig:bispectrum}. 
Demanding that perturbative unitarity is not violated in the subhorizon regime \eqref{eqn:subhorizon} improves the volume of the $68 \%$ confidence interval in $( f_{NL}^{\rm equil} , f_{NL}^{\rm orth} )$ by a factor of $\approx 3$. 

For comparison, Simons Observatory (due to begin taking data in early 2020s) has a goal sensitivity of $\sigma(f_{NL}^{\rm equil})=\pm 24$ and $\sigma(f_{NL}^{\rm orth})\pm 13$ (compared with Planck's $\pm 47$ and $\pm 24$) \cite{Ade:2018sbj}. This corresponds to improving the volume of the $68\%$ confidence interval by a factor of $\approx 3.6$, versus Planck 2018 data alone. 
Similarly, the later CMB-S4
%(late 2020s at least)
experiment aims to reduce the uncertainty to $\sigma(f_{NL}^{\rm equil})=\pm 21$ and $\sigma(f_{NL}^{\rm orth})\pm 9$ \cite{Abazajian:2016yjj, Abazajian:2019eic}, reducing Planck's 68$\%$ confidence interval by a factor of
%\footnote{ 
%This is consistent with the product of the relative improvements in $f_{NL}^{\rm equil}$ and $f_{NL}^{\rm orth}$ quoted separately in Table 2-2 of \cite{Abazajian:2016yjj}. 
%} 
$\approx 6.0$. 
The provides context for the improvement that we have achieved here using the Planck 2018 data supplemented with theoretical priors (the most conservative unitarity prior shown in Figure~\ref{fig:5} corresponds to a factor of $\approx 3$ improvement). Going forwards, the theoretical priors identified here can also be combined with future data sets to further improve our understanding of primordial non-Gaussianity. 
% subsection priors_on_cmb_data (end)

%%%%%%%%%%%%%%%%%%%%%%%%%%%%%%%%
\bibliographystyle{JHEP}
\bibliography{multifield_positivity}

\providecommand{\href}[2]{#2}\begingroup\raggedright\begin{thebibliography}{10}

\bibitem{Chew}
G.~F. Chew, \emph{{$S$-matrix theory of strong interactions}}.
\newblock Benjamin, New York, 1961.

\bibitem{Eden}
{R. J. Eden, P. V. Landshoff, D. I. Olive, and J. C. Polkinghorne},
  \emph{{$S$-matrix theory of strong interactions'}}.
\newblock Cambridge University Press, 1966.

\bibitem{Lee:1977eg}
B.~W. Lee, C.~Quigg and H.~Thacker, \emph{{Weak Interactions at Very
  High-Energies: The Role of the Higgs Boson Mass}},
  \href{http://dx.doi.org/10.1103/PhysRevD.16.1519}{\emph{Phys. Rev. D}
  {\bfseries 16} (1977) 1519}.

\bibitem{Cheung:2007st}
C.~Cheung, P.~Creminelli, A.~L. Fitzpatrick, J.~Kaplan and L.~Senatore,
  \emph{{The Effective Field Theory of Inflation}},
  \href{http://dx.doi.org/10.1088/1126-6708/2008/03/014}{\emph{JHEP} {\bfseries
  03} (2008) 014}, [\href{https://arxiv.org/abs/0709.0293}{{\ttfamily
  0709.0293}}].

\bibitem{Chen:2006nt}
X.~Chen, M.-x. Huang, S.~Kachru and G.~Shiu, \emph{{Observational signatures
  and non-Gaussianities of general single field inflation}},
  \href{http://dx.doi.org/10.1088/1475-7516/2007/01/002}{\emph{JCAP} {\bfseries
  01} (2007) 002}, [\href{https://arxiv.org/abs/hep-th/0605045}{{\ttfamily
  hep-th/0605045}}].

\bibitem{Cheung:2007sv}
C.~Cheung, A.~L. Fitzpatrick, J.~Kaplan and L.~Senatore, \emph{{On the
  consistency relation of the 3-point function in single field inflation}},
  \href{http://dx.doi.org/10.1088/1475-7516/2008/02/021}{\emph{JCAP} {\bfseries
  0802} (2008) 021}, [\href{https://arxiv.org/abs/0709.0295}{{\ttfamily
  0709.0295}}].

\bibitem{Huang:2006eha}
X.~Chen, M.-x. Huang and G.~Shiu, \emph{{The Inflationary Trispectrum for
  Models with Large Non-Gaussianities}},
  \href{http://dx.doi.org/10.1103/PhysRevD.74.121301}{\emph{Phys.\ Rev.\ D}
  {\bfseries 74} (2006) 121301},
  [\href{https://arxiv.org/abs/hep-th/0610235}{{\ttfamily hep-th/0610235}}].

\bibitem{Aghanim:2018eyx}
{\scshape Planck} collaboration, N.~Aghanim et~al., \emph{{Planck 2018 results.
  VI. Cosmological parameters}},
  \href{https://arxiv.org/abs/1807.06209}{{\ttfamily 1807.06209}}.

\bibitem{Birrell:1982ix}
N.~Birrell and P.~Davies, \emph{{Quantum Fields in Curved Space}}.
\newblock Cambridge Monographs on Mathematical Physics. Cambridge Univ. Press,
  Cambridge, UK, 2, 1984,
  \href{http://dx.doi.org/10.1017/CBO9780511622632}{10.1017/CBO9780511622632}.

\bibitem{Shandera:2008ai}
S.~Shandera, \emph{{The structure of correlation functions in single field
  inflation}}, \href{http://dx.doi.org/10.1103/PhysRevD.79.123518}{\emph{Phys.
  Rev.} {\bfseries D79} (2009) 123518},
  [\href{https://arxiv.org/abs/0812.0818}{{\ttfamily 0812.0818}}].

\bibitem{Leblond:2008gg}
L.~Leblond and S.~Shandera, \emph{{Simple Bounds from the Perturbative Regime
  of Inflation}},
  \href{http://dx.doi.org/10.1088/1475-7516/2008/08/007}{\emph{JCAP} {\bfseries
  0808} (2008) 007}, [\href{https://arxiv.org/abs/0802.2290}{{\ttfamily
  0802.2290}}].

\bibitem{ArmendarizPicon:2008yv}
C.~Armendariz-Picon, M.~Fontanini, R.~Penco and M.~Trodden, \emph{{Where does
  Cosmological Perturbation Theory Break Down?}},
  \href{http://dx.doi.org/10.1088/0264-9381/26/18/185002}{\emph{Class. Quant.
  Grav.} {\bfseries 26} (2009) 185002},
  [\href{https://arxiv.org/abs/0805.0114}{{\ttfamily 0805.0114}}].

\bibitem{Baumann:2011su}
D.~Baumann and D.~Green, \emph{{Equilateral Non-Gaussianity and New Physics on
  the Horizon}},
  \href{http://dx.doi.org/10.1088/1475-7516/2011/09/014}{\emph{JCAP} {\bfseries
  1109} (2011) 014}, [\href{https://arxiv.org/abs/1102.5343}{{\ttfamily
  1102.5343}}].

\bibitem{Assassi:2013gxa}
V.~Assassi, D.~Baumann, D.~Green and L.~McAllister, \emph{{Planck-Suppressed
  Operators}},
  \href{http://dx.doi.org/10.1088/1475-7516/2014/01/033}{\emph{JCAP} {\bfseries
  01} (2014) 033}, [\href{https://arxiv.org/abs/1304.5226}{{\ttfamily
  1304.5226}}].

\bibitem{Cannone:2014qna}
D.~Cannone, N.~Bartolo and S.~Matarrese, \emph{{Perturbative Unitarity of
  Inflationary Models with Features}},
  \href{http://dx.doi.org/10.1103/PhysRevD.89.127301}{\emph{Phys. Rev. D}
  {\bfseries 89} (2014) 127301},
  [\href{https://arxiv.org/abs/1402.2258}{{\ttfamily 1402.2258}}].

\bibitem{Adshead:2014sga}
P.~Adshead and W.~Hu, \emph{{Bounds on nonadiabatic evolution in single-field
  inflation}}, \href{http://dx.doi.org/10.1103/PhysRevD.89.083531}{\emph{Phys.
  Rev. D} {\bfseries 89} (2014) 083531},
  [\href{https://arxiv.org/abs/1402.1677}{{\ttfamily 1402.1677}}].

\bibitem{Koehn:2015vvy}
M.~Koehn, J.-L. Lehners and B.~Ovrut, \emph{{Nonsingular bouncing cosmology:
  Consistency of the effective description}},
  \href{http://dx.doi.org/10.1103/PhysRevD.93.103501}{\emph{Phys. Rev. D}
  {\bfseries 93} (2016) 103501},
  [\href{https://arxiv.org/abs/1512.03807}{{\ttfamily 1512.03807}}].

\bibitem{deRham:2017aoj}
C.~de~Rham and S.~Melville, \emph{{Unitary null energy condition violation in
  P(X) cosmologies}},
  \href{http://dx.doi.org/10.1103/PhysRevD.95.123523}{\emph{Phys. Rev.}
  {\bfseries D95} (2017) 123523},
  [\href{https://arxiv.org/abs/1703.00025}{{\ttfamily 1703.00025}}].

\bibitem{Georgi:1989xy}
H.~Georgi, \emph{{Vector Realization of Chiral Symmetry}},
  \href{http://dx.doi.org/10.1016/0550-3213(90)90210-5}{\emph{Nucl. Phys.}
  {\bfseries B331} (1990) 311--330}.

\bibitem{Manohar:2018aog}
A.~V. Manohar, \emph{{Introduction to Effective Field Theories}},  in
  \emph{{Les Houches summer school}: {EFT in Particle Physics and Cosmology}},
  4, 2018.
\newblock \href{https://arxiv.org/abs/1804.05863}{{\ttfamily 1804.05863}}.

\bibitem{Adshead:2017srh}
P.~Adshead, C.~P. Burgess, R.~Holman and S.~Shandera, \emph{{Power-counting
  during single-field slow-roll inflation}},
  \href{http://dx.doi.org/10.1088/1475-7516/2018/02/016}{\emph{JCAP} {\bfseries
  1802} (2018) 016}, [\href{https://arxiv.org/abs/1708.07443}{{\ttfamily
  1708.07443}}].

\bibitem{Babic:2019ify}
I.~Babic, C.~P. Burgess and G.~Geshnizjani, \emph{{Keeping an Eye on DBI:
  Power-counting for small-$c_s$ Cosmology}},
  \href{https://arxiv.org/abs/1910.05277}{{\ttfamily 1910.05277}}.

\bibitem{Colladay:1998fq}
D.~Colladay and V.~Kostelecky, \emph{{Lorentz violating extension of the
  standard model}},
  \href{http://dx.doi.org/10.1103/PhysRevD.58.116002}{\emph{Phys. Rev. D}
  {\bfseries 58} (1998) 116002},
  [\href{https://arxiv.org/abs/hep-ph/9809521}{{\ttfamily hep-ph/9809521}}].

\bibitem{Kostelecky:2003fs}
V.~Kostelecky, \emph{{Gravity, Lorentz violation, and the standard model}},
  \href{http://dx.doi.org/10.1103/PhysRevD.69.105009}{\emph{Phys. Rev. D}
  {\bfseries 69} (2004) 105009},
  [\href{https://arxiv.org/abs/hep-th/0312310}{{\ttfamily hep-th/0312310}}].

\bibitem{Kostelecky:2000mm}
V.~Kostelecky and R.~Lehnert, \emph{{Stability, causality, and Lorentz and CPT
  violation}}, \href{http://dx.doi.org/10.1103/PhysRevD.63.065008}{\emph{Phys.
  Rev. D} {\bfseries 63} (2001) 065008},
  [\href{https://arxiv.org/abs/hep-th/0012060}{{\ttfamily hep-th/0012060}}].

\bibitem{Giudice:2007fh}
G.~F. Giudice, C.~Grojean, A.~Pomarol and R.~Rattazzi, \emph{{The
  Strongly-Interacting Light Higgs}},
  \href{http://dx.doi.org/10.1088/1126-6708/2007/06/045}{\emph{JHEP} {\bfseries
  06} (2007) 045}, [\href{https://arxiv.org/abs/hep-ph/0703164}{{\ttfamily
  hep-ph/0703164}}].

\bibitem{Manohar:1983md}
A.~Manohar and H.~Georgi, \emph{{Chiral Quarks and the Nonrelativistic Quark
  Model}}, \href{http://dx.doi.org/10.1016/0550-3213(84)90231-1}{\emph{Nucl.
  Phys.} {\bfseries B234} (1984) 189--212}.

\bibitem{Cohen:1997rt}
A.~G. Cohen, D.~B. Kaplan and A.~E. Nelson, \emph{{Counting 4 pis in strongly
  coupled supersymmetry}},
  \href{http://dx.doi.org/10.1016/S0370-2693(97)00995-7}{\emph{Phys. Lett.}
  {\bfseries B412} (1997) 301--308},
  [\href{https://arxiv.org/abs/hep-ph/9706275}{{\ttfamily hep-ph/9706275}}].

\bibitem{Jenkins:2013sda}
E.~E. Jenkins, A.~V. Manohar and M.~Trott, \emph{{Naive Dimensional Analysis
  Counting of Gauge Theory Amplitudes and Anomalous Dimensions}},
  \href{http://dx.doi.org/10.1016/j.physletb.2013.09.020}{\emph{Phys. Lett.}
  {\bfseries B726} (2013) 697--702},
  [\href{https://arxiv.org/abs/1309.0819}{{\ttfamily 1309.0819}}].

\bibitem{Buchalla:2013eza}
G.~Buchalla, O.~Catá and C.~Krause, \emph{{On the Power Counting in Effective
  Field Theories}},
  \href{http://dx.doi.org/10.1016/j.physletb.2014.02.015}{\emph{Phys. Lett.}
  {\bfseries B731} (2014) 80--86},
  [\href{https://arxiv.org/abs/1312.5624}{{\ttfamily 1312.5624}}].

\bibitem{Weinberg:1978kz}
S.~Weinberg, \emph{{Phenomenological Lagrangians}},
  \href{http://dx.doi.org/10.1016/0378-4371(79)90223-1}{\emph{Physica}
  {\bfseries A96} (1979) 327--340}.

\bibitem{Burgess:2007pt}
C.~P. Burgess, \emph{{Introduction to Effective Field Theory}},
  \href{http://dx.doi.org/10.1146/annurev.nucl.56.080805.140508}{\emph{Ann.
  Rev. Nucl. Part. Sci.} {\bfseries 57} (2007) 329--362},
  [\href{https://arxiv.org/abs/hep-th/0701053}{{\ttfamily hep-th/0701053}}].

\bibitem{Goon:2016ihr}
G.~Goon, K.~Hinterbichler, A.~Joyce and M.~Trodden, \emph{{Aspects of Galileon
  Non-Renormalization}},
  \href{http://dx.doi.org/10.1007/JHEP11(2016)100}{\emph{JHEP} {\bfseries 11}
  (2016) 100}, [\href{https://arxiv.org/abs/1606.02295}{{\ttfamily
  1606.02295}}].

\bibitem{Gwyn:2012mw}
R.~Gwyn, G.~A. Palma, M.~Sakellariadou and S.~Sypsas, \emph{{Effective field
  theory of weakly coupled inflationary models}},
  \href{http://dx.doi.org/10.1088/1475-7516/2013/04/004}{\emph{JCAP} {\bfseries
  04} (2013) 004}, [\href{https://arxiv.org/abs/1210.3020}{{\ttfamily
  1210.3020}}].

\bibitem{Pajer:2016ieg}
E.~Pajer, G.~L. Pimentel and J.~V.~S. Van~Wijck, \emph{{The Conformal Limit of
  Inflation in the Era of CMB Polarimetry}},
  \href{http://dx.doi.org/10.1088/1475-7516/2017/06/009}{\emph{JCAP} {\bfseries
  06} (2017) 009}, [\href{https://arxiv.org/abs/1609.06993}{{\ttfamily
  1609.06993}}].

\bibitem{Creminelli:2006xe}
P.~Creminelli, M.~A. Luty, A.~Nicolis and L.~Senatore, \emph{{Starting the
  Universe: Stable Violation of the Null Energy Condition and Non-standard
  Cosmologies}},
  \href{http://dx.doi.org/10.1088/1126-6708/2006/12/080}{\emph{JHEP} {\bfseries
  12} (2006) 080}, [\href{https://arxiv.org/abs/hep-th/0606090}{{\ttfamily
  hep-th/0606090}}].

\bibitem{Piazza:2013coa}
F.~Piazza and F.~Vernizzi, \emph{{Effective Field Theory of Cosmological
  Perturbations}},
  \href{http://dx.doi.org/10.1088/0264-9381/30/21/214007}{\emph{Class. Quant.
  Grav.} {\bfseries 30} (2013) 214007},
  [\href{https://arxiv.org/abs/1307.4350}{{\ttfamily 1307.4350}}].

\bibitem{Delacretaz:2015edn}
L.~V. Delacretaz, T.~Noumi and L.~Senatore, \emph{{Boost Breaking in the EFT of
  Inflation}},
  \href{http://dx.doi.org/10.1088/1475-7516/2017/02/034}{\emph{JCAP} {\bfseries
  1702} (2017) 034}, [\href{https://arxiv.org/abs/1512.04100}{{\ttfamily
  1512.04100}}].

\bibitem{Shapere:2012nq}
A.~Shapere and F.~Wilczek, \emph{{Classical Time Crystals}},
  \href{http://dx.doi.org/10.1103/PhysRevLett.109.160402}{\emph{Phys. Rev.
  Lett.} {\bfseries 109} (2012) 160402},
  [\href{https://arxiv.org/abs/1202.2537}{{\ttfamily 1202.2537}}].

\bibitem{Wilczek:2012jt}
F.~Wilczek, \emph{{Quantum Time Crystals}},
  \href{http://dx.doi.org/10.1103/PhysRevLett.109.160401}{\emph{Phys. Rev.
  Lett.} {\bfseries 109} (2012) 160401},
  [\href{https://arxiv.org/abs/1202.2539}{{\ttfamily 1202.2539}}].

\bibitem{Castillo:2013sfa}
E.~Castillo, B.~Koch and G.~Palma, \emph{{On the integration of fields and
  quanta in time dependent backgrounds}},
  \href{http://dx.doi.org/10.1007/JHEP05(2014)111}{\emph{JHEP} {\bfseries 05}
  (2014) 111}, [\href{https://arxiv.org/abs/1312.3338}{{\ttfamily 1312.3338}}].

\bibitem{Baumann:2014cja}
D.~Baumann, D.~Green and R.~A. Porto, \emph{{B-modes and the Nature of
  Inflation}},
  \href{http://dx.doi.org/10.1088/1475-7516/2015/01/016}{\emph{JCAP} {\bfseries
  1501} (2015) 016}, [\href{https://arxiv.org/abs/1407.2621}{{\ttfamily
  1407.2621}}].

\bibitem{Richman:1984gh}
J.~D. Richman, \emph{{An Experimenter's Guide to the Helicity Formalism}}, .

\bibitem{Bartolo:2010di}
N.~Bartolo, M.~Fasiello, S.~Matarrese and A.~Riotto, \emph{{Large
  non-Gaussianities in the Effective Field Theory Approach to Single-Field
  Inflation: the Trispectrum}},
  \href{http://dx.doi.org/10.1088/1475-7516/2010/09/035}{\emph{JCAP} {\bfseries
  1009} (2010) 035}, [\href{https://arxiv.org/abs/1006.5411}{{\ttfamily
  1006.5411}}].

\bibitem{Smith:2015uia}
K.~M. Smith, L.~Senatore and M.~Zaldarriaga, \emph{{Optimal analysis of the CMB
  trispectrum}},  \href{https://arxiv.org/abs/1502.00635}{{\ttfamily
  1502.00635}}.

\bibitem{Akrami:2019izv}
{\scshape Planck} collaboration, Y.~Akrami et~al., \emph{{Planck 2018 results.
  IX. Constraints on primordial non-Gaussianity}},
  \href{https://arxiv.org/abs/1905.05697}{{\ttfamily 1905.05697}}.

\bibitem{Ade:2018sbj}
{\scshape Simons Observatory} collaboration, P.~Ade et~al., \emph{{The Simons
  Observatory: Science goals and forecasts}},
  \href{http://dx.doi.org/10.1088/1475-7516/2019/02/056}{\emph{JCAP} {\bfseries
  1902} (2019) 056}, [\href{https://arxiv.org/abs/1808.07445}{{\ttfamily
  1808.07445}}].

\bibitem{Abazajian:2016yjj}
{\scshape CMB-S4} collaboration, K.~N. Abazajian et~al., \emph{{CMB-S4 Science
  Book, First Edition}},  \href{https://arxiv.org/abs/1610.02743}{{\ttfamily
  1610.02743}}.

\bibitem{Abazajian:2019eic}
K.~Abazajian et~al., \emph{{CMB-S4 Science Case, Reference Design, and Project
  Plan}},  \href{https://arxiv.org/abs/1907.04473}{{\ttfamily 1907.04473}}.

\bibitem{Baumann:2015nta}
D.~Baumann, D.~Green, H.~Lee and R.~A. Porto, \emph{{Signs of Analyticity in
  Single-Field Inflation}},
  \href{http://dx.doi.org/10.1103/PhysRevD.93.023523}{\emph{Phys. Rev.}
  {\bfseries D93} (2016) 023523},
  [\href{https://arxiv.org/abs/1502.07304}{{\ttfamily 1502.07304}}].

\bibitem{Gubitosi:2012hu}
G.~Gubitosi, F.~Piazza and F.~Vernizzi, \emph{{The Effective Field Theory of
  Dark Energy}},
  \href{http://dx.doi.org/10.1088/1475-7516/2013/02/032}{\emph{JCAP} {\bfseries
  02} (2013) 032}, [\href{https://arxiv.org/abs/1210.0201}{{\ttfamily
  1210.0201}}].

\bibitem{Nicolis:2015sra}
A.~Nicolis, R.~Penco, F.~Piazza and R.~Rattazzi, \emph{{Zoology of condensed
  matter: Framids, ordinary stuff, extra-ordinary stuff}},
  \href{http://dx.doi.org/10.1007/JHEP06(2015)155}{\emph{JHEP} {\bfseries 06}
  (2015) 155}, [\href{https://arxiv.org/abs/1501.03845}{{\ttfamily
  1501.03845}}].

\bibitem{Grall:2020ibl}
T.~Grall, S.~Jazayeri and D.~Stefanyszyn, \emph{{The Cosmological Phonon:
  Symmetries and Amplitudes on Sub-Horizon Scales}},
  \href{https://arxiv.org/abs/2005.12937}{{\ttfamily 2005.12937}}.

\bibitem{Adams:2006sv}
A.~Adams, N.~Arkani-Hamed, S.~Dubovsky, A.~Nicolis and R.~Rattazzi,
  \emph{{Causality, analyticity and an IR obstruction to UV completion}},
  \href{http://dx.doi.org/10.1088/1126-6708/2006/10/014}{\emph{JHEP} {\bfseries
  10} (2006) 014}, [\href{https://arxiv.org/abs/hep-th/0602178}{{\ttfamily
  hep-th/0602178}}].

\bibitem{Grall2}
T.~Grall and S.~Melville, \emph{{Positivity Bounds with Spontaneously Broken
  Boosts}}.
\newblock To appear.

\bibitem{Maldacena:2011nz}
J.~M. Maldacena and G.~L. Pimentel, \emph{{On graviton non-Gaussianities during
  inflation}}, \href{http://dx.doi.org/10.1007/JHEP09(2011)045}{\emph{JHEP}
  {\bfseries 09} (2011) 045},
  [\href{https://arxiv.org/abs/1104.2846}{{\ttfamily 1104.2846}}].

\bibitem{Raju:2012zr}
S.~Raju, \emph{{New Recursion Relations and a Flat Space Limit for AdS/CFT
  Correlators}},
  \href{http://dx.doi.org/10.1103/PhysRevD.85.126009}{\emph{Phys. Rev.}
  {\bfseries D85} (2012) 126009},
  [\href{https://arxiv.org/abs/1201.6449}{{\ttfamily 1201.6449}}].

\bibitem{Arkani-Hamed:2015bza}
N.~Arkani-Hamed and J.~Maldacena, \emph{{Cosmological Collider Physics}},
  \href{https://arxiv.org/abs/1503.08043}{{\ttfamily 1503.08043}}.

\bibitem{Arkani-Hamed:2018kmz}
N.~Arkani-Hamed, D.~Baumann, H.~Lee and G.~L. Pimentel, \emph{{The Cosmological
  Bootstrap: Inflationary Correlators from Symmetries and Singularities}},
  \href{https://arxiv.org/abs/1811.00024}{{\ttfamily 1811.00024}}.

\end{thebibliography}\endgroup
%%%%%%%%%%%%%%%%%%%%%%%%%%%%%%%%

\end{document}